\documentclass[11pt]{article}
\usepackage[utf8]{inputenc}
\usepackage{amsfonts}
\usepackage{amsmath}
\usepackage{amssymb}
\usepackage{indentfirst}
\usepackage{graphicx}
\usepackage[colorlinks]{hyperref}
\usepackage{cite}
\usepackage{colortbl}

\usepackage{microtype}
\usepackage[normalem]{ulem}

\numberwithin{equation}{section}
\allowdisplaybreaks

\begin{document}

\baselineskip=18pt 
\baselineskip 0.6cm
\begin{titlepage}
\vskip 4cm

\begin{center}
\textbf{\LARGE{Non-relativistic and Ultra-relativistic Expansions of Three-dimensional Spin-3 Gravity Theories}}
\par\end{center}{\LARGE \par}

\begin{center}
	\vspace{1cm}
	\textbf{Patrick Concha}$^{\ast}$,
	\textbf{Carla Henríquez-Baez}$^{\ast}$
	\textbf{Evelyn Rodríguez}$^{\ast}$,
	\small
	\\[6mm]
	$^{\ast}$\textit{Departamento de Matemática y Física Aplicadas, }\\
	\textit{ Universidad Católica de la Santísima Concepción, }\\
\textit{ Alonso de Ribera 2850, Concepción, Chile.}
	\\[5mm]
	\footnotesize
	\texttt{patrick.concha@ucsc.cl},
	\texttt{carlah.baez@gmail.com},
	\texttt{erodriguez@ucsc.cl}
	\par\end{center}
\vskip 20pt
\centerline{{\bf Abstract}}
\medskip
\noindent  
In this paper, we present novel and known non-relativistic and ultra-relativistic spin-3 algebras, by considering the Lie algebra expansion method. We start by applying the expansion procedure using different semigroups to the spin-3 extension of the AdS algebra, leading to spin-3 extensions of known non-relativistic and ultra-relativistic algebras. We then generalize the procedure considering an infinite-dimensional semigroup, which allows to obtain a spin-3 extension of two new infinite families of the Newton-Hooke type and AdS Carroll type. We also present the construction of the gravity theories based on the aforementioned algebras. In particular, the expansion method based on semigroups also allows to derive the (non-degenerate) invariant bilinear forms, ensuring the proper construction of the Chern-Simons gravity actions. Interestingly, in the vanishing cosmological constant limit we recover the spin-3 extensions of the infinite-dimensional Galilean and infinite-dimensional Carroll gravity theories.
\end{titlepage}\newpage {\baselineskip=12pt \tableofcontents{}}

\section{Introduction}\label{sec1}
Non- and ultra-relativistic symmetries have received a growing interest due to their utilities in diverse physical theories. On one hand, non-relativistic symmetries appear in the context of holography \cite{Son:2008ye,Balasubramanian:2008dm,Kachru:2008yh,Taylor:2008tg,Bagchi:2009my,Hartnoll:2009sz,Bagchi:2009pe,Christensen:2013lma,Christensen:2013rfa,Hartong:2014oma,Hartong:2014pma,Zaanen:2015oix}, Ho\v{r}ava-Lifshitz gravity \cite{Horava:2009uw,Hartong:2015wxa,Hartong:2015zia,Taylor:2015glc,Hartong:2016yrf,Devecioglu:2018apj}, effective field theory description of the quantum Hall effect \cite{Hoyos:2011ez,Son:2013rqa,Abanov:2014ula,Geracie:2014nka,Gromov:2015fda}, among others. On the other hand, ultra-relativistic symmetries of the Carrollian type have found recent applications in the study of tachyon condensation \cite{Gibbons:2002tv}, warped conformal field theories \cite{Hofman:2014loa}, tensionless strings \cite{Bagchi:2013bga, Bagchi:2015nca, Bagchi:2016yyf, Bagchi:2017cte, Bagchi:2018wsn}, holography in asymptotically flat space-times \cite{Barnich:2010eb,Barnich:2012aw,Bagchi:2010zz,Hartong:2015xda,Hartong:2015usd,Bagchi:2016bcd,Donnay:2022aba,Saha:2022gjw}, asymptotic symmetries \cite{Perez:2021abf,Perez:2022jpr,Fuentealba:2022gdx} and in the context of soft hair on black hole horizons \cite{Hawking:2016msc,Hawking:2016sgy}.

The formulation of a non-relativistic gravity theory requires to consider the Newton-Cartan formalism \cite{Cartan1,Cartan2}. Newton-Cartan gravity theories along its diverse extensions have been studied in different contexts in \cite{Duval:1983pb,Duval:1984cj,Duval:2009vt,Andringa:2010it,Banerjee:2014nja,Banerjee:2016laq,Bergshoeff:2017dqq,Aviles:2019xed,Chernyavsky:2019hyp,Concha:2019lhn,Harmark:2019upf,Hansen:2020pqs,Ergen:2020yop,Kasikci:2020qsj}. In three spacetime dimensions, the construction of a gauge-invariant action for a given non-relativistic symmetry is possible through the Chern-Simons (CS) formalism. The CS formalism not only offers us a straightforward way to construct an action for a given symmetry but also can be seen as a toy model to approach higher-dimensional gravity models. Nevertheless, in the non-relativistic limit ($c\rightarrow\infty$) there might appear infinities and degeneracy. Such difficulties can be avoided by considering $U(1)$-enlargements of the relativistic symmetry. Although the non-relativistic limit of the Poincaré algebra leads to the Galilei algebra \cite{Bacry:1968zf}, it is necessary to add two $\mathfrak{u}\left(1\right)$ generators to Poincaré in order to avoid degeneracy and construct a proper non-relativistic CS action. In such case, the non-relativistic symmetry corresponds to an extension of the so-called Bargmann algebra \cite{Grigore:1993fz,Bose:1994sj,Duval:2000xr,Jackiw:2000tz,Papageorgiou:2009zc} which consist in a central extension of the Galilei one. Further extensions of the extended Bargmann gravity theory have been largely explored by diverse authors in the context of enlarged symmetries \cite{Aviles:2018jzw,Penafiel:2019czp,Bergshoeff:2020fiz,Concha:2020sjt,Concha:2020ebl} and supersymmetry \cite{Andringa:2013mma,Bergshoeff:2015ija,Bergshoeff:2016lwr,Ozdemir:2019orp,deAzcarraga:2019mdn,Ozdemir:2019tby,Concha:2019mxx,Concha:2020eam,Concha:2021llq}.

At the ultra-relativistic level, Carroll gravity theories have been discussed in \cite{Bergshoeff:2017btm,Matulich:2019cdo,Henneaux:2021yzg,Hansen:2021fxi,Bagchi:2022owq,Ekiz:2022wbi}. A CS formulation of Carroll gravity was first discussed in \cite{Matulich:2019cdo} and then extended to supersymmetry in \cite{Ravera:2019ize,Ali:2019jjp}. Unlike non-relativistic CS gravity theories, Carroll and AdS Carroll CS actions do not suffer from degeneracy in the ultra-relativistic limit ($c\rightarrow 0$) and then do not require the introduction of additional generators in the ultra-relativistic algebra\footnote{The degeneracy may occur for enlarged symmetries. As it was noticed in \cite{Concha:2021jnn}, a Carrollian version of the Maxwell gravity requires to extend the Maxwellian Carroll algebra in order to avoid degeneracy.}. More recently, Carrollian versions of the Jackiw-Teitelboim (super)gravity model have been presented in \cite{Grumiller:2020elf,Gomis:2020wxp,Ravera:2022buz}.

On the other hand, the exploration of theories involving massless higher-spin fields coupled to gravity has been studied in \cite{Vasiliev:1990en}. In the last years, gauge theories with spin $>2$ have received a growing interest mainly due to its applications in the AdS/CFT correspondence context \cite{Sundborg:2000wp,Klebanov:2002ja,Sezgin:2002rt,Sorokin:2004ie,Bekaert:2012ux,Gaberdiel:2012uj,Giombi:2012ms,Gaberdiel:2014cha,Rahman:2015pzl,Giombi:2016ejx}. Moreover in three spacetime dimensions, a consistent coupling of massless higher-spin fields to gravity is possible considering a CS action for the $\mathfrak{sl}\left(n,\mathbb{R}\right)\times\mathfrak{sl}\left(n,\mathbb{R}\right)$ algebra \cite{Blencowe:1988gj,Bergshoeff:1989ns,Henneaux:2010xg}. Such theory contains rich boundary dynamics whose asymptotic symmetry is given by two copies of the $\mathcal{W}_{n}$ algebra \cite{Campoleoni:2010zq,Gaberdiel:2010ar,Campoleoni:2011hg} being an infinite-dimensional extension of the relativistic $\mathfrak{sl}\left(n,\mathbb{R}\right)$ algebra. Interestingly, as it was shown in \cite{Campoleoni:2010zq,Caroca:2017izc}, the CS formalism allows to consider a finite number of interacting higher-spin fields. 

Higher-spin gravity models share many properties with the pure gravity theory, as black hole solutions \cite{Gutperle:2011kf,Perez:2012cf,Perez:2013xi,Compere:2013gja,Bunster:2014mua,Banados:2015tft,Grumiller:2016kcp,Banados:2016hze}, conical singularities \cite{Castro:2016ehj,Campoleoni:2013iha}, among others. More recently, a first approach to explore non- and ultra-relativistic versions of spin-3 gravity theories in absence of cosmological constant was presented in \cite{Bergshoeff:2016soe}. However, the spin-3 Galilei symmetry suffers from degeneracy which can be overcome by considering a spin-3 extension of the extended Bargmann algebra. Nonetheless, the spin-3 Carroll algebra admits non-degenerate invariant bilinear trace as it spin-2 version. Although the spin-3 versions of the kinematical algebras with cosmological constant have been presented in \cite{Bergshoeff:2016soe}, a consistent spin-3 gravity action for such symmetries remains unknown. 

In this paper, motivated by the several applications of the higher-spin theories as well as non- and ultra-relativistic symmetries, we explore the construction of novel non- and ultra-relativistic spin-3 gravity theories in three spacetime dimensions. We first present new non- and ultra-relativistic spin-3 algebras in presence of a cosmological constant. To this end, we apply the Lie algebra expansion method\footnote{The algebra expansion procedure has been first introduced in \cite{Hatsuda:2001pp} and then subsequently developed in \cite{deAzcarraga:2002xi,deAzcarraga:2007et} considering the Maurer-Cartan forms.} based on semigroups \cite{Izaurieta:2006zz,Caroca:2011qs,Andrianopoli:2013ooa,Artebani:2016gwh,Ipinza:2016bfc,Penafiel:2016ufo,Inostroza:2018gzd} to the relativistic spin-3 AdS algebra. Here, the semigroup acts as a non- or ultra-relativistic expansion of a relativistic algebra. In the flat limit, we show that the spin-3 algebras obtained in \cite{Bergshoeff:2016soe} are recovered. We then extend our procedure to spin-3 extensions of post-Newtonian and Extended Carroll symmetries. We generalize our results by presenting a spin-3 infinite-dimensional Newton-Hooke and spin-3 infinite-dimensional AdS Carroll algebras. Remarkably, in the vanishing cosmological constant limit they reproduce a spin-3 extension of the infinite-dimensional Galilean and infinite-dimensional Carroll algebras discussed in \cite{Gomis:2019fdh,Gomis:2019nih,Gomis:2022spp,Concha:2022you}. It is important to clarify that the infinite-dimensional extensions of non- and ultra-relativistic spin-3 symmetries obtained here are not the respective asymptotic symmetries of a non- or ultra-relativistic spin-3 CS gravity theory defined on three spacetime dimensions. Nonetheless, the expansion procedure considered here could be useful to address such issue analogously to the one considered in \cite{Caroca:2019dds} for obtaining supersymmetric extension of known asymptotic symmetries. However, we leave this problem for a future work and only discuss the general idea in the concluding section.

The construction of gauge-invariant CS action for the finite and infinite-dimensional spin-3 non- and ultra-relativistic symmetries is also presented. In particular, our results can be seen as the spin-3 version of several known non- and ultra-relativistic gravity theories. Indeed one can argue that we obtain the spin-3 extensions of the extended Newton-Hooke gravity \cite{Aldrovandi:1998im,Gibbons:2003rv,Brugues:2006yd,Alvarez:2007fw,Papageorgiou:2010ud,Duval:2011mi,Hartong:2016yrf,Duval:2016tzi}, enhanced Bargmann-Newton-Hooke gravity \cite{Gomis:2019nih,Concha:2019dqs,Bergshoeff:2020fiz}, AdS Carroll gravity \cite{Matulich:2019cdo,Ali:2019jjp,Ravera:2019ize} and their respective flat limits \cite{Bergshoeff:2016lwr,Ozdemir:2019orp,Hartong:2015xda}.

The paper is organized as follows: In section \ref{sec2} we briefly review the three-dimensional spin-3 AdS gravity theory along its vanishing cosmological constant limit. Sections \ref{sec3}, \ref{sec4} and \ref{sec5} contains our mains results. In section \ref{sec3} we present new and known non- and ultra-relativistic spin-3 algebras by applying the semigroup expansion method to the spin-3 AdS algebra. In section \ref{sec4}, we present the explicit construction of CS gravity actions based on the non-relativistic spin-3 symmetries previously obtained. In Section \ref{sec5} we study the construction of Carrollian spin-3 gravity theories. Section \ref{concl} is devoted to some concluding remarks and discussion about future applications of our results and method. Additional contents regarding alternative non- and ultra-relativistic spin-3 symmetries appears in Appendix \ref{App1} and \ref{App2}.

Note added: while this manuscript was in the process of typesetting, it came to our knowledge
the ref. \cite{Caroca:2022byi}, which possesses some overlap with particular cases of our results.

\section{Three-dimensional Spin-3 AdS Gravity theory and its Flat Limit}\label{sec2}

A spin-3 extension of AdS gravity defined in three spacetime dimensions can be written as a CS theory considering the $\mathfrak{sl}\left(3,\mathbb{R}\right)\oplus\mathfrak{sl}\left(3,\mathbb{R}\right)$ Lie algebra \cite{Henneaux:2010xg,Campoleoni:2010zq}. The $\mathfrak{sl}\left(3,\mathbb{R}\right)$ algebra consists of spin-2 generators $\texttt{M}_a$ and spin-3 generators $\texttt{T}_{ab}$ which satisfy the following commutation relations:
\begin{eqnarray}
\left[\texttt{M}_{A},\texttt{M}_B\right]&=&\epsilon_{ABC}\texttt{M}^{C}\,,\notag\\
\left[\texttt{M}_A,\texttt{T}_{BC}\right]&=&\epsilon^{M}_{\ A\left(B\right.} \texttt{T}_{\left.C\right)M}\,,\notag\\
\left[\texttt{T}_{AB},\texttt{T}_{CD}\right]&=&\sigma\, \eta_{\left(A\left(C\right.\right.}\epsilon_{\left.\left.D\right)B\right)M}\texttt{J}^{M}\,, \label{sl3}
\end{eqnarray}
where $A=0,1,2$ are Lorentz indices which are lowered and raised with the Minkowski metric $\eta_{AB}=\left(-1,1,1\right)$ and $\epsilon_{ABC}$ is the three-dimensional Levi Civita tensor which satisfies $\epsilon_{012}=-\epsilon^{012}=1$. Here $\sigma$ is a normalization constant where $\sigma<0$ corresponds to the $\mathfrak{sl}\left(3,\mathbb{R}\right)$ algebra, while $\sigma>0$ corresponds to the $\mathfrak{su}\left(1,2\right)$ algebra. The spin-3 AdS gravity requires two copies of the $\mathfrak{sl}\left(3,\mathbb{R}\right)$ algebra which can be written as a spin-3 extension of the AdS algebra. Such symmetry, which has been denoted as $\mathfrak{hs_{3}AdS}$ in \cite{Bergshoeff:2016soe}, contains spin-2 generators $\{\hat{\texttt{J}}_{A},\hat{\texttt{P}}_{A}\}$ and spin-3 generators $\{\hat{\texttt{J}}_{AB},\hat{\texttt{P}}_{AB}\}$. The $\mathfrak{hs_{3}AdS}$ generators satisfy the commutation relations given by \cite{Henneaux:2010xg,Campoleoni:2010zq,Bergshoeff:2016soe,Caroca:2017izc}:
\begin{align}
\left[\hat{\texttt{J}}_{A},\hat{\texttt{J}}_B\right]&=\epsilon_{ABC}\hat{\texttt{J}}^{C}\,,\qquad \qquad \qquad \qquad \ \ &\left[\hat{\texttt{J}}_{A},\hat{\texttt{P}}_B\right]&=\epsilon_{ABC}\hat{\texttt{P}}^{C}\,, \notag\\
\left[\hat{\texttt{P}}_{A},\hat{\texttt{P}}_B\right]&=\frac{1}{\ell^{2}}\epsilon_{ABC}\hat{\texttt{J}}^{C}\,, \qquad \qquad \qquad \quad &\left[\hat{\texttt{J}}_A,\hat{\texttt{J}}_{BC}\right]&=\epsilon^{M}_{\ A\left(B\right.} \hat{\texttt{J}}_{\left.C\right)M}\,, \notag\\
\left[\hat{\texttt{J}}_A,\hat{\texttt{P}}_{BC}\right]&=\epsilon^{M}_{\ A\left(B\right.} \hat{\texttt{P}}_{\left.C\right)M}\,, \qquad \qquad \qquad \ \  &\left[\hat{\texttt{P}}_A,\hat{\texttt{P}}_{BC}\right]&=\frac{1}{\ell^{2}}\epsilon^{M}_{\ A\left(B\right.} \hat{\texttt{J}}_{\left.C\right)M}\,, \notag\\ 
\left[\hat{\texttt{P}}_A,\hat{\texttt{J}}_{BC}\right]&=\epsilon^{M}_{\ A\left(B\right.} \hat{\texttt{P}}_{\left.C\right)M}\,, \qquad \qquad \qquad
&\left[\hat{\texttt{J}}_{AB},\hat{\texttt{J}}_{CD}\right]&=-\, \eta_{\left(A\left(C\right.\right.}\epsilon_{\left.\left.D\right)B\right)M}\hat{\texttt{J}}^{M}\,, \notag\\
\left[\hat{\texttt{J}}_{AB},\hat{\texttt{P}}_{CD}\right]&=-\, \eta_{\left(A\left(C\right.\right.}\epsilon_{\left.\left.D\right)B\right)M}\hat{\texttt{P}}^{M}\,, \qquad \ \,
&\left[\hat{\texttt{P}}_{AB},\hat{\texttt{P}}_{CD}\right]&=-\frac{1}{\ell^{2}}\, \eta_{\left(A\left(C\right.\right.}\epsilon_{\left.\left.D\right)B\right)M}\hat{\texttt{J}}^{M}\,.
\label{AdS3}
\end{align}
Here, the $\ell$ parameter is related to the cosmological constant through $\Lambda \propto - \frac{1}{\ell^2}$. In the vanishing cosmological constant limit $\ell\rightarrow\infty$, the algebra reduces to the spin-3 extension of the Poincaré one \cite{Campoleoni:2010zq} \footnote{also denoted as $\mathfrak{hs_{3}poi}$ algebra}. One can notice that the spin-2 generators are the usual AdS generators which can be rewritten as the $\mathfrak{sl}\left(2,\mathbb{R}\right)\oplus\mathfrak{sl}\left(2,\mathbb{R}\right)$ algebra. On the other hand, the algebra \eqref{AdS3} can be recovered from the $\mathfrak{sl}\left(3,\mathbb{R}\right)\oplus\mathfrak{sl}\left(3,\mathbb{R}\right)$ structure after considering the redefinition of the generators as
\begin{align}
\hat{\texttt{J}}_{A}&=\texttt{M}_{A}+\bar{\texttt{M}}_{A}\,,  \qquad    &\hat{\texttt{P}}_{A}&=\frac{1}{\ell}\left(\texttt{M}_{A}-\bar{\texttt{M}}_{A}\right)\,, \notag \\
\hat{\texttt{J}}_{AB}&=\texttt{T}_{AB}+\bar{\texttt{T}}_{AB}\,,  \qquad  &\hat{\texttt{P}}_{AB}&=\frac{1}{\ell}\left(\texttt{T}_{AB}-\bar{\texttt{T}}_{AB}\right)\,, \label{rdef1}
\end{align}
and setting $\sigma=\bar{\sigma}=-1$. The $\mathfrak{hs_{3}AdS}$ algebra admits the following non-degenerate invariant bilinear form \cite{Caroca:2017izc}:
\begin{align}
\langle\hat{\texttt{J}}_{A}\hat{\texttt{J}}_{B}\rangle&=\hat{\alpha}_{0}\eta_{AB}\,,\qquad \qquad  &\langle\hat{\texttt{J}}_{AB}\hat{\texttt{J}}_{CD}\rangle&=\hat{\alpha}_{0}\left(\eta_{A\left(C\right.}\eta_{\left.D\right)B}-\frac{2}{3}\eta_{AB}\eta_{CD}\right)\,, \notag \\
\langle\hat{\texttt{J}}_{A}\hat{\texttt{P}}_{B}\rangle&=\hat{\alpha}_{1}\eta_{AB}\,,\qquad \qquad  &\langle\hat{\texttt{J}}_{AB}\hat{\texttt{P}}_{CD}\rangle&=\hat{\alpha}_{1}\left(\eta_{A\left(C\right.}\eta_{\left.D\right)B}-\frac{2}{3}\eta_{AB}\eta_{CD}\right)\,, \notag \\
\langle\hat{\texttt{P}}_{A}\hat{\texttt{P}}_{B}\rangle&=\frac{\hat{\alpha}_{0}}{\ell^2}\eta_{AB}\,,\qquad \qquad   &\langle\hat{\texttt{P}}_{AB}\hat{\texttt{P}}_{CD}\rangle&=\frac{\hat{\alpha}_{0}}{\ell^2}\left(\eta_{A\left(C\right.}\eta_{\left.D\right)B}-\frac{2}{3}\eta_{AB}\eta_{CD}\right)\,, \label{IT1}
\end{align}
where $\hat{\alpha}_{0}$ is related to an exotic sector of the spin-3 AdS gravity analogously to the one appearing in the AdS gravity theory \cite{Witten:1988hc,Zanelli:2005sa}. Hence, the most general quadratic Casimir invariant is \cite{Caroca:2017izc}
\begin{eqnarray}
C&=&\hat{\alpha}_{0}\left(\hat{\texttt{J}}_{A}\hat{\texttt{J}}^{A}+\frac{1}{2}\hat{\texttt{J}}_{AB}\hat{\texttt{J}}^{AB}+\frac{1}{\ell^2}\left[\hat{\texttt{P}}_{A}\hat{\texttt{P}}^{A}+\frac{1}{2}\hat{\texttt{P}}_{AB}\hat{\texttt{P}}^{AB}\right]\right) \notag \\
&&+\,\hat{\alpha_1}\left(\hat{\texttt{J}}_{A}\hat{\texttt{P}}^{A}+\frac{1}{2}\hat{\texttt{J}}_{AB}\hat{\texttt{P}}^{AB}\right)\,. 
\end{eqnarray}
A CS action gauge-invariant under the $\mathfrak{hs_{3}AdS}$ algebra can be obtained by considering the gauge connection one-form for the spin-3 extension of the AdS symmetry:
\begin{equation}
    A=W^{A}\hat{\texttt{J}}_{A}+E^{A}\hat{\texttt{P}}_{A}+W^{AB}\hat{\texttt{J}}_{AB}+E^{AB}\hat{\texttt{P}}_{AB}\,,
\end{equation}
and the non-vanishing components of the invariant tensor \eqref{IT1} in the general expression of the CS form,
\begin{eqnarray}
I_{CS}=\frac{k}{4\pi}\int\langle AdA+\frac{2}{3}A^3\rangle\,,\label{CS}
\end{eqnarray}
with $k$ being the CS level of the theory which is related to the gravitational constant $G$ through $k=1/(4G)$. Thus, one finds the following CS action for the $\mathfrak{hs_{3}AdS}$ algebra
\begin{align}
I_{\mathfrak{hs_{3}AdS}}=\frac{k}{4\pi}\int& \hat{\alpha}_0 \left[ W^A d W_A + \frac{1}{3}\epsilon^{ABC}W_A W_B W_C+2W^{AB}dW_{AB}+4\epsilon_{ABC}W^{A}W^{BD}W^{C}_{\ D}\right.\notag \\
&+\left.\frac{2}{\ell^2}E^{AB}\left(dE_{AB}+\epsilon_{CD\left(A\right|}W^{C}E_{\left|B\right)}^{\ D}+2\epsilon_{CD\left(A\right|}E^{C}W_{\left|B\right)}^{\ D}\right)+\frac{1}{\ell^2}T^A E_A
    \right] \notag \\
& + 2\hat{\alpha}_1 \left[  E_{A}\left(R^{A} +2\epsilon_{ABC}W^{BD}W^{C}_{\ D}\right)+2E^{AB}\left(dW_{AB}+\epsilon_{CD\left(A\right|}W^{C}W_{\left|B\right)}^{\ D}\right)\right.\notag \\
&+ \left.\frac{1}{6\ell^{2}}\left( \epsilon^{ABC} E_A E_B E_C+12 E^{A}E^{BD}E^{C}_{\ D}\right)  \right]\,, \label{CSADS3}
\end{align}
where
\begin{eqnarray}
 R^{A}&=&dW^{A}-\frac{1}{2}\epsilon^{ABC}W_{B}W_{C} \,,\notag\\
 T^{A}&=& dE^{A}-\epsilon^{ABC}W_{B}E_{C} \,,\label{curvatures}
\end{eqnarray}
are the Lorentz curvature and torsion curvature two-forms, respectively. Here the wedge product between forms are not written explicitly. Each sector of the CS gravity action is gauge invariant under the $\mathfrak{hs_{3}AdS}$ algebra. In particular, the term proportional to $\hat{\alpha}_0$ can be seen as a spin-3 extension of the exotic gravity Lagrangian \cite{Witten:1988hc,Zanelli:2005sa}. On the other hand, the spin-3 extension of the Einstein-Hilbert CS term plus the cosmological constant appear along $\hat{\alpha}_1$. Both sectors allow us to define the most general CS action for the $\mathfrak{hs_{3}AdS}$ algebra being related to a spin-3 extension of the Pontryagin and Euler density, respectively \cite{Troncoso:1999pk}. One can notice that the three-dimensional Poincaré CS gravity action coupled to spin-3 gauge fields appears in the vanishing cosmological constant limit $\ell\rightarrow\infty$. The study of spin-3 gravity in three-dimensional flat space has been largely studied in \cite{Afshar:2013vka,Grumiller:2014lna,Gary:2014ppa,Matulich:2014hea}. Since the invariant tensor \eqref{IT1} is non-degenerate, the equations of motion of the $\mathfrak{hs_{3}AdS}$ theory are given by the vanishing of the curvature two-forms:
\begin{eqnarray}
  0&=&R^{A}\left(W\right)\equiv dW^{A}-\frac{1}{2}\epsilon^{ABC}\left(W_{B}W_{C}+\frac{1}{\ell^2}E_{B}E_{C}\right)-2\epsilon^{ABC}\left(W_{BD}W_{C}^{\ D}+\frac{1}{\ell^2}E_{BD}E_{C}^{\ D}\right) \,,\notag\\
 0&=&R^{A}\left(E\right)\equiv dE^{A}-\epsilon^{ABC}W_{B}E_{C}-4\epsilon^{ABC}E_{BD}W_{C}^{\ D}\,,\notag\\
 0&=&R^{AB}\left(W\right)\equiv dW^{AB}-\epsilon^{CD\left(A\right|}W_{C}W_{D}^{\ \left|B\right)}-\frac{1}{\ell^2}\epsilon^{CD\left(A\right|}E_{C}E_{D}^{\ \left|B\right)}\,,\notag \\
 0&=&R^{AB}\left(E\right)\equiv dE^{AB}-\epsilon^{CD\left(A\right|}W_{C}E_{D}^{\ \left|B\right)}-\epsilon^{CD\left(A\right|}E_{C}W_{D}^{\ \left|B\right)}\,. \label{AdS3Curv}
\end{eqnarray}
Analogously to minimal supergravity \cite{Achucarro:1986uwr}, we have contributions of the spin-3 gauge fields in the equations for the spin-2 curvatures $R^{A}\left(W\right)$ and $R^{A}\left(E\right)$. Such behavior is inherited from the commutation relations \eqref{AdS3}.

\section{Non-relativistic and Ultra-relativistic Spin-3 Algebras}\label{sec3}

A spin-3 extension of the infinite family of non-relativistic algebras of the Newton-Hooke type and AdS Carrol type discussed in \cite{Gomis:2019nih} can be obtained considering an S-expansion of the $\mathfrak{hs_{3}AdS}$ algebra. First, we will approach finite cases in order to obtain spin-3 extensions of known non-relativistic and ultra-relativistic algebras. We will then generalize our procedure to infinite-dimensional non-relativistic and ultra-relativistic algebras by considering an infinite-dimensional semigroup. Both families of algebras are new and contain a cosmological constant inherited from the expansion of the spin-3 AdS algebra. Interestingly, in the vanishing cosmological constant limit, we obtain spin-3 extensions of the infinite-dimensional Galilean \cite{Hansen:2019vqf,Gomis:2019fdh} and infinite-dimensional Carroll algebras \cite{Gomis:2019nih}. In particular, the spin-3 extension of the extended Bargmann algebra, corresponding to $\mathfrak{hs_{3}barg1}$ in \cite{Bergshoeff:2016soe}, appears as a flat limit of the spin-3 extended Newton-Hooke algebra obtained here. As we shall see in the next section, the expansion method based on semigroups allows us to obtain non-relativistic and ultra-relativistic spin-3 algebras having a non-degenerate invariant bilinear form, and thus ensuring the proper construction of a CS action.

In order to get both non-relativistic and ultra-relativistic spin-3 AdS algebras, we first consider a decomposition of the relativistic $A$ index in time and space components $A=\left(0,a\right)$ with $a=1,2$. Then, the commutation relations of the $\mathfrak{hs_{3}AdS}$ algebra \eqref{AdS3} can be rewritten as
\begin{align}
\left[\hat{\texttt{J}},\hat{\texttt{G}}_a\right]&=\epsilon_{ab}\hat{\texttt{G}}_{b}\,,\quad \quad   &\left[\hat{\texttt{G}}_{a},\hat{\texttt{G}}_{b}\right]&=-\epsilon_{ab}\hat{\texttt{J}}\,,\quad  &\left[\hat{\texttt{H}},\hat{\texttt{G}}_a\right]&=\epsilon_{ab}\hat{\texttt{P}}_{b}\,, \notag\\
\left[\hat{\texttt{J}},\hat{\texttt{P}}_a\right]&=\epsilon_{ab}\hat{\texttt{P}}_{b}\,,\quad \quad   &\left[\hat{\texttt{G}}_{a},\hat{\texttt{P}}_{b}\right]&=-\epsilon_{ab}\hat{\texttt{H}}\,,\quad  &\left[\hat{\texttt{H}},\hat{\texttt{P}}_a\right]&=\frac{1}{\ell^2}\epsilon_{ab}\hat{\texttt{G}}_{b}\,, \notag\\
\left[\hat{\texttt{J}},\hat{\texttt{J}}_a\right]&=\epsilon_{ab}\hat{\texttt{J}}_{b}\,,\quad \quad   &\left[\hat{\texttt{P}}_{a},\hat{\texttt{P}}_{b}\right]&=-\frac{1}{\ell^2}\epsilon_{ab}\hat{\texttt{J}}\,,\quad   &\left[\hat{\texttt{H}},\hat{\texttt{J}}_a\right]&=\epsilon_{ab}\hat{\texttt{H}}_{b}\,, \notag\\
\left[\hat{\texttt{J}},\hat{\texttt{H}}_a\right]&=\epsilon_{ab}\hat{\texttt{H}}_{b}\,,\quad \quad   &\left[\hat{\texttt{G}}_{a},\hat{\texttt{J}}_{b}\right]&=-\left(\epsilon_{am}\hat{\texttt{G}}_{mb}+\epsilon_{ab}\hat{\texttt{G}}_{mm}\right)\,,\quad  &\left[\hat{\texttt{H}},\hat{\texttt{H}}_a\right]&=\frac{1}{\ell^2}\epsilon_{ab}\hat{\texttt{J}}_{b}\,, \notag\\
\left[\hat{\texttt{J}},\hat{\texttt{G}}_{ab}\right]&=-\epsilon_{m\left(a\right.}\hat{\texttt{G}}_{\left.b\right)m}\,, \quad \quad &\left[\hat{\texttt{G}}_{a},\hat{\texttt{H}}_{b}\right]&=-\left(\epsilon_{am}\hat{\texttt{P}}_{mb}+\epsilon_{ab}\hat{\texttt{P}}_{mm}\right)\,,\quad  &\left[\hat{\texttt{H}},\hat{\texttt{G}}_{ab}\right]&=-\epsilon_{m\left(a\right.}\hat{\texttt{P}}_{\left.b\right)m}\,, \notag \\
\left[\hat{\texttt{J}},\hat{\texttt{P}}_{ab}\right]&=-\epsilon_{m\left(a\right.}\hat{\texttt{P}}_{\left.b\right)m}\,, \quad \quad &\left[\hat{\texttt{P}}_{a},\hat{\texttt{J}}_{b}\right]&=-\left(\epsilon_{am}\hat{\texttt{P}}_{mb}+\epsilon_{ab}\hat{\texttt{P}}_{mm}\right)\,,\quad  &\left[\hat{\texttt{H}},\hat{\texttt{P}}_{ab}\right]&=-\frac{1}{\ell^2}\epsilon_{m\left(a\right.}\hat{\texttt{G}}_{\left.b\right)m}\,, \notag \\
\left[\hat{\texttt{G}}_a,\hat{\texttt{G}}_{bc}\right]&=-\epsilon_{a\left(b\right.}\hat{\texttt{J}}_{\left.c\right)}\,, \quad \quad &\left[\hat{\texttt{P}}_{a},\hat{\texttt{H}}_{b}\right]&=-\frac{1}{\ell^2}\left(\epsilon_{am}\hat{\texttt{G}}_{mb}+\epsilon_{ab}\hat{\texttt{G}}_{mm}\right)\,,\quad  &\left[\hat{\texttt{P}}_a,\hat{\texttt{G}}_{bc}\right]&=-\epsilon_{a\left(b\right.}\hat{\texttt{H}}_{\left.c\right)}\,, \notag \\
\left[\hat{\texttt{G}}_a,\hat{\texttt{P}}_{bc}\right]&=-\epsilon_{a\left(b\right.}\hat{\texttt{H}}_{\left.c\right)}\,, \quad \quad &\left[\hat{\texttt{J}}_{a},\hat{\texttt{J}}_{b}\right]&=\epsilon_{ab}\hat{\texttt{J}}\,,\quad  &\left[\hat{\texttt{P}}_a,\hat{\texttt{P}}_{bc}\right]&=-\frac{1}{\ell^2}\epsilon_{a\left(b\right.}\hat{\texttt{J}}_{\left.c\right)}\,, \notag \\
\left[\hat{\texttt{J}}_a,\hat{\texttt{G}}_{bc}\right]&=\delta_{a\left(b\right.}\epsilon_{\left.c\right)m}\hat{\texttt{G}}_{m}\,, \quad \quad &\left[\hat{\texttt{J}}_{a},\hat{\texttt{H}}_{b}\right]&=\epsilon_{ab}\hat{\texttt{H}}\,,\quad  &\left[\hat{\texttt{H}}_a,\hat{\texttt{G}}_{bc}\right]&=\delta_{a\left(b\right.}\epsilon_{\left.c\right)m}\hat{\texttt{P}}_{m}\,, \notag \\
\left[\hat{\texttt{J}}_a,\hat{\texttt{P}}_{bc}\right]&=\delta_{a\left(b\right.}\epsilon_{\left.c\right)m}\hat{\texttt{P}}_{m}\,, \quad \quad &\left[\hat{\texttt{H}}_{a},\hat{\texttt{H}}_{b}\right]&=\frac{1}{\ell^2}\epsilon_{ab}\hat{\texttt{J}}\,, \quad &\left[\hat{\texttt{H}}_a,\hat{\texttt{P}}_{bc}\right]&=\frac{1}{\ell^2}\delta_{a\left(b\right.}\epsilon_{\left.c\right)m}\hat{\texttt{G}}_{m}\,, \notag \\
\left[\hat{\texttt{G}}_{ab},\hat{\texttt{G}}_{cd}\right]&=\delta_{\left(a\right.\left(c\right.}\epsilon_{\left.d\right)\left.b\right)}\hat{\texttt{J}}\,, \quad \quad &\left[\hat{\texttt{G}}_{ab},\hat{\texttt{P}}_{cd}\right]&=\delta_{\left(a\right.\left(c\right.}\epsilon_{\left.d\right)\left.b\right)}\hat{\texttt{H}}\,,\quad  &\left[\hat{\texttt{P}}_{ab},\hat{\texttt{P}}_{cd}\right]&=\frac{1}{\ell^2}\delta_{\left(a\right.\left(c\right.}\epsilon_{\left.d\right)\left.b\right)}\hat{\texttt{J}}\,, \label{AdS3b}
\end{align}
where we have relabelled the spin-3 AdS generators as in \cite{Bergshoeff:2016soe},
\begin{align}
  \hat{\texttt{J}}&=\hat{\texttt{J}}_0\,, \qquad &\hat{\texttt{G}}_a&=\hat{\texttt{J}}_{a}\,, \qquad &\hat{\texttt{H}}&=\hat{\texttt{P}}_0\,, \qquad &\hat{\texttt{P}}_a&=\hat{\texttt{P}}_{a}\,, \qquad \notag \\
  \hat{\texttt{J}}_a&=\hat{\texttt{J}}_{0a}\,, \qquad &\hat{\texttt{G}}_{ab}&=\hat{\texttt{J}}_{ab}\,, \qquad &\hat{\texttt{H}}_a&=\hat{\texttt{P}}_{0a}\,, \qquad &\hat{\texttt{P}}_{ab}&=\hat{\texttt{P}}_{ab}\,. \label{relabel}
\end{align}
Moreover, for the Levi-Civita symbol, we have considered $\epsilon_{ab}=\epsilon_{0ab}$, $\epsilon^{ab}=\epsilon^{0ab}$ and $\epsilon_{ab}\epsilon^{bc}=-\delta_{a}^{\ c}$. Let us note that the generators $\hat{\texttt{G}}_{00}$ and $\hat{\texttt{P}}_{00}$ are not independent due to the tracelessness condition $\eta^{AB}\hat{\texttt{J}}_{AB}=0=\eta^{AB}\hat{\texttt{P}}_{AB}$.
\subsection{Non-relativistic Expansions of Spin-3 AdS Algebra}
Non-relativistic expansions of the Spin-3 AdS algebra require to consider a particular subspace decomposition of the relativistic algebra. Let $V_0$ and $V_1$ be two subspaces of the $\mathfrak{hs_{3}AdS}$ algebra given by
\begin{align}
   V_0&=\{\hat{\texttt{J}},\hat{\texttt{H}},\hat{\texttt{J}}_{a},\hat{\texttt{H}}_{a}\} \,,\notag \\
   V_1&=\{\hat{\texttt{G}}_{a},\hat{\texttt{P}}_{a},\hat{\texttt{G}}_{ab},\hat{\texttt{P}}_{ab}\}\,,\label{sd}
\end{align}
which satisfies a $\mathbb{Z}_2$ graded-Lie algebra,
\begin{align}
[V_0,V_0]&\subset V_0\,, \quad &[V_0,V_1]&\subset V_1\,, \quad &[V_1,V_1]&\subset V_0\,.\label{sd2}
\end{align}
Moreover the non-relativistic expansion based on semigroups requires to consider a particular semigroup denoted as $S_{E}^{\left(N\right)}$ which contains $N+1$ elements \cite{Gomis:2019nih,Concha:2020tqx,Concha:2021jos,Concha:2021llq,Concha:2022you}. Then we have to apply a subset decomposition $S=S_0\cup S_1$ which has to satisfy the same structure than the subspace decomposition \eqref{sd}, namely
\begin{align}
S_0\cdot S_0&\subset S_0\,, \quad &S_0\cdot S_1&\subset S_1\,, \quad &S_1\cdot S_1&\subset S_0\,.\label{rc}
\end{align}
Such condition on the subset decomposition of the semigroup is called resonant condition \cite{Izaurieta:2006zz}. Then a resonant expansion of the relativistic $\mathfrak{hs_{3}AdS}$ algebra is given by $\mathcal{G}_{R}=\left(S_{0}\times V_{0}\right)\oplus \left(S_{1}\times V_{1}\right)$. In addition, we shall require to impose the $0_S$-reduction condition $0_S T_A =0$ with $0_S$ being the zero element of the semigroup. Then, the $0_S$-reduced resonant expansion of the $\mathfrak{hs_{3}AdS}$ algebra will correspond to a non-relativistic expansion. 

The choice of the semigroup is not arbitrary but it is due to the fact that a resonant $S_{E}^{\left(N\right)}$-expansion can be seen as a generalized Inönü-Wigner contraction \cite{Izaurieta:2006zz}, similarly to the contraction used to apply a non-relativistic limit.  One can identify the elements of the $S_{E}^{\left(N\right)}$ semigroup as powers of an arbitrary variable $\varepsilon$ such that $\lambda_{i}=\varepsilon^{i}$. Then, an $S$-expansion using $S_{E}^{\left(N\right)}$ as the relevant semigroup represents an expansion to order $O\left(\varepsilon^{N+1}\right)$. Such semigroup allows us to get all the expanded algebras obtained through the Maurer-Cartan (MC) forms
power-series expansion \cite{deAzcarraga:2002xi,deAzcarraga:2007et}. However, a non-relativistic algebra appears from a relativistic one when the previous considerations for the relativistic subspaces and semigroup subsets are taking into account. In particular, the resonant $S_{E}^{\left(1\right)}$-expansion coincides with the non-relativistic contraction. As was noticed in \cite{Gomis:2019nih}, the use of $S_{E}^{\left(N\right)}$ will generate a family of generalised non-relativistic algebras from a relativistic one for all the possible values of N which, in our case, would corresponds to spin-3 extensions of the results obtained in \cite{Gomis:2019nih}. Such method offers us a large family of non-relativistic symmetries which the usual non-relativistic contraction procedure cannot reproduce. Interestingly, the $S_{E}^{\left(N\right)}$-expansion can be extended to the ultra-relativistic realm by considering a different subspace decomposition of the relativistic algebra. It would be interesting to explore the relation between the $S_{E}$-expansion and the cube summarizing the sequential limits
for the kinematical algebras \cite{Bacry:1968zf}. Such analysis could allows us to extend the cube to infinite-dimensional symmetries starting from finite relativistic ones considering $S_{E}^{\left(\infty\right)}$. 

Before approaching the arbitrary $N$ case, we will focus first in $N=2$ and $N=4$ which reproduce a spin-3 extended Newton-Hooke and spin-3 extended Post-Newtonian algebras, respectively.
\subsubsection{Spin-3 Extended Newton-Hooke Algebra}\label{sec311}
Let $S_{E}^{\left(2\right)}=\{\lambda_0,\lambda_1,\lambda_2,\lambda_3\}$ be the relevant semigroup where $\lambda_3=0_S$ is the zero element of the semigroup satisfying $0_S \lambda_i=0_S$ for $i=0,\ldots,3$. The elements of the semigroup $S_{E}^{\left(2\right)}$ satisfy the following multiplication law
\begin{equation}
\begin{tabular}{l|llll}
$\lambda _{3}$ & $\lambda _{3}$ & $\lambda _{3}$ & $\lambda _{3}$ & $\lambda
_{3}$ \\
$\lambda _{2}$ & $\lambda _{2}$ & $\lambda _{3}$ & $\lambda _{3}$ & $\lambda
_{3}$ \\
$\lambda _{1}$ & $\lambda _{1}$ & $\lambda _{2}$ & $\lambda _{3}$ & $\lambda
_{3}$ \\
$\lambda _{0}$ & $\lambda _{0}$ & $\lambda _{1}$ & $\lambda _{2}$ & $\lambda
_{3}$ \\ \hline
& $\lambda _{0}$ & $\lambda _{1}$ & $\lambda _{2}$ & $\lambda _{3}$%
\end{tabular}
\label{mlSE4}
\end{equation}
Let us consider now a subset decomposition $S=S_0\cup S_1$ with
$S_0=\{\lambda_0,\lambda_2,\lambda_3\}$ and $S_1=\{\lambda_1,\lambda_3\}$. One can notice that such decomposition satisfies the resonant condition \eqref{rc}. After applying a resonant $S_{E}^{\left(2\right)}$-expansion and extracting a $0_S$-reduction we get a non-relativistic version of the $\mathfrak{hs_{3}AdS}$ algebra whose expanded generators are related to the relativistic ones through the semigroup elements as follows:
\begin{equation}
    \begin{tabular}{lll}
\multicolumn{1}{l|}{$\lambda_3$} & \multicolumn{1}{|l}{\cellcolor[gray]{0.8}} & \multicolumn{1}{|l|}{\cellcolor[gray]{0.8}} \\ \hline
\multicolumn{1}{l|}{$\lambda_2$} & \multicolumn{1}{|l}{$\texttt{S},\ \texttt{M},\ \, \texttt{S}_{a},\ \texttt{M}_{a},\,$} & \multicolumn{1}{|l|}{\cellcolor[gray]{0.8}} \\ \hline
\multicolumn{1}{l|}{$\lambda_1$} & \multicolumn{1}{|l}{\cellcolor[gray]{0.8}} & \multicolumn{1}{|l|}{$\texttt{G}_a,\,\texttt{P}_a,\,\texttt{G}_{ab},\,\texttt{P}_{ab},\,$} \\ \hline
\multicolumn{1}{l|}{$\lambda_0$} & \multicolumn{1}{|l}{$ \texttt{J},\,\ \texttt{H},\ \texttt{J}_{a},\ \texttt{H}_{a},\,$} & \multicolumn{1}{|l|}{\cellcolor[gray]{0.8}} \\ \hline
\multicolumn{1}{l|}{} & \multicolumn{1}{|l}{$\hat{\texttt{J}},\ \, \hat{\texttt{H}},\  \hat{\texttt{J}}_{a},\ \hat{\texttt{H}}_{a},\,$} & \multicolumn{1}{|l|}{$\hat{\texttt{G}}_{a},\ \hat{\texttt{P}}_{a},\,\hat{\texttt{G}}_{ab},\,\hat{\texttt{P}}_{ab}$}
\end{tabular}\label{Sexp}%
\end{equation}
The commutation relations for such generators are obtained considering the commutation relations of the relativistic $\mathfrak{hs_{3}AdS}$ algebra and the multiplication law of the semigroup $S_{E}^{\left(3\right)}$. Indeed, the expanded generators satisfy the following commutators:
\begin{align}
\left[\texttt{J},\texttt{G}_a\right]&=\epsilon_{ab}\texttt{G}_{b}\,,    &\left[\texttt{G}_{a},\texttt{G}_{b}\right]&=-\epsilon_{ab}\texttt{S}\,, &\left[\texttt{H},\texttt{G}_a\right]&=\epsilon_{ab}\texttt{P}_{b}\,, \notag\\
\left[\texttt{J},\texttt{P}_a\right]&=\epsilon_{ab}\texttt{P}_{b}\,,  &\left[\texttt{G}_{a},\texttt{P}_{b}\right]&=-\epsilon_{ab}\texttt{M}\,, &\left[\texttt{H},\texttt{P}_a\right]&=\frac{1}{\ell^2}\epsilon_{ab}\texttt{G}_{b}\,, \notag\\
\left[\texttt{J},\texttt{J}_a\right]&=\epsilon_{ab}\texttt{J}_{b}\,,  &\left[\texttt{P}_{a},\texttt{P}_{b}\right]&=-\frac{1}{\ell^2}\epsilon_{ab}\texttt{S}\,,  &\left[\texttt{H},\texttt{J}_a\right]&=\epsilon_{ab}\texttt{H}_{b}\,, \notag\\
\left[\texttt{J},\texttt{H}_a\right]&=\epsilon_{ab}\texttt{H}_{b}\,,  &\left[\texttt{G}_{a},\texttt{J}_{b}\right]&=-\left(\epsilon_{am}\texttt{G}_{mb}+\epsilon_{ab}\texttt{G}_{mm}\right)\,,  &\left[\texttt{H},\texttt{H}_a\right]&=\frac{1}{\ell^2}\epsilon_{ab}\texttt{J}_{b}\,, \notag\\
\left[\texttt{J},\texttt{S}_a\right]&=\epsilon_{ab}\texttt{S}_{b}\,, &\left[\texttt{G}_{a},\texttt{H}_{b}\right]&=-\left(\epsilon_{am}\texttt{P}_{mb}+\epsilon_{ab}\texttt{P}_{mm}\right)\,,  &\left[\texttt{H},\texttt{S}_a\right]&=\epsilon_{ab}\texttt{M}_{b}\,, \notag\\
\left[\texttt{J},\texttt{M}_a\right]&=\epsilon_{ab}\texttt{M}_{b}\,,  &\left[\texttt{P}_{a},\texttt{J}_{b}\right]&=-\left(\epsilon_{am}\texttt{P}_{mb}+\epsilon_{ab}\texttt{P}_{mm}\right)\,,  &\left[\texttt{H},\texttt{M}_a\right]&=\frac{1}{\ell^2}\epsilon_{ab}\texttt{S}_{b}\,, \notag\\
\left[\texttt{S},\texttt{J}_a\right]&=\epsilon_{ab}\texttt{S}_{b}\,,   &\left[\texttt{P}_{a},\texttt{H}_{b}\right]&=-\frac{1}{\ell^2}\left(\epsilon_{am}\texttt{G}_{mb}+\epsilon_{ab}\texttt{G}_{mm}\right)\,,  &\left[\texttt{M},\texttt{J}_a\right]&=\epsilon_{ab}\texttt{M}_{b}\,, \notag\\
\left[\texttt{S},\texttt{H}_a\right]&=\epsilon_{ab}\texttt{M}_{b}\,,    &\left[\texttt{J}_{a},\texttt{J}_{b}\right]&=\epsilon_{ab}\texttt{J}\,,  &\left[\texttt{M},\texttt{H}_a\right]&=\frac{1}{\ell^2}\epsilon_{ab}\texttt{S}_{b}\,, \notag\\
\left[\texttt{J},\texttt{G}_{ab}\right]&=-\epsilon_{m\left(a\right.}\texttt{G}_{\left.b\right)m}\,,  &\left[\texttt{J}_{a},\texttt{H}_{b}\right]&=\epsilon_{ab}\texttt{H}\,,  &\left[\texttt{H},\texttt{G}_{ab}\right]&=-\epsilon_{m\left(a\right.}\texttt{P}_{\left.b\right)m}\,, \notag \\
\left[\texttt{J},\texttt{P}_{ab}\right]&=-\epsilon_{m\left(a\right.}\texttt{P}_{\left.b\right)m}\,,  &\left[\texttt{H}_{a},\texttt{H}_{b}\right]&=\frac{1}{\ell^2}\epsilon_{ab}\texttt{J}\,, &\left[\texttt{H},\texttt{P}_{ab}\right]&=-\frac{1}{\ell^2}\epsilon_{m\left(a\right.}\texttt{G}_{\left.b\right)m}\,, \notag \\
\left[\texttt{G}_a,\texttt{G}_{bc}\right]&=-\epsilon_{a\left(b\right.}\texttt{S}_{\left.c\right)}\,,  &\left[\texttt{J}_{a},\texttt{S}_{b}\right]&=\epsilon_{ab}\texttt{S}\,,  &\left[\texttt{P}_a,\texttt{G}_{bc}\right]&=-\epsilon_{a\left(b\right.}\texttt{M}_{\left.c\right)}\,, \notag \\
\left[\texttt{G}_a,\texttt{P}_{bc}\right]&=-\epsilon_{a\left(b\right.}\texttt{M}_{\left.c\right)}\,,  &\left[\texttt{J}_{a},\texttt{M}_{b}\right]&=\epsilon_{ab}\texttt{M}\,,  &\left[\texttt{P}_a,\texttt{P}_{bc}\right]&=-\frac{1}{\ell^2}\epsilon_{a\left(b\right.}\texttt{S}_{\left.c\right)}\,, \notag \\
\left[\texttt{J}_a,\texttt{G}_{bc}\right]&=\delta_{a\left(b\right.}\epsilon_{\left.c\right)m}\texttt{G}_{m}\,, &\left[\texttt{H}_{a},\texttt{S}_{b}\right]&=\epsilon_{ab}\texttt{M}\,, &\left[\texttt{H}_a,\texttt{G}_{bc}\right]&=\delta_{a\left(b\right.}\epsilon_{\left.c\right)m}\texttt{P}_{m}\,, \notag \\
\left[\texttt{J}_a,\texttt{P}_{bc}\right]&=\delta_{a\left(b\right.}\epsilon_{\left.c\right)m}\texttt{P}_{m}\,,  &\left[\texttt{H}_{a},\texttt{M}_{b}\right]&=\frac{1}{\ell^2}\epsilon_{ab}\texttt{S}\,,  &\left[\texttt{H}_a,\texttt{P}_{bc}\right]&=\frac{1}{\ell^2}\delta_{a\left(b\right.}\epsilon_{\left.c\right)m}\texttt{G}_{m}\,, \notag \\
\left[\texttt{G}_{ab},\texttt{G}_{cd}\right]&=\delta_{\left(a\right.\left(c\right.}\epsilon_{\left.d\right)\left.b\right)}\texttt{S}\,,  &\left[\texttt{G}_{ab},\texttt{P}_{cd}\right]&=\delta_{\left(a\right.\left(c\right.}\epsilon_{\left.d\right)\left.b\right)}\texttt{M}\,,  &\left[\texttt{P}_{ab},\texttt{P}_{cd}\right]&=\frac{1}{\ell^2}\delta_{\left(a\right.\left(c\right.}\epsilon_{\left.d\right)\left.b\right)}\texttt{S}\,.\label{enh3}
\end{align}
The expanded algebra \eqref{enh3} is a spin-3 extension of the extended Newton-Hooke algebra \cite{Aldrovandi:1998im,Gibbons:2003rv,Brugues:2006yd,Alvarez:2007fw,Papageorgiou:2010ud,Duval:2011mi,Hartong:2016yrf,Duval:2016tzi} and corresponds to the non-relativistic counterpart of the $\mathfrak{hs_{3}AdS}$ algebra. One can see that the extended Newton-Hooke algebra spanned by $\{\texttt{J},\texttt{G}_{a},\texttt{S},\texttt{H},\texttt{P}_{a},\texttt{M}\}$ appears as a spin-2 subalgebra. The present non-relativistic algebra will be denoted as $\mathfrak{hs_{3}enh}$ and can be seen as an extension of the $\mathfrak{hs_{3}nh1}$ algebra defined in \cite{Bergshoeff:2016soe}. As we shall see, such extension is required to avoid degeneracy in the invariant tensor. Interestingly, in the vanishing cosmological constant limit $\ell\rightarrow\infty$ we obtain a spin-3 extension of the extended Bargmann ($\mathfrak{hs_{3}ebarg}$) algebra \footnote{Let us note that the spin-3 extension of the extended Bargmann algebra obtained as a flat limit of the $\mathfrak{hs_{3}enh}$ algebra coincides with the $\mathfrak{hs_{3}ebarg1}$ algebra of \cite{Bergshoeff:2016soe}.}. Both $\mathfrak{hs_{3}enh}$ and $\mathfrak{hs_{3}ebarg}$ admit a non-degenerate bilinear trace allowing us to construct a proper non-relativistic CS action. Remarkably, the $\mathfrak{hs_{3}ebarg}$ algebra can alternatively be recovered by applying a resonant $S_{E}^{\left(2\right)}$-expansion to the spin-3 Poincaré algebra following the same procedure used here.

Let us note that the $\mathfrak{hs_{3}enh}$ algebra can be rewritten as two copies of a spin-3 extension of the so-called Nappi-Witten algebra \cite{Nappi:1993ie,Figueroa-OFarrill:1999cmq}:
\begin{align}
\left[\texttt{J}^{\pm},\texttt{G}^{\pm}_a\right]&=\epsilon_{ab}\texttt{G}^{\pm}_{b}\,,\quad \quad   &\left[\texttt{G}^{\pm}_a,\texttt{G}^{\pm}_{bc}\right]&=-\epsilon_{a\left(b\right.}\texttt{S}^{\pm}_{\left.c\right)}\,,\quad  &\left[\texttt{G}^{\pm}_{a},\texttt{G}^{\pm}_{b}\right]&=-\epsilon_{ab}\texttt{S}^{\pm}\,, \notag\\
\left[\texttt{J}^{\pm},\texttt{J}^{\pm}_a\right]&=\epsilon_{ab}\texttt{J}^{\pm}_{b}\,,\quad \quad   &\left[\texttt{J}^{\pm}_a,\texttt{G}^{\pm}_{bc}\right]&=\delta_{a\left(b\right.}\epsilon_{\left.c\right)m}\texttt{G}^{\pm}_{m}\,,\quad  &\left[\texttt{J}^{\pm}_{a},\texttt{J}^{\pm}_{b}\right]&=\epsilon_{ab}\texttt{J}^{\pm}\,, \notag\\
\left[\texttt{J}^{\pm},\texttt{G}^{\pm}_{ab}\right]&=-\epsilon_{m\left(a\right.}\texttt{G}^{\pm}_{\left.b\right)m} \,,\quad \quad   &\left[\texttt{G}^{\pm}_{ab},\texttt{G}^{\pm}_{cd}\right]&=\delta_{\left(a\right.\left(c\right.}\epsilon_{\left.d\right)\left.b\right)}\texttt{S}^{\pm}\,,\quad &\left[\texttt{G}^{\pm}_{a},\texttt{J}^{\pm}_{b}\right]&=-\left(\epsilon_{am}\texttt{G}^{\pm}_{mb}+\epsilon_{ab}\texttt{G}^{\pm}_{mm}\right)\,, \notag\\
\left[\texttt{J}^{\pm},\texttt{S}^{\pm}_{a}\right]&=\epsilon_{ab}\texttt{S}^{\pm}_{b}\,, \quad \quad &\left[\texttt{S}^{\pm},\texttt{J}^{\pm}_{a}\right]&=\epsilon_{ab}\texttt{S}^{\pm}_{b}\,, \quad \quad &\left[\texttt{J}^{\pm}_{a},\texttt{S}^{\pm}_{b}\right]&=\epsilon_{ab}\texttt{S}^{\pm}\,. \label{hs3nw}
\end{align}
The two copies of the spin-3 Nappi-Witten algebra, which we have denoted as $\mathfrak{hs_{3}nw}$, appear after considering the redefinition of the $\mathfrak{hs_{3}enh}$ generators as follows:
\begin{align}
    \texttt{J}&=\texttt{J}^{+}+\texttt{J}^{-}\,, \quad \quad &\texttt{G}_{a}&=\texttt{G}^{+}_{a}-\texttt{G}^{-}_{a}\,, \quad \quad &\texttt{H}&=\frac{1}{\ell}\left(\texttt{J}^{+}-\texttt{J}^{-}\right)\,, \notag\\
    \texttt{S}&=\texttt{S}^{+}+\texttt{S}^{-}\,, \quad \quad &\texttt{P}_{a}&=\frac{1}{\ell}\left(\texttt{G}^{+}_{a}+\texttt{G}^{-}_{a}\right)\,, \quad \quad & \texttt{M}&=\frac{1}{\ell}\left(\texttt{S}^{+}-\texttt{S}^{-}\right)\,,\notag\\
    \texttt{G}_{ab}&=\texttt{G}^{+}_{ab}+\texttt{G}^{-}_{ab}\,, \quad \quad &\texttt{J}_{a}&=\texttt{J}^{+}_{a}-\texttt{J}^{-}_{a}\,, \quad \quad &\texttt{P}_{ab}&=\frac{1}{\ell}\left(\texttt{G}^{+}_{ab}-\texttt{G}^{-}_{ab}\right)\,, \notag \\
    \texttt{S}_{a}&=\texttt{S}^{+}_{a}-\texttt{S}^{-}_{a}\,, \quad \quad &\texttt{H}_{a}&=\frac{1}{\ell}\left(\texttt{J}^{+}_{a}+\texttt{J}^{-}_{a}\right)\,, \quad \quad &\texttt{M}_{a}&=\frac{1}{\ell}\left(\texttt{S}^{+}_{a}+\texttt{S}^{-}_{a}\right)\,. \label{rdef2}
\end{align}
One can notice that the spin-2 subalgebra spanned by $\{\texttt{J}^{\pm},\texttt{G}^{\pm}_{a},\texttt{S}^{\pm}\}$ in \eqref{hs3nw} defines two copies of the usual Nappi-Witten algebra. The Nappi-Witten algebra can be seen as the central extension of the Euclidean algebra in two spacetime dimensions and result to be useful in the description of the Quantum Hall effect \cite{Duval:2000xr,Horvathy:2004am,Salgado-Rebolledo:2021wtf}. Its spin-3 extension is new and could be useful to derive novel non-relativistic spin-3 algebras following the procedure used in \cite{Concha:2019lhn,Penafiel:2019czp,Concha:2020sjt,Concha:2020eam}. Although the basis based on spin-3 Nappi-Witten algebra \eqref{hs3nw} seems more practical to construct a CS action, it is the basis \eqref{Sexp} which will allows us to construct a CS action for the $\mathfrak{hs_{3}enh}$ algebra with a proper flat limit reproducing the $\mathfrak{hs_{3}ebarg}$ CS gravity action.
\subsubsection{Post-Newtonian extension of the Spin-3 Extended Newton-Hooke Algebra}
A bigger semigroup allows us to derive a larger non-relativistic version of the $\mathfrak{hs_{3}ADS}$ algebra. Indeed, let us consider $S_{E}^{\left(4\right)}=\{\lambda_0,\lambda_1,\lambda_2,\lambda_3,\lambda_4,\lambda_5\}$ as the relevant semigroup where $\lambda_5=0_S$ is the zero element of the semigroup and whose elements satisfy the following multiplication law 
\begin{equation}
\lambda _{\alpha }\lambda _{\beta }=\left\{ 
\begin{array}{lcl}
\lambda _{\alpha +\beta }\,\,\,\, & \mathrm{if}\,\,\,\,\alpha +\beta \leq
5\,, &  \\ 
\lambda _{N+1}\,\,\, & \mathrm{if}\,\,\,\,\alpha +\beta >5\,, & 
\end{array}%
\right.   \label{mlSE2}
\end{equation}%
After applying a resonant $S_{E}^{\left(4\right)}$-expansion and extracting a $0_S$-reduction we get a non-relativistic expanded algebra whose generators are related to the relativistic ones through the semigroup elements as follows:
\begin{equation}
    \begin{tabular}{lll}
\multicolumn{1}{l|}{$\lambda_5$} & \multicolumn{1}{|l}{\cellcolor[gray]{0.8}} & \multicolumn{1}{|l|}{\cellcolor[gray]{0.8}} \\ \hline
\multicolumn{1}{l|}{$\lambda_4$} & \multicolumn{1}{|l}{$\texttt{Z},\ \texttt{Y},\ \, \texttt{Z}_{a},\ \texttt{Y}_{a},\,$} & \multicolumn{1}{|l|}{\cellcolor[gray]{0.8}} \\ \hline
\multicolumn{1}{l|}{$\lambda_3$} & \multicolumn{1}{|l}{\cellcolor[gray]{0.8}} & \multicolumn{1}{|l|}{$\texttt{B}_a,\,\texttt{T}_a,\,\texttt{S}_{ab},\,\texttt{M}_{ab},\,$} \\ \hline
\multicolumn{1}{l|}{$\lambda_2$} & \multicolumn{1}{|l}{$\texttt{S},\ \texttt{M},\ \, \texttt{S}_{a},\ \texttt{M}_{a},\,$} & \multicolumn{1}{|l|}{\cellcolor[gray]{0.8}} \\ \hline
\multicolumn{1}{l|}{$\lambda_1$} & \multicolumn{1}{|l}{\cellcolor[gray]{0.8}} & \multicolumn{1}{|l|}{$\texttt{G}_a,\,\texttt{P}_a,\,\texttt{G}_{ab},\,\texttt{P}_{ab},\,$} \\ \hline
\multicolumn{1}{l|}{$\lambda_0$} & \multicolumn{1}{|l}{$ \texttt{J},\,\ \texttt{H},\ \texttt{J}_{a},\ \texttt{H}_{a},\,$} & \multicolumn{1}{|l|}{\cellcolor[gray]{0.8}} \\ \hline
\multicolumn{1}{l|}{} & \multicolumn{1}{|l}{$\hat{\texttt{J}},\ \, \hat{\texttt{H}},\  \hat{\texttt{J}}_{a},\ \hat{\texttt{H}}_{a},\,$} & \multicolumn{1}{|l|}{$\hat{\texttt{G}}_{a},\ \hat{\texttt{P}}_{a},\,\hat{\texttt{G}}_{ab},\,\hat{\texttt{P}}_{ab}$}
\end{tabular}\label{Sexp2}%
\end{equation}
One can see that the expanded generators satisfy \eqref{enh3} along the following commutation relations:
\begin{align}
\left[\texttt{J},\texttt{B}_a\right]&=\epsilon_{ab}\texttt{B}_{b}\,,\quad \quad   &\left[\texttt{G}_{a},\texttt{B}_{b}\right]&=-\epsilon_{ab}\texttt{Z}\,,\quad  &\left[\texttt{H},\texttt{B}_a\right]&=\epsilon_{ab}\texttt{T}_{b}\,, \notag\\
\left[\texttt{J},\texttt{T}_a\right]&=\epsilon_{ab}\texttt{T}_{b}\,,\quad \quad   &\left[\texttt{G}_{a},\texttt{T}_{b}\right]&=-\epsilon_{ab}\texttt{Y}\,,\quad  &\left[\texttt{H},\texttt{T}_a\right]&=\frac{1}{\ell^2}\epsilon_{ab}\texttt{B}_{b}\,, \notag\\
\left[\texttt{J},\texttt{Z}_a\right]&=\epsilon_{ab}\texttt{Z}_{b}\,,\quad \quad   &\left[\texttt{P}_{a},\texttt{B}_{b}\right]&=-\epsilon_{ab}\texttt{Y}\,,\quad   &\left[\texttt{H},\texttt{Z}_a\right]&=\epsilon_{ab}\texttt{Y}_{b}\,, \notag\\
\left[\texttt{J},\texttt{Y}_a\right]&=\epsilon_{ab}\texttt{Y}_{b}\,,\quad \quad   &\left[\texttt{P}_{a},\texttt{T}_{b}\right]&=-\frac{1}{\ell^2}\epsilon_{ab}\texttt{Z}\,,\quad   &\left[\texttt{H},\texttt{Y}_a\right]&=\frac{1}{\ell^2}\epsilon_{ab}\texttt{Z}_{b}\,, \notag\\
\left[\texttt{S},\texttt{G}_a\right]&=\epsilon_{ab}\texttt{B}_{b}\,,\quad \quad   &\left[\texttt{G}_{a},\texttt{S}_{b}\right]&=-\left(\epsilon_{am}\texttt{S}_{mb}+\epsilon_{ab}\texttt{S}_{mm}\right)\,,\quad  &\left[\texttt{M},\texttt{G}_a\right]&=\epsilon_{ab}\texttt{T}_{b}\,, \notag\\
\left[\texttt{S},\texttt{P}_a\right]&=\epsilon_{ab}\texttt{T}_{b}\,,\quad \quad   &\left[\texttt{G}_{a},\texttt{M}_{b}\right]&=-\left(\epsilon_{am}\texttt{M}_{mb}+\epsilon_{ab}\texttt{M}_{mm}\right)\,,\quad  &\left[\texttt{M},\texttt{P}_a\right]&=\frac{1}{\ell^2}\epsilon_{ab}\texttt{B}_{b}\,, \notag\\
\left[\texttt{Z},\texttt{J}_a\right]&=\epsilon_{ab}\texttt{Z}_{b}\,,\quad \quad   &\left[\texttt{P}_{a},\texttt{S}_{b}\right]&=-\left(\epsilon_{am}\texttt{M}_{mb}+\epsilon_{ab}\texttt{M}_{mm}\right)\,,\quad  &\left[\texttt{Y},\texttt{J}_a\right]&=\epsilon_{ab}\texttt{Y}_{b}\,, \notag\\
\left[\texttt{Z},\texttt{H}_a\right]&=\epsilon_{ab}\texttt{Y}_{b}\,,\quad \quad   &\left[\texttt{P}_{a},\texttt{M}_{b}\right]&=-\frac{1}{\ell^2}\left(\epsilon_{am}\texttt{S}_{mb}+\epsilon_{ab}\texttt{S}_{mm}\right)\,,\quad  &\left[\texttt{Y},\texttt{H}_a\right]&=\frac{1}{\ell^2}\epsilon_{ab}\texttt{Z}_{b}\,, \notag\\
\left[\texttt{J},\texttt{S}_{ab}\right]&=-\epsilon_{m\left(a\right.}\texttt{S}_{\left.b\right)m}\,, \quad \quad &\left[\texttt{B}_{a},\texttt{J}_{b}\right]&=-\left(\epsilon_{am}\texttt{S}_{mb}+\epsilon_{ab}\texttt{S}_{mm}\right)\,,\quad  &\left[\texttt{H},\texttt{S}_{ab}\right]&=-\epsilon_{m\left(a\right.}\texttt{M}_{\left.b\right)m}\,, \notag \\
\left[\texttt{J},\texttt{M}_{ab}\right]&=-\epsilon_{m\left(a\right.}\texttt{M}_{\left.b\right)m}\,, \quad \quad &\left[\texttt{B}_{a},\texttt{H}_{b}\right]&=-\left(\epsilon_{am}\texttt{M}_{mb}+\epsilon_{ab}\texttt{M}_{mm}\right)\,,\quad  &\left[\texttt{H},\texttt{M}_{ab}\right]&=-\frac{1}{\ell^2}\epsilon_{m\left(a\right.}\texttt{S}_{\left.b\right)m}\,, \notag \\
\left[\texttt{S},\texttt{G}_{ab}\right]&=-\epsilon_{m\left(a\right.}\texttt{S}_{\left.b\right)m}\,, \quad \quad &\left[\texttt{T}_{a},\texttt{J}_{b}\right]&=-\left(\epsilon_{am}\texttt{M}_{mb}+\epsilon_{ab}\texttt{M}_{mm}\right)\,,\quad  &\left[\texttt{M},\texttt{G}_{ab}\right]&=-\epsilon_{m\left(a\right.}\texttt{M}_{\left.b\right)m}\,, \notag \\
\left[\texttt{S},\texttt{P}_{ab}\right]&=-\epsilon_{m\left(a\right.}\texttt{M}_{\left.b\right)m}\,, \quad \quad &\left[\texttt{T}_{a},\texttt{H}_{b}\right]&=-\frac{1}{\ell^2}\left(\epsilon_{am}\texttt{S}_{mb}+\epsilon_{ab}\texttt{S}_{mm}\right)\,,\quad  &\left[\texttt{M},\texttt{P}_{ab}\right]&=-\frac{1}{\ell^2}\epsilon_{m\left(a\right.}\texttt{S}_{\left.b\right)m}\,, \notag \\
\left[\texttt{S},\texttt{S}_a\right]&=\epsilon_{ab}\texttt{Z}_{b}\,, \quad \quad &\left[\texttt{J}_{a},\texttt{Z}_{b}\right]&=\epsilon_{ab}\texttt{Z}\,,\quad  &\left[\texttt{M},\texttt{S}_a\right]&=\epsilon_{ab}\texttt{Y}_{b}\,, \notag \\
\left[\texttt{S},\texttt{M}_a\right]&=\epsilon_{ab}\texttt{Y}_{b}\,, \quad \quad &\left[\texttt{J}_{a},\texttt{Y}_{b}\right]&=\epsilon_{ab}\texttt{Y}\,,\quad  &\left[\texttt{M},\texttt{M}_a\right]&=\frac{1}{\ell^2}\epsilon_{ab}\texttt{Z}_{b}\,, \notag \\
\left[\texttt{G}_a,\texttt{S}_{bc}\right]&=-\epsilon_{a\left(b\right.}\texttt{Z}_{\left.c\right)}\,, \quad \quad &\left[\texttt{H}_{a},\texttt{Z}_{b}\right]&=\epsilon_{ab}\texttt{Y}\,,\quad  &\left[\texttt{P}_a,\texttt{S}_{bc}\right]&=-\epsilon_{a\left(b\right.}\texttt{Y}_{\left.c\right)}\,, \notag \\
\left[\texttt{G}_a,\texttt{M}_{bc}\right]&=-\epsilon_{a\left(b\right.}\texttt{Y}_{\left.c\right)}\,, \quad \quad &\left[\texttt{H}_{a},\texttt{Y}_{b}\right]&=\frac{1}{\ell^2}\epsilon_{ab}\texttt{Z}\,,\quad  &\left[\texttt{P}_a,\texttt{M}_{bc}\right]&=-\frac{1}{\ell^2}\epsilon_{a\left(b\right.}\texttt{Z}_{\left.c\right)}\,, \notag \\
\left[\texttt{B}_a,\texttt{G}_{bc}\right]&=-\epsilon_{a\left(b\right.}\texttt{Z}_{\left.c\right)}\,, \quad \quad &\left[\texttt{S}_{a},\texttt{S}_{b}\right]&=\epsilon_{ab}\texttt{Z}\,,\quad  &\left[\texttt{T}_a,\texttt{G}_{bc}\right]&=-\epsilon_{a\left(b\right.}\texttt{Y}_{\left.c\right)}\,, \notag \\
\left[\texttt{B}_a,\texttt{P}_{bc}\right]&=-\epsilon_{a\left(b\right.}\texttt{Y}_{\left.c\right)}\,, \quad \quad &\left[\texttt{S}_{a},\texttt{M}_{b}\right]&=\epsilon_{ab}\texttt{Y}\,,\quad  &\left[\texttt{T}_a,\texttt{P}_{bc}\right]&=-\frac{1}{\ell^2}\epsilon_{a\left(b\right.}\texttt{Z}_{\left.c\right)}\,, \notag \\
\left[\texttt{S}_a,\texttt{G}_{bc}\right]&=\delta_{a\left(b\right.}\epsilon_{\left.c\right)m}\texttt{B}_{m}\,, \quad \quad &\left[\texttt{M}_{a},\texttt{M}_{b}\right]&=\frac{1}{\ell^2}\epsilon_{ab}\texttt{Z}\,,\quad  &\left[\texttt{M}_a,\texttt{G}_{bc}\right]&=\delta_{a\left(b\right.}\epsilon_{\left.c\right)m}\texttt{T}_{m}\,, \notag \\
\left[\texttt{S}_a,\texttt{P}_{bc}\right]&=\delta_{a\left(b\right.}\epsilon_{\left.c\right)m}\texttt{T}_{m}\,, \quad \quad &\left[\texttt{G}_{ab},\texttt{S}_{cd}\right]&=\delta_{\left(a\right.\left(c\right.}\epsilon_{\left.d\right)\left.b\right)}\texttt{Z}\,,\quad  &\left[\texttt{M}_a,\texttt{P}_{bc}\right]&=\frac{1}{\ell^2}\delta_{a\left(b\right.}\epsilon_{\left.c\right)m}\texttt{B}_{m}\,, \notag \\
\left[\texttt{J}_a,\texttt{S}_{bc}\right]&=\delta_{a\left(b\right.}\epsilon_{\left.c\right)m}\texttt{B}_{m}\,, \quad \quad &\left[\texttt{G}_{ab},\texttt{M}_{cd}\right]&=\delta_{\left(a\right.\left(c\right.}\epsilon_{\left.d\right)\left.b\right)}\texttt{Y}\,,\quad  &\left[\texttt{H}_a,\texttt{S}_{bc}\right]&=\delta_{a\left(b\right.}\epsilon_{\left.c\right)m}\texttt{T}_{m}\,, \notag \\
\left[\texttt{J}_a,\texttt{M}_{bc}\right]&=\delta_{a\left(b\right.}\epsilon_{\left.c\right)m}\texttt{T}_{m}\,, \quad \quad &\left[\texttt{P}_{ab},\texttt{S}_{cd}\right]&=\delta_{\left(a\right.\left(c\right.}\epsilon_{\left.d\right)\left.b\right)}\texttt{Y}\,,\quad  &\left[\texttt{H}_a,\texttt{M}_{bc}\right]&=\frac{1}{\ell^2}\delta_{a\left(b\right.}\epsilon_{\left.c\right)m}\texttt{B}_{m}\,, \notag \\
\left[\texttt{P}_{ab},\texttt{M}_{cd}\right]&=\frac{1}{\ell^2}\delta_{\left(a\right.\left(c\right.}\epsilon_{\left.d\right)\left.b\right)}\texttt{Z}\,,\label{pn3}
\end{align}
which appear by combining the multiplication law of the semigroup $S_{E}^{\left(4\right)}$ with the original commutators of the relativistic $\mathfrak{hs_{3}AdS}$ algebra. The expanded non-relativistic algebra \eqref{pn3} can be seen as a post-Newtonian extension of the $\mathfrak{hs_{3}enh}$ algebra and corresponds to a spin-3 extension of the so-called enhanced Bargmann-Newton-Hooke symmetry \cite{Bergshoeff:2020fiz}\footnote{Also denoted as exotic Newtonian algebra in \cite{Concha:2019dqs} or Newton-Hooke version of the extended Post-Newtonian algebra in \cite{Gomis:2019nih}}. The non-relativistic algebra obtained \eqref{pn3} will be denoted as $\mathfrak{hs_{3}pne}$ algebra and contains the enhanced Bargmann-Newton-Hooke algebra as a spin-2 subalgebra spanned by $\{\texttt{J},\texttt{G}_{a},\texttt{S},\texttt{H},\texttt{P}_{a},\texttt{M},\texttt{Z},\texttt{B}_{a},\texttt{Y},\texttt{T}_{a}\}$. Its spin-3 extension requires to introduce additional spin-3 generators $\{\texttt{Z}_{a},\texttt{Y}_{a},\texttt{S}_{ab},\texttt{M}_{ab}\}$ besides those appearing in the $\mathfrak{hs_{3}enh}$ algebra. In the vanishing cosmological constant limit $\ell\rightarrow\infty$, we obtain a spin-3 extension of the extended Newtonian algebra \cite{Ozdemir:2019orp} which we shall denote as $\mathfrak{hs_{3}enewt}$. Alternatively, the $\mathfrak{hs_{3}enewt}$ algebra can be obtained by considering a resonant expansion of the spin-3 Poincaré algebra with $S_{E}^{\left(4\right)}$ being the relevant semigroup.

Let us note that for $\texttt{Z}=\texttt{Y}=\texttt{Z}_{a}=\texttt{Y}_{a}=0$, the non-relativistic algebra \eqref{pn3} reduces to a spin-3 extension of the Newton-Hooke version of the Newtonian algebra which appears as the symmetry that leaves invariant the Newtonian gravity action \cite{Hansen:2018ofj}. In such case, the flat limit $\ell\rightarrow\infty$ reproduces a spin-3 extension of the Newtonian algebra. As we shall see, analogously to its spin-2 version, the additional generators $\{\texttt{Z},\texttt{Y},\texttt{Z}_{a},\texttt{Y}_{a}\}$ are required in order to define non-degenerate invariant trace. However, unlike its spin-2 version, $\texttt{Z}$ and $\texttt{Y}$ are no longer central charges but their commutators with spin-3 generators $\{\texttt{J}_{a},\texttt{H}_{a}\}$ are proportional to the respective spin-3 generators $\{\texttt{Z}_{a},\texttt{Y}_{a}\}$.

On the other hand, similarly to the $\mathfrak{hs_{3}enh}$ case, the $\mathfrak{hs_{3}pne}$ algebra \eqref{pn3} can be rewritten as two copies of a spin-3 extension of the enhanced Nappi-Witten algebra discussed in \cite{Bergshoeff:2020fiz,Concha:2020ebl,Concha:2022you}. Indeed, two copies of the spin-3 enhanced Nappi-Witten structure, which we shall denote as $\mathfrak{hs_{3}enw}$, is revealed after considering the redefinition of the $\mathfrak{hs_{3}pne}$ generators as in \eqref{rdef2} along:
\begin{align}
    \texttt{Z}&=\texttt{B}^{+}+\texttt{B}^{-}\,, \quad \quad &\texttt{B}_{a}&=\texttt{B}^{+}_{a}-\texttt{B}^{-}_{a}\,, \quad \quad &\texttt{Y}&=\frac{1}{\ell}\left(\texttt{B}^{+}-\texttt{B}^{-}\right)\,, \notag\\
    \texttt{S}_{ab}&=\texttt{S}^{+}_{ab}+\texttt{S}^{-}_{ab}\,, \quad \quad &\texttt{T}_{a}&=\frac{1}{\ell}\left(\texttt{B}^{+}_{a}+\texttt{B}^{-}_{a}\right)\,, \quad \quad &\texttt{M}_{ab}&=\frac{1}{\ell}\left(\texttt{S}^{+}_{ab}-\texttt{S}^{-}_{ab}\right)\,, \notag \\
    \texttt{Z}_{a}&=\texttt{Z}^{+}_{a}-\texttt{Z}^{-}_{a}\,, \quad \quad &\texttt{Y}_{a}&=\frac{1}{\ell}\left(\texttt{Z}^{+}_{a}+\texttt{Z}^{-}_{a}\right)\,, \label{rdef3}
\end{align}
where $\{\texttt{J}^{\pm},\texttt{G}_{a}^{\pm},\texttt{S}^{\pm},\texttt{G}_{ab}^{\pm},\texttt{J}_{a}^{\pm},\texttt{S}_{a}^{\pm},\texttt{S}_{ab}^{\pm},\texttt{B}^{\pm},\texttt{B}_{a}^{\pm},\texttt{Z}_{a}^{\pm}\}$ satisfies the usual $\mathfrak{hs_{3}nw}$ algebra \eqref{hs3nw} along the following commutation relations:
\begin{align}
\left[\texttt{J}^{\pm},\texttt{B}^{\pm}_a\right]&=\epsilon_{ab}\texttt{B}^{\pm}_{b}\,,\quad \quad   &\left[\texttt{G}^{\pm}_a,\texttt{S}^{\pm}_{bc}\right]&=-\epsilon_{a\left(b\right.}\texttt{Z}^{\pm}_{\left.c\right)}\,,\quad  &\left[\texttt{G}^{\pm}_{a},\texttt{B}^{\pm}_{b}\right]&=-\epsilon_{ab}\texttt{B}^{\pm}\,, \notag\\
\left[\texttt{J}^{\pm},\texttt{Z}^{\pm}_a\right]&=\epsilon_{ab}\texttt{Z}^{\pm}_{b}\,,\quad \quad   &\left[\texttt{J}^{\pm}_a,\texttt{S}^{\pm}_{bc}\right]&=\delta_{a\left(b\right.}\epsilon_{\left.c\right)m}\texttt{B}^{\pm}_{m}\,,\quad  &\left[\texttt{J}^{\pm}_{a},\texttt{Z}^{\pm}_{b}\right]&=\epsilon_{ab}\texttt{B}^{\pm}\,, \notag\\
\left[\texttt{J}^{\pm},\texttt{G}^{\pm}_{ab}\right]&=-\epsilon_{m\left(a\right.}\texttt{G}^{\pm}_{\left.b\right)m} \,,\quad \quad   &\left[\texttt{B}^{\pm}_a,\texttt{G}^{\pm}_{bc}\right]&=-\epsilon_{a\left(b\right.}\texttt{Z}^{\pm}_{\left.c\right)}\,,\quad &\left[\texttt{G}^{\pm}_{a},\texttt{S}^{\pm}_{b}\right]&=-\left(\epsilon_{am}\texttt{S}^{\pm}_{mb}+\epsilon_{ab}\texttt{S}^{\pm}_{mm}\right)\,, \notag\\
\left[\texttt{S}^{\pm},\texttt{G}^{\pm}_{ab}\right]&=-\epsilon_{m\left(a\right.}\texttt{S}^{\pm}_{\left.b\right)m}\,, \quad \quad &\left[\texttt{S}^{\pm}_a,\texttt{G}^{\pm}_{bc}\right]&=-\epsilon_{a\left(b\right.}\texttt{B}^{\pm}_{\left.c\right)}\,, \quad \quad &\left[\texttt{J}^{\pm}_{a},\texttt{B}^{\pm}_{b}\right]&=-\left(\epsilon_{am}\texttt{S}^{\pm}_{mb}+\epsilon_{ab}\texttt{S}^{\pm}_{mm}\right)\,. \notag\\
\left[\texttt{J}^{\pm},\texttt{S}^{\pm}_{ab}\right]&=-\epsilon_{m\left(a\right.}\texttt{S}^{\pm}_{\left.b\right)m} \,,\quad \quad &\left[\texttt{S}^{\pm},\texttt{G}^{\pm}_a\right]&=\epsilon_{ab}\texttt{B}^{\pm}_{b}\,,\quad \quad  &\left[\texttt{B}^{\pm},\texttt{J}^{\pm}_a\right]&=\epsilon_{ab}\texttt{Z}^{\pm}_{b}\,.\label{hs3enw}
\end{align}
The commutation relations given by \eqref{hs3nw} and \eqref{hs3enw} define two copies of the $\mathfrak{hs_{3}enw}$ algebra. One can note that the spin-2 subalgebra given by $\{\texttt{J}^{\pm},\texttt{G}_{a}^{\pm},\texttt{S}^{\pm},\texttt{B}_{a}^{\pm},\texttt{B}^{\pm}\}$ satisfies two copies of the enhanced Nappi-witten algebra where one copy has been useful to obtain novel Newtonian symmetries \cite{Concha:2020ebl}. In this direction, the $\mathfrak{hs_{3}enw}$ algebra could be useful to derive novel spin-3 Newtonian algebras following the procedure used in spin-2 Newtonian gravity theories.  For instance the Maxwellian version of the extended Newtonian (MENt) algebra appears as an $S_{E}^{\left(2\right)}$-expansion of the enhanced Nappi-Witten one \cite{Concha:2020ebl}. Then, one could expect that a spin-3 extension of the MENt algebra appears as $S_{E}^{\left(2\right)}$-expansion of the $\mathfrak{hs_{3}enw}$ algebra obtained here.
\subsubsection{Infinite-dimensional Spin-3 Newton-Hooke Algebra}
A generalized spin-3 Newton-Hooke algebra can be obtained considering the $S_{E}^{\left(N\right)}$ semigroup whose elements satisfy
\begin{equation}
\lambda _{\alpha }\lambda _{\beta }=\left\{ 
\begin{array}{lcl}
\lambda _{\alpha +\beta }\,\,\,\, & \mathrm{if}\,\,\,\,\alpha +\beta \leq
N+1\,, &  \\ 
\lambda _{N+1}\,\,\, & \mathrm{if}\,\,\,\,\alpha +\beta >N+1\,, & 
\end{array}%
\right.   \label{mlSEN}
\end{equation}%
where $\lambda_{N+1}=0_S$ is the zero element of the semigroup. Let us consider now a subset decomposition of the semigroup $S_{E}^{\left(N\right)}=S_{0}\cup S_{1}$ with
\begin{eqnarray}
S_{0} &=&\left\{ \lambda _{2m},\ \text{with }m=0,\ldots ,\left[ \frac{N}{2}%
\right] \right\} \cup \{\lambda _{N+1}\}\,, \notag \\
S_{1} &=&\left\{ \lambda _{2m+1},\ \text{with }m=0,\ldots ,\left[ \frac{N+1}{%
2}\right] \right\} \cup \{\lambda _{N+1}\}\,, \label{sdN}
\end{eqnarray}%
where $[\ldots ]$ denotes the integer part. One can notice that the subset decomposition \eqref{sdN} is resonant since it satisfies the same algebraic structure than the subspace decomposition \eqref{sd2}. Then, an expanded algebra is obtained after considering a resonant $S_{E}^{\left(N\right)}$-expansion and applying a $0_S$-reduction. One can show that the expanded generators
\begin{align}
\texttt{J}^{(m)} &=\lambda _{2m}\hat{\texttt{J}}\,, \quad 
&\texttt{G}_{a}^{(m)} &=\lambda _{2m+1}\hat{\texttt{G}}_{a}\,,  \quad &\texttt{G}_{ab}^{(m)} &=\lambda _{2m+1}\hat{\texttt{G}}_{ab}\,, \quad &\texttt{J}_{a}^{(m)} &=\lambda _{2m}\hat{\texttt{J}}_{a}\,,\notag \\
\texttt{H}^{(m)} &=\lambda _{2m}\hat{\texttt{H}}\,,  \quad
&\texttt{P}_{a}^{(m)} &=\lambda_{2m+1}\hat{\texttt{P}}_{a}\,, \quad &\texttt{P}_{ab}^{(m)} &=\lambda _{2m+1}\hat{\texttt{P}}_{ab}\,, \quad &\texttt{H}_{a}^{(m)} &=\lambda _{2m}\hat{\texttt{H}}_{a}\,. \label{expgen}
\end{align}
satisfy a generalized spin-2 Newton-Hooke algebra:
\begin{align}
    \left[\texttt{J}^{(m)},\texttt{G}_{a}^{(n)}\right]&=\epsilon_{ab}\texttt{G}_{b}^{(m+n)}\,, &\left[\texttt{G}_{a}^{(m)},\texttt{G}_{b}^{(n)}\right]&=-\epsilon_{ab}\texttt{J}^{(m+n+1)}\,, \quad &\left[\texttt{H}^{(m)},\texttt{G}_{a}^{(n)}\right]&=\epsilon_{ab}\texttt{P}_{b}^{(m+n)}\,, \notag\\
    \left[\texttt{J}^{(m)},\texttt{P}_a^{(n)}\right]&=\epsilon_{ab}\texttt{P}_{b}^{(m+n)}\,,  &\left[\texttt{G}_{a}^{(m)},\texttt{P}_{b}^{(n)}\right]&=-\epsilon_{ab}\texttt{H}^{(m+n+1)}\,, \quad &\left[\texttt{H}^{(m)},\texttt{P}_a^{(n)}\right]&=\frac{1}{\ell^2}\epsilon_{ab}\texttt{G}_{b}^{(m+n)}\,, \notag\\
    \left[\texttt{P}_{a}^{(m)},\texttt{P}_{b}^{(n)}\right]&=-\frac{1}{\ell^2}\epsilon_{ab}\texttt{J}^{(m+n+1)}\,, \label{idNH}
\end{align}
along its spin-3 extensions:
\begin{align}
\left[\texttt{J}^{(m)},\texttt{J}_a^{(n)}\right]&=\epsilon_{ab}\texttt{J}_{b}^{(m+n)}\,,   &\left[\texttt{H}^{(m)},\texttt{J}_a^{(n)}\right]&=\epsilon_{ab}\texttt{H}_{b}^{(m+n)}\,, \notag\\
\left[\texttt{J}^{(m)},\texttt{H}_a^{(n)}\right]&=\epsilon_{ab}\texttt{H}_{b}^{(m+n)}\,,   &\left[\texttt{H}^{(m)},\texttt{H}_a^{(n)}\right]&=\frac{1}{\ell^2}\epsilon_{ab}\texttt{J}_{b}^{(m+n)}\,, \notag\\
\left[\texttt{J}^{(m)},\texttt{G}_{ab}^{(n)}\right]&=-\epsilon_{c\left(a\right.}\texttt{G}_{\left.b\right)c}^{(m+n)}\,,  &\left[\texttt{H}^{(m)},\texttt{G}_{ab}^{(n)}\right]&=-\epsilon_{c\left(a\right.}\texttt{P}_{\left.b\right)c}^{(m+n)}\,, \notag \\
\left[\texttt{J}^{(m)},\texttt{P}_{ab}^{(n)}\right]&=-\epsilon_{c\left(a\right.}\texttt{P}_{\left.b\right)c}^{(m+n)}\,,   &\left[\texttt{H}^{(m)},\texttt{P}_{ab}^{(n)}\right]&=-\frac{1}{\ell^2}\epsilon_{c\left(a\right.}\texttt{G}_{\left.b\right)c}^{(m+n)}\,, \notag \\
\left[\texttt{G}_a^{(m)},\texttt{G}_{bc}^{(n)}\right]&=-\epsilon_{a\left(b\right.}\texttt{J}_{\left.c\right)}^{(m+n+1)}\,,  &\left[\texttt{P}_a^{(m)},\texttt{G}_{bc}^{(n)}\right]&=-\epsilon_{a\left(b\right.}\texttt{H}_{\left.c\right)}^{(m+n+1)}\,, \notag \\
\left[\texttt{G}_{a}^{(m)},\texttt{J}_{b}^{(n)}\right]&=-\left(\epsilon_{ac}\texttt{G}_{cb}^{(m+n)}+\epsilon_{ab}\texttt{G}_{cc}^{(m+n)}\right)\,, &\left[\texttt{G}_a^{(m)},\texttt{P}_{bc}^{(n)}\right]&=-\epsilon_{a\left(b\right.}\texttt{H}_{\left.c\right)}^{(m+n+1)}\,, \notag \\
\left[\texttt{P}_{a}^{(m)},\texttt{J}_{b}^{(n)}\right]&=-\left(\epsilon_{ac}\texttt{P}_{cb}^{(m+n)}+\epsilon_{ab}\texttt{P}_{cc}^{(m+n)}\right)\,, &\left[\texttt{P}_a^{(m)},\texttt{P}_{bc}^{(n)}\right]&=-\frac{1}{\ell^2}\epsilon_{a\left(b\right.}\texttt{J}_{\left.c\right)}^{(m+n+1)}\,, \notag\\
\left[\texttt{G}_{a}^{(m)},\texttt{H}_{b}^{(n)}\right]&=-\left(\epsilon_{ac}\texttt{P}_{cb}^{(m+n)}+\epsilon_{ab}\texttt{P}_{cc}^{(m+n)}\right)\,, &\left[\texttt{J}_a^{(m)},\texttt{G}_{bc}^{(n)}\right]&=\delta_{a\left(b\right.}\epsilon_{\left.c\right)d}\texttt{G}_{d}^{(m+n)}\,,  \notag \\
\left[\texttt{P}_{a}^{(m)},\texttt{H}_{b}^{(n)}\right]&=-\frac{1}{\ell^2}\left(\epsilon_{ac}\texttt{G}_{cb}^{(m+n)}+\epsilon_{ab}\texttt{G}_{cc}^{(m+n)}\right)\,,  &\left[\texttt{H}_a^{(m)},\texttt{G}_{bc}^{(n)}\right]&=\delta_{a\left(b\right.}\epsilon_{\left.c\right)d}\texttt{P}_{d}^{(m+n)}\,, \notag \\
\left[\texttt{J}_a^{(m)},\texttt{P}_{bc}^{(n)}\right]&=\delta_{a\left(b\right.}\epsilon_{\left.c\right)d}\texttt{P}_{d}^{(m+n)}\,,   &\left[\texttt{H}_a^{(m)},\texttt{P}_{bc}^{(n)}\right]&=\frac{1}{\ell^2}\delta_{a\left(b\right.}\epsilon_{\left.c\right)d}\texttt{G}_{d}^{(m+n)}\,, \notag \\
\left[\texttt{G}_{ab}^{(m)},\texttt{G}_{cd}^{(n)}\right]&=\delta_{\left(a\right.\left(c\right.}\epsilon_{\left.d\right)\left.b\right)}\texttt{J}^{(m+n+1)}\,,  &\left[\texttt{G}_{ab}^{(m)},\texttt{P}_{cd}^{(n)}\right]&=\delta_{\left(a\right.\left(c\right.}\epsilon_{\left.d\right)\left.b\right)}\texttt{H}^{(m+n+1)}\,, \notag\\ \left[\texttt{P}_{ab}^{(m)},\texttt{P}_{cd}^{(n)}\right]&=\frac{1}{\ell^2}\delta_{\left(a\right.\left(c\right.}\epsilon_{\left.d\right)\left.b\right)}\texttt{J}^{(m+n+1)}\,, 
&\left[\texttt{J}_{a}^{(m)},\texttt{J}_{b}^{(n)}\right]&=\epsilon_{ab}\texttt{J}^{(m+n)}\,,\notag\\ \left[\texttt{J}_{a}^{(m)},\texttt{H}_{b}^{(n)}\right]&=\epsilon_{ab}\texttt{H}^{(m+n)}\,, 
&\left[\texttt{H}_{a}^{(m)},\texttt{H}_{b}^{(n)}\right]&=\frac{1}{\ell^2}\epsilon_{ab}\texttt{J}^{(m+n)}\,. \label{idNH3}
\end{align}
The new non-relativistic algebra \eqref{idNH}-\eqref{idNH3} defines a generalized spin-3 Newton-Hooke algebra and will be denoted as $\mathfrak{hs_{3}nh}^{\left(N\right)}$. Let us note that the $0_S$-reduction condition implies $T_{A}^{\left(N+1\right)}=\lambda_{N+1}T_A=0$ which abelianize a large sector of the expanded algebra. Indeed, for $\left(m+n\right)\geq N+1$ the commutators vanish. For an infinite $N$, the semigroup does not contain zero element and the infinite-dimensional expanded algebra does not admit abelian commutators in \eqref{idNH}-\eqref{idNH3}. In such case, the vanishing cosmological constant limit $\ell\rightarrow\infty$ leads to a spin-3 extension of the infinite-dimensional Galilei algebra discussed in \cite{Gomis:2019fdh,Gomis:2019nih,Gomis:2022spp,Concha:2022you}.

For $N=1$, we obtain a spin-3 extension of the Newton-Hooke algebra which is the simplest finite structure with spin-3 that we can obtain by expanding the $\mathfrak{hs_{3}ADS}$ algebra. In the flat limit, the $\mathfrak{hs_{3}nh}^{\left(1\right)}$ reduces to a spin-3 Galilei algebra \footnote{The $\mathfrak{hs_{3}nh}^{\left(1\right)}$ algebra obtained here along its flat limit correspond to the respective $\mathfrak{hs_{3}nh1}$ and $\mathfrak{hs_{3}gal1}$ algebras introduced in \cite{Bergshoeff:2016soe}}. Although both spin-3 Newton-Hooke and spin-3 Galilei algebras are well-defined they do not admit a non-degenerate bilinear invariant trace. As we shall see, even values of $N$ provides us with non-relativistic symmetries admitting non-degenerate invariant tensor allowing us to construct proper CS non-relativistic actions in three spacetime dimensions. In particular, for $N=2$ and $N=4$, we recover the $\mathfrak{hs_{3}enh}$ and $\mathfrak{hs_{3}pne}$ algebras obtained previously. The particular case $N=3$ reproduces the spin-3 extension of the Newton-Hooke version of the Newtonian algebra appearing in \cite{Hansen:2018ofj} which appears as a particular subcase of the $\mathfrak{hs_{3}pne}$ algebra. The following table summarize the non-relativistic versions of the spin-3 AdS algebra along its post-Newtonian extensions and flat limit.
\begin{equation}
    \begin{tabular}{|m{6em}||c|c|c|c|c|c|}
\hline
 Relativistic spin-3 algebra & $N=1$ & $N=2$ & $N=3$ & $N=4$ & $\cdots$ & $N$ \\ \hline\hline
$\mathfrak{hs_{3}AdS}$ & \,$\mathfrak{hs_{3}nh}$ \, & \, $\mathfrak{hs_{3}enh}$ \,  & \, $\mathfrak{hs_{3}nhNewt}$ \, & \, $\mathfrak{hs_{3}pne}$ \, &  $\cdots$ & \, $\mathfrak{hs_{3}nh}^{\left(N\right)}$\,  \\ \hline
$\mathfrak{hs_{3}poi}$ &  $\mathfrak{hs_{3}gal}$ & \ $\mathfrak{hs_{3}ebarg}$\ & $\mathfrak{hs_{3}Newt}$ & $\mathfrak{hs_{3}eNewt}$ & $\cdots$ & \,$\mathfrak{hs_{3}gal}^{\left(N\right)}$\, \\ \hline
\end{tabular}\notag
\end{equation}
Here, the second line can be obtained as a resonant $S_{E}^{\left(N\right)}$-expansion of the relativistic spin-3 Poincaré algebra. For each particular value of $N$, the $\mathfrak{hs_{3}gal}^{\left(N\right)}$ algebra appears as a vanishing cosmological constant limit $\ell\rightarrow\infty$ of the $\mathfrak{hs_{3}nh}^{\left(N\right)}$ algebra.

Interestingly, the $\mathfrak{hs_{3}nh}^{\left(N\right)}$ algebra can be rewritten as two copies of a generalized spin-3 Nappi-Witten \eqref{hs3nw}:
\begin{align}
\left[\texttt{J}^{(m)\pm},\texttt{G}^{(n)\pm}_a\right]&=\epsilon_{ab}\texttt{G}^{(m+n)\pm}_{b}\,,  &\left[\texttt{J}^{(m)\pm},\texttt{J}^{(n)\pm}_a\right]&=\epsilon_{ab}\texttt{J}^{(m+n)\pm}_{b}\,,\notag \\
\left[\texttt{G}^{(m)\pm}_{a},\texttt{G}^{(n)\pm}_{b}\right]&=-\epsilon_{ab}\texttt{J}^{(m+n+1)\pm}\,,
&\left[\texttt{G}^{(m)\pm}_a,\texttt{G}^{(n)\pm}_{bc}\right]&=-\epsilon_{a\left(b\right.}\texttt{J}^{(m+n+1)\pm}_{\left.c\right)}\,,\notag\\  \left[\texttt{J}^{(m)\pm}_a,\texttt{G}^{(n)\pm}_{bc}\right]&=\delta_{a\left(b\right.}\epsilon_{\left.c\right)d}\texttt{G}^{(m+n)\pm}_{d}\,, &\left[\texttt{J}^{(m)\pm}_{a},\texttt{J}^{(n)\pm}_{b}\right]&=\epsilon_{ab}\texttt{J}^{(m+n)\pm}\,, \notag\\
\left[\texttt{J}^{(m)\pm},\texttt{G}^{(n)\pm}_{ab}\right]&=-\epsilon_{c\left(a\right.}\texttt{G}^{(m+n)\pm}_{\left.b\right)c} \,,\quad \quad   &\left[\texttt{G}^{(m)\pm}_{ab},\texttt{G}^{(n)\pm}_{cd}\right]&=\delta_{\left(a\right.\left(c\right.}\epsilon_{\left.d\right)\left.b\right)}\texttt{J}^{(m+n+1)\pm}\,,\notag\\ \left[\texttt{G}^{(m)}_{a},\texttt{J}^{(n)\pm}_{b}\right]&=-\left(\epsilon_{ac}\texttt{G}^{(m+n)\pm}_{cb}+\epsilon_{ab}\texttt{G}^{(m+n)\pm}_{cc}\right)\,. \label{hs3nwN}
\end{align}
These two copies, which we shall denote as $\mathfrak{hs_{3}nw}^{\left(N\right)}$, are obtained after considering the redefinition of the $\mathfrak{hs_{3}nh}^{\left(N\right)}$ generators as
\begin{align}
    \texttt{J}^{(m)}&=\texttt{J}^{(m)+}+\texttt{J}^{(m)-}\,, \quad \quad &\texttt{G}_{a}^{(m)}&=\texttt{G}^{(m)+}_{a}-\texttt{G}^{(m)-}_{a}\,, \notag\\ \texttt{H}^{(m)}&=\frac{1}{\ell}\left(\texttt{J}^{(m)+}-\texttt{J}^{(m)-}\right)\,, \quad \quad &\texttt{P}_{a}^{(m)}&=\frac{1}{\ell}\left(\texttt{G}^{(m)+}_{a}+\texttt{G}^{(m)-}_{a}\right)\,, \notag\\ 
    \texttt{G}_{ab}^{(m)}&=\texttt{G}^{(m)+}_{ab}+\texttt{G}^{(m)-}_{ab}\,, \quad \quad &\texttt{J}_{a}^{(m)}&=\texttt{J}^{(m)+}_{a}-\texttt{J}^{(m)-}_{a}\,, \notag\\ \texttt{P}_{ab}^{(m)}&=\frac{1}{\ell}\left(\texttt{G}^{(m)+}_{ab}-\texttt{G}^{(m)-}_{ab}\right)\,, \quad \quad
     &\texttt{H}_{a}^{(m)}&=\frac{1}{\ell}\left(\texttt{J}^{(m)+}_{a}+\texttt{J}^{(m)-}_{a}\right)\,. \label{rdef4}
\end{align}
For $N=2$ and $N=4$, the $\mathfrak{hs_{3}nw}^{\left(N\right)}$ algebra reproduces the spin-3 Nappi-Witten \eqref{hs3nw} and its enhanced version \eqref{hs3enw}, respectively. One can note that the spin-2 subalgebra spanned by the generators $\{\texttt{J}^{(m)\pm},\texttt{G}_{a}^{(m)\pm}\}$ define two copies of the generalized Nappi-Witten algebra discussed in \cite{Gomis:2019nih,Concha:2022you}.

It is important to mention that the spin-3 extension of the infinite-dimensional Newton-Hooke algebra is not unique. Indeed an alternative spin-3 infinite-dimensional Newton-Hooke algebra can be obtained considering a different subspace decomposition of the $\mathfrak{hs_{3}AdS}$ algebra. Further details can be found in Appendix \ref{App1}. Although the alternative spin-3 non-relativistic family also admits non-degenerate invariant tensors, we shall only explore the construction of CS gravity action based on the spin-3 non-relativistic symmetries discussed in section \ref{sec3}.

\subsection{Ultra-relativistic Expansions of Spin-3 AdS Algebra}
Analogously to the previous NR expansions, the
ultra-relativistic expansions of the Spin-3 AdS algebra require to consider a subspace decomposition of the relativistic algebra. In this case, the subspaces $V_0$ and $V_1$ are given by 
\begin{align}
   V_0&=\{\hat{\texttt{J}},\hat{\texttt{P}}_{a},\hat{\texttt{J}}_{a},\hat{\texttt{P}}_{ab}\} \,,\notag \\
   V_1&=\{\hat{\texttt{G}}_{a},\hat{\texttt{H}},\hat{\texttt{H}}_{a},\hat{\texttt{G}}_{ab}\}\,,\label{uvsd}
\end{align}
which is also a $\mathbb{Z}_2$ graded-Lie algebra satisfying \eqref{sd2}. Here, we can see that the generators $\hat{\texttt{P}}_{a}$ and $\hat{\texttt{H}}$, as well as the Spin-3 generators $\hat{\texttt{P}}_{ab}$ and $\hat{\texttt{H}}_{a}$, have been interchanged with respect to the subspace decomposition considered in the NR expansion \eqref{sd}.
In what follows, we shall define the ultra-relativistic expansions of the Spin-3 AdS algebra, by choosing the same semigroup than in the NR expansions, namely $S_{E}^{(N)}$. Before considering the general $N$ case, we will approach first the $N=1$ and $N=2$ cases, reproducing the Spin-3 extension of the AdS Carroll and extended AdS Carroll algebras, respectively.

\subsubsection{Spin-3 AdS Carroll Algebra}
Let $S_{E}^{\left(1\right)}=\{\lambda_0,\lambda_1,\lambda_2\}$ be the relevant semigroup where $\lambda_2=0_S$ is the zero element of the semigroup satisfying $0_S \lambda_i=0_S$ for $i=0,\ldots,2$. The elements of the semigroup $S_{E}^{\left(2\right)}$ satisfy the following multiplication law
\begin{equation}
\begin{tabular}{l|lll}
$\lambda _{2}$ & $\lambda _{2}$ & $\lambda _{2}$ & $\lambda _{2}$ \\
$\lambda _{1}$ & $\lambda _{1}$ & $\lambda _{2}$ & $\lambda _{2}$ \\
$\lambda _{0}$ & $\lambda _{0}$ & $\lambda _{1}$ & $\lambda _{2}$  \\ \hline
& $\lambda _{0}$ & $\lambda _{1}$ & $\lambda _{2}$
\end{tabular}
\label{se1}
\end{equation}
A spin-3 version of the AdS-Carroll algebra appears after applying a resonant $S_{E}^{\left(1\right)}$-expansion of the $\mathfrak{hs_{3}AdS}$ algebra and extracting a $0_S$-reduction. In particular, the spin-3 AdS-Carroll generators are related to the relativistic ones through the semigroup elements as follows:
\begin{equation}
    \begin{tabular}{lll}
\multicolumn{1}{l|}{$\lambda_2$} & \multicolumn{1}{|l|}{\cellcolor[gray]{0.8}} & \multicolumn{1}{|l|}{\cellcolor[gray]{0.8}} \\ \hline
\multicolumn{1}{l|}{$\lambda_1$} & \multicolumn{1}{|l}{\cellcolor[gray]{0.8}} & \multicolumn{1}{|l|}{$\texttt{G}_a,\,\texttt{H},\,\texttt{H}_{a},\,\texttt{G}_{ab},\,$} \\ \hline
\multicolumn{1}{l|}{$\lambda_0$} & \multicolumn{1}{|l}{$ \texttt{J},\,\ \texttt{P}_{a},\ \texttt{J}_{a},\ \texttt{P}_{ab},\,$} & \multicolumn{1}{|l|}{\cellcolor[gray]{0.8}} \\ \hline
\multicolumn{1}{l|}{} & \multicolumn{1}{|l}{$\hat{\texttt{J}},\ \, \hat{\texttt{P}}_{a},\  \hat{\texttt{J}}_{a},\ \hat{\texttt{P}}_{ab},\,$} & \multicolumn{1}{|l|}{$\hat{\texttt{G}}_{a},\ \hat{\texttt{H}},\,\hat{\texttt{H}}_{a},\,\hat{\texttt{G}}_{ab}$}
\end{tabular}\label{Sexpuv}
\end{equation}
The commutation relations for the UR expanded generators are obtained considering the commutation relations of the relativistic $\mathfrak{hs_{3}AdS}$ algebra and the multiplication law of the semigroup $S_{E}^{\left(2\right)}$. Indeed, the expanded generators satisfy the following commutators: 
 
\begin{align}
    \left[\texttt{J},\texttt{G}_{a}\right]&=\epsilon_{ab}\texttt{G}_{b}\,, & \left[\texttt{G}_{a},\texttt{P}_{b}\right]&=-\epsilon_{ab}\texttt{H}\,, & 
    \left[\texttt{H},\texttt{P}_{a}\right]&=\frac{1}{\ell^{2}}\epsilon_{ab}\texttt{G}_{b}\,,\notag \\
     \left[\texttt{J},\texttt{P}_{a}\right]&=\epsilon_{ab}\texttt{P}_{b}\,, & 
     \left[\texttt{P}_{a},\texttt{P}_{b}\right]&=-\frac{1}{\ell^{2}}\epsilon_{ab}\texttt{J}\,, & 
     \left[\texttt{H},\texttt{J}_{a}\right]&=\epsilon_{ab}\texttt{H}_{b}\,, \notag\\
      \left[\texttt{J},\texttt{H}_{a}\right]&=\epsilon_{ab}\texttt{H}_{b}\,, & 
      \left[\texttt{G}_{a},\texttt{J}_{b}\right]&=-\left(\epsilon_{am}\texttt{G}_{mb}+\epsilon_{ab}\texttt{G}_{mm}\right)\,, & 
      \left[\texttt{H},\texttt{P}_{ab}\right]&=-\frac{1}{\ell^2}\epsilon_{m\left(a\right.}\texttt{G}_{\left.b\right)m}\,,\notag \\
      \left[\texttt{J},\texttt{J}_{a}\right]&=\epsilon_{ab}\texttt{J}_{b}\,, & 
      \left[\texttt{P}_{a},\texttt{H}_{b}\right]&=-\frac{1}{\ell^2}\left(\epsilon_{am}\texttt{G}_{mb}+\epsilon_{ab}\texttt{G}_{mm}\right)\,, & \left[\texttt{P}_{a},\texttt{P}_{bc}\right]&=-\frac{1}{\ell^2}\epsilon_{a\left(b\right.}\texttt{J}_{\left.c\right)}\,,\notag \\
      \left[\texttt{J},\texttt{P}_{ab}\right]&=-\epsilon_{m\left(a\right.}\texttt{P}_{\left.b\right)m}\,, & \left[\texttt{P}_{a},\texttt{J}_{b}\right]&=-\left(\epsilon_{am}\texttt{P}_{mb}+\epsilon_{ab}\texttt{P}_{mm}\right)\,, & \left[\texttt{P}_{a},\texttt{G}_{bc}\right]&=-\epsilon_{a\left(b\right.}\texttt{H}_{\left.c\right)}\,, \notag \\
      \left[\texttt{J},\texttt{G}_{ab}\right]&=-\epsilon_{m\left(a\right.}\texttt{G}_{\left.b\right)m}\,, &  \left[\texttt{J}_{a},\texttt{H}_{b}\right]&=\epsilon_{ab}\texttt{H}\,, & \left[\texttt{J}_{a},\texttt{P}_{bc}\right]&=\delta_{a\left(b\right.}\epsilon_{\left.c\right)m}\texttt{P}_{m}\,, \notag \\
       \left[\texttt{G}_{a},\texttt{P}_{bc}\right]&=-\epsilon_{a\left(b\right.}\texttt{H}_{\left.c\right)}\,, &  \left[\texttt{J}_{a},\texttt{J}_{b}\right]&=\epsilon_{ab}\texttt{J}\,, &
       \left[\texttt{J}_{a},\texttt{G}_{bc}\right]&=\delta_{a\left(b\right.}\epsilon_{\left.c\right)m}\texttt{G}_{m}\,, \notag \\
       \left[\texttt{H}_{a},\texttt{P}_{bc}\right]&=\frac{1}{\ell^2}\delta_{a\left(b\right.}\epsilon_{\left.c\right)m}\texttt{G}_{m}\,, &  \left[\texttt{G}_{ab},\texttt{P}_{cd}\right]&=\delta_{\left(a\right.\left(c\right.}\epsilon_{\left.d\right)\left.b\right)}\texttt{H}\,, & \left[\texttt{P}_{ab},\texttt{P}_{cd}\right]&=\frac{1}{\ell^2}\delta_{\left(a\right.\left(c\right.}\epsilon_{\left.d\right)\left.b\right)}\texttt{J}\,.\label{sp3adscar1}
\end{align}
This algebra corresponds to an spin-3 extension of the AdS Carroll symmetry which we denote as $\mathfrak{hs_{3}adscar}$\footnote{The $\mathfrak{hs_{3}adscar}$ algebra obtained here coincides with the $\mathfrak{hs_{3}ppoi1}$ presented in \cite{Bergshoeff:2016soe}}. Naturally, the AdS Carroll algebra spanned by $\{\texttt{J},\texttt{G}_{a},\texttt{H},\texttt{P}_{a}\}$ is a spin-2 subalgebra. Note that the flat limit $\ell\rightarrow\infty$ applied to \eqref{sp3adscar1} leads to one of the higher spin versions of the Carroll algebra found in \cite{Bergshoeff:2016soe}, which was denoted as $\mathfrak{hs_3car1}$ algebra. Interestingly, the spin-3 extension of the Carroll algebra can alternatively be obtained by considering the S-expansion of the spin-3 Poincaré algebra following the same procedure discussed here.

Although, both $\mathfrak{hs_{3}adscar}$ and $\mathfrak{hs_{3}nh}$ are obtained from the spin-3 AdS algebra considering $S_{E}^{\left(2\right)}$ as the relevant semigroup, they are not isomorphic and cannot be related through a mapping. As we shall see in section \ref{sec5}, the spin-3 AdS Carroll admits a non-degenerate invariant tensor allowing us to construct a well-defined Chern-Simons gravity action. This is not the case for the spin-3 Newton-Hooke which suffers from degeneracy.
 
\subsubsection{Spin-3 Extended AdS Carroll Algebra}
A Spin-3 extended AdS Carroll algebra appears from the $\mathfrak{hs_{3}AdS}$ algebra considering $S_{E}^{\left(2\right)}$ as the relevant semigroup whose elements satisfy \eqref{mlSE4}. Indeed, after applying a resonant $S_{E}^{\left(2\right)}$-expansion and extracting a $0_S$-reduction we get \eqref{sp3adscar1} along with
\begin{align}
    \left[\texttt{J},\texttt{T}_{a}\right]&=\epsilon_{ab}\texttt{T}_{b}\,, & \left[\texttt{G}_{a},\texttt{G}_{b}\right]&=-\epsilon_{ab}\texttt{S}\,, & 
    \left[\texttt{H},\texttt{H}_{a}\right]&=\frac{1}{\ell^{2}}\epsilon_{ab}\texttt{S}_{b}\,,\notag \\
     \left[\texttt{J},\texttt{S}_{a}\right]&=\epsilon_{ab}\texttt{S}_{b}\,, & 
     \left[\texttt{P}_{a},\texttt{T}_{b}\right]&=-\frac{1}{\ell^{2}}\epsilon_{ab}\texttt{S}\,, & 
     \left[\texttt{H},\texttt{G}_{a}\right]&=\epsilon_{ab}\texttt{T}_{b}\,, \notag\\
      \left[\texttt{S},\texttt{J}_{a}\right]&=\epsilon_{ab}\texttt{S}_{b}\,, & 
      \left[\texttt{G}_{a},\texttt{H}_{b}\right]&=-\left(\epsilon_{am}\texttt{M}_{mb}+\epsilon_{ab}\texttt{M}_{mm}\right)\,, & 
      \left[\texttt{H},\texttt{G}_{ab}\right]&=-\epsilon_{m\left(a\right.}\texttt{M}_{\left.b\right)m}\,,\notag \\
      \left[\texttt{S},\texttt{P}_{a}\right]&=\epsilon_{ab}\texttt{T}_{b}\,, & 
      \left[\texttt{P}_{a},\texttt{S}_{b}\right]&=-\left(\epsilon_{am}\texttt{M}_{mb}+\epsilon_{ab}\texttt{M}_{mm}\right)\,, & \left[\texttt{G}_{a},\texttt{G}_{bc}\right]&=-\epsilon_{a\left(b\right.}\texttt{S}_{\left.c\right)}\,,\notag \\
      \left[\texttt{J},\texttt{M}_{ab}\right]&=-\epsilon_{m\left(a\right.}\texttt{M}_{\left.b\right)m}\,, & \left[\texttt{T}_{a},\texttt{J}_{b}\right]&=-\left(\epsilon_{am}\texttt{M}_{mb}+\epsilon_{ab}\texttt{M}_{mm}\right)\,, & \left[\texttt{T}_{a},\texttt{P}_{bc}\right]&=-\frac{1}{\ell^2}\epsilon_{a\left(b\right.}\texttt{S}_{\left.c\right)}\,, \notag \\
      \left[\texttt{S},\texttt{P}_{ab}\right]&=-\epsilon_{m\left(a\right.}\texttt{M}_{\left.b\right)m}\,, &  \left[\texttt{J}_{a},\texttt{S}_{b}\right]&=\epsilon_{ab}\texttt{S}\,, & \left[\texttt{J}_{a},\texttt{M}_{bc}\right]&=\delta_{a\left(b\right.}\epsilon_{\left.c\right)m}\texttt{T}_{m}\,, \notag \\
       \left[\texttt{P}_{a},\texttt{M}_{bc}\right]&=-\frac{1}{\ell^2}\epsilon_{a\left(b\right.}\texttt{S}_{\left.c\right)}\,, &  \left[\texttt{H}_{a},\texttt{H}_{b}\right]&=\frac{1}{\ell^2}\epsilon_{ab}\texttt{S}\,, &
       \left[\texttt{H}_{a},\texttt{G}_{bc}\right]&=\delta_{a\left(b\right.}\epsilon_{\left.c\right)m}\texttt{T}_{m}\,, \notag \\
       \left[\texttt{S}_{a},\texttt{P}_{bc}\right]&=\delta_{a\left(b\right.}\epsilon_{\left.c\right)m}\texttt{T}_{m}\,, &  \left[\texttt{G}_{ab},\texttt{G}_{cd}\right]&=\delta_{\left(a\right.\left(c\right.}\epsilon_{\left.d\right)\left.b\right)}\texttt{S}\,, & \left[\texttt{P}_{ab},\texttt{M}_{cd}\right]&=\frac{1}{\ell^2}\delta_{\left(a\right.\left(c\right.}\epsilon_{\left.d\right)\left.b\right)}\texttt{S}\,.\label{sp3adscar}
\end{align}
where the spin-3 extended AdS Carroll generators are related to the relativistic ones through the semigroup elements as follows:
\begin{equation}
    \begin{tabular}{lll}
\multicolumn{1}{l|}{$\lambda_3$} & \multicolumn{1}{|l}{\cellcolor[gray]{0.8}} & \multicolumn{1}{|l|}{\cellcolor[gray]{0.8}} \\ \hline
\multicolumn{1}{l|}{$\lambda_2$} & \multicolumn{1}{|l}{$\texttt{S},\ \texttt{T}_{a},\ \, \texttt{S}_{a},\ \texttt{M}_{ab},\,$} & \multicolumn{1}{|l|}{\cellcolor[gray]{0.8}} \\ \hline
\multicolumn{1}{l|}{$\lambda_1$} & \multicolumn{1}{|l}{\cellcolor[gray]{0.8}} & \multicolumn{1}{|l|}{$\texttt{G}_a,\,\texttt{H},\,\texttt{H}_{a},\,\texttt{G}_{ab},\,$} \\ \hline
\multicolumn{1}{l|}{$\lambda_0$} & \multicolumn{1}{|l}{$ \texttt{J},\,\ \texttt{P}_{a},\ \texttt{J}_{a},\ \texttt{P}_{ab},\,$} & \multicolumn{1}{|l|}{\cellcolor[gray]{0.8}} \\ \hline
\multicolumn{1}{l|}{} & \multicolumn{1}{|l}{$\hat{\texttt{J}},\ \, \hat{\texttt{P}}_{a},\  \hat{\texttt{J}}_{a},\ \hat{\texttt{P}}_{ab},\,$} & \multicolumn{1}{|l|}{$\hat{\texttt{G}}_{a},\ \hat{\texttt{H}},\,\hat{\texttt{H}}_{a},\,\hat{\texttt{G}}_{ab}$}
\end{tabular}\label{Sexp2b}%
\end{equation}
The previous algebra corresponds to an spin-3 extension of the extended AdS Carroll symmetry, which we will denote as $\mathfrak{hs_{3}eadscar}$. In the vanishing cosmological constant limit $\ell\rightarrow\infty$, we obtain an extended version of the spin-3 Carroll algebra which we have denoted as $\mathfrak{hs_{3}ecar}$. The algebra \eqref{sp3adscar} along its flat limit correspond to spin-3 extensions of the respective extended AdS Carroll and extended Carroll ones discussed in \cite{Gomis:2019nih}. Let us note that the spin-3 extended Carroll algebra can alternatively been recovered from the expansion of the relativistic spin-3 AdS one considering $S_{E}^{(2)}$ as the relevant semigroup.

It is important to clarify that both spin-3 extended AdS Carroll and spin-3 extended Newton-Hooke algebras appear from the relativistic $\mathfrak{hs_{3}AdS}$ algebra considering the same semigroup $S_{E}^{\left(2\right)}$. However, the expansion produces quite different algebras due to the interchanging of relativistic generators between the subspaces $V_0$ and $V_1$, namely
\begin{align}
    H&\longleftrightarrow P_a & H_a&\longleftrightarrow P_{ab}\,.
\end{align}
However, such particularity does not imply that the spin-3 extended AdS Carroll and spin-3 extended Newton-Hooke algebra are related through an interchanging of generators or change of basis. This can be seen from the dimensions of the non-relativistic and ultra-relativistic algebras. In fact, the $\mathfrak{hs_{3}enh}$ algebra has 22 generators, while the spin-3 extended AdS-Carroll algebra contains 24 generators. This difference appears in the dimensions of the relativistic $V_0$ and $V_1$ in the non-relativistic and ultra-relativistic expansion.

\subsubsection{Infinite-dimensional Spin-3 AdS Carroll Algebra}
Let us consider now $S_{E}^{\left(N\right)}=\{\lambda_0,\lambda_1,\lambda_2,\cdots,\lambda_N,\lambda_{N+1}\}$ as the relevant semigroup in order to obtain a generalized spin-3 AdS Carroll algebra. The elements of the $S_{E}^{\left(N\right)}$ semigroup satisfy the multiplication law \eqref{mlSEN} and contains $\lambda_{N+1}$ as the zero element of the semigroup. Let $S_{E}^{\left(N\right)}=S_0\cup S_1$ be a resonant decomposition given by \eqref{sdN}. After considering a resonant $S_{E}^{\left(N\right)}$-expansion of the spin-3 AdS algebra and applying a $0_S$-reduction we get:
\begin{align}
     \left[\texttt{J}^{(m)},\texttt{G}_{a}^{(n)}\right]&=\epsilon_{ab}\texttt{G}_{b}^{(m+n)}\,, & \left[\texttt{G}_{a}^{(m)},\texttt{G}_{b}^{(n)}\right]&=-\epsilon_{ab}\texttt{J}^{(m+n+1)}\,, & \left[\texttt{H}^{(m)},\texttt{G}_{a}^{(n)}\right]&=\epsilon_{ab}\texttt{P}_{b}^{(m+n+1)} \notag\\
     \left[\texttt{J}^{(m)},\texttt{P}_{a}^{(n)}\right]&=\epsilon_{ab}\texttt{P}_{b}^{(m+n)}\,,& \left[\texttt{G}_{a}^{(m)},\texttt{P}_{b}^{(n)}\right]&=-\epsilon_{ab}\texttt{H}^{(m+n)}\,, & \left[\texttt{H}^{(m)},\texttt{P}_{a}^{(n)}\right]&=\frac{1}{\ell^{2}}\epsilon_{ab}\texttt{G}_{b}^{(m+n)}\notag\\
     \left[\texttt{P}_{a}^{(m)},\texttt{P}_{b}^{(n)}\right]&=-\frac{1}{\ell^{2}}\epsilon_{ab}\texttt{J}^{(m+n)}\,,\label{adscar2N}
\end{align}
along with:
\begin{align}
      \left[\texttt{J}^{(m)},\texttt{J}_{a}^{(n)}\right]&=\epsilon_{ab}\texttt{J}_{b}^{(m+n)}\,,&
      \left[\texttt{J}^{(m)},\texttt{H}_{a}^{(n)}\right]&=\epsilon_{ab}\texttt{H}_{b}^{(m+n)}\,, \notag\\ \left[\texttt{H}^{(m)},\texttt{J}_{a}^{(n)}\right]&=\epsilon_{ab}\texttt{H}_{b}^{(m+n)}\,, &
      \left[\texttt{H}^{(m)},\texttt{H}_{a}^{(n)}\right]&=\frac{1}{\ell^2}\epsilon_{ab}\texttt{J}_{b}^{(m+n+1)}\,,\notag\\
       \left[\texttt{G}_{a}^{(m)},\texttt{J}_{b}^{(n)}\right]&=-\left(\epsilon_{ac}\texttt{G}_{cb}^{(m+n)}+\epsilon_{ab}\texttt{G}_{cc}^{(m+n)}\right)\,, &
      \left[\texttt{J}_{a}^{(m)},\texttt{G}_{bc}^{(n)}\right]&=\delta_{a\left(b\right.}\epsilon_{\left.c\right)d}\texttt{G}_{d}^{(m+n)}\,, \notag\\ \left[\texttt{G}_{a}^{(m)},\texttt{H}_{b}^{(n)}\right]&=-\left(\epsilon_{ac}\texttt{P}_{cb}^{(m+n+1)}+\epsilon_{ab}\texttt{P}_{cc}^{(m+n+1)}\right) \,, &
      \left[\texttt{J}_{a}^{(m)},\texttt{P}_{bc}^{(n)}\right]&=\delta_{a\left(b\right.}\epsilon_{\left.c\right)d}\texttt{P}_{d}^{(m+n)}\,, \notag\\ \left[\texttt{P}_{a}^{(m)},\texttt{J}_{b}^{(n)}\right]&=-\left(\epsilon_{ac}\texttt{P}_{cb}^{(m+n)}+\epsilon_{ab}\texttt{P}_{cc}^{(m+n)}\right)\,, &
       \left[\texttt{H}_{a}^{(m)},\texttt{G}_{bc}^{(n)}\right]&=\delta_{a\left(b\right.}\epsilon_{\left.c\right)d}\texttt{P}_{d}^{(m+n+1)}\,, \notag\\ \left[\texttt{P}_{a}^{(m)},\texttt{H}_{b}^{(n)}\right]&=-\frac{1}{\ell^2}\left(\epsilon_{ac}\texttt{G}_{cb}^{(m+n)}+\epsilon_{ab}\texttt{G}_{cc}^{(m+n)}\right)\,, &
       \left[\texttt{H}_{a}^{(m)},\texttt{P}_{bc}^{(n)}\right]&=\frac{1}{\ell^2}\delta_{a\left(b\right.}\epsilon_{\left.c\right)d}\texttt{G}_{d}^{(m+n)}\,, \notag\\ \left[\texttt{G}_{a}^{(m)},\texttt{G}_{bc}^{(n)}\right]&=-\epsilon_{a\left(b\right.}\texttt{J}_{\left.c\right)}^{(m+n+1)}\,,  &
       \left[\texttt{J}^{(m)},\texttt{G}_{ab}^{(n)}\right]&=-\epsilon_{c\left(a\right.}\texttt{G}_{\left.b\right)c}^{(m+n)}\,, \notag\\ \left[\texttt{G}_{a}^{(m)},\texttt{P}_{bc}^{(n)}\right]&=-\epsilon_{a\left(b\right.}\texttt{H}_{\left.c\right)}^{(m+n)}\,, &
       \left[\texttt{J}^{(m)},\texttt{P}_{ab}^{(n)}\right]&=-\epsilon_{c\left(a\right.}\texttt{P}_{\left.b\right)c}^{(m+n)}\,, \notag\\ \left[\texttt{P}_{a}^{(m)},\texttt{G}_{bc}^{(n)}\right]&=-\epsilon_{a\left(b\right.}\texttt{H}_{\left.c\right)}^{(m+n)}\,, &
       \left[\texttt{H}^{(m)},\texttt{G}_{ab}^{(n)}\right]&=-\epsilon_{c\left(a\right.}\texttt{P}_{\left.b\right)c}^{(m+n+1)}\,, \notag\\ \left[\texttt{P}_{a}^{(m)},\texttt{P}_{bc}^{(n)}\right]&=-\frac{1}{\ell^2}\epsilon_{a\left(b\right.}\texttt{J}_{\left.c\right)}^{(m+n)}\,, &
       \left[\texttt{H}^{(m)},\texttt{P}_{ab}^{(n)}\right]&=-\frac{1}{\ell^2}\epsilon_{c\left(a\right.}\texttt{G}_{\left.b\right)c}^{(m+n)}\,, \notag\\ \left[\texttt{J}_{a}^{(m)},\texttt{J}_{b}^{(n)}\right]&=\epsilon_{ab}\texttt{J}^{(m+n)}\,, &
       \left[\texttt{G}_{ab}^{(m)},\texttt{G}_{cd}^{(n)}\right]&=\delta_{\left(a\right.\left(c\right.}\epsilon_{\left.d\right)\left.b\right)}\texttt{J}^{(m+n+1)}\,, \notag\\ \left[\texttt{J}_{a}^{(m)},\texttt{H}_{b}^{(n)}\right]&=\epsilon_{ab}\texttt{H}^{(m+n)} \,, &
       \left[\texttt{G}_{ab}^{(m)},\texttt{P}_{cd}^{(n)}\right]&=\delta_{\left(a\right.\left(c\right.}\epsilon_{\left.d\right)\left.b\right)}\texttt{H}^{(m+n)}\,, \notag\\ \left[\texttt{H}_{a}^{(m)},\texttt{H}_{b}^{(n)}\right]&=\frac{1}{\ell^2}\epsilon_{ab}\texttt{J}^{(m+n+1)} \,, &
        \left[\texttt{P}_{ab}^{(m)},\texttt{P}_{cd}^{(n)}\right]&=\frac{1}{\ell^2}\delta_{\left(a\right.\left(c\right.}\epsilon_{\left.d\right)\left.b\right)}\texttt{J}^{(m+n)}\,.\label{adscar3N}
\end{align}
where the expanded generators are related to the relativistic spin-3 AdS ones through the semigroup elements as
\begin{align}
\texttt{J}^{(m)} &=\lambda _{2m}\hat{\texttt{J}}\,, \quad 
&\texttt{G}_{a}^{(m)} &=\lambda _{2m+1}\hat{\texttt{G}}_{a}\,,  \quad &\texttt{G}_{ab}^{(m)} &=\lambda _{2m+1}\hat{\texttt{G}}_{ab}\,, \quad &\texttt{J}_{a}^{(m)} &=\lambda _{2m}\hat{\texttt{J}}_{a}\,,\notag \\
\texttt{H}^{(m)} &=\lambda _{2m+1}\hat{\texttt{H}}\,,  \quad
&\texttt{P}_{a}^{(m)} &=\lambda_{2m}\hat{\texttt{P}}_{a}\,, \quad &\texttt{P}_{ab}^{(m)} &=\lambda _{2m}\hat{\texttt{P}}_{ab}\,, \quad &\texttt{H}_{a}^{(m)} &=\lambda _{2m+1}\hat{\texttt{H}}_{a}\,. \label{expgencar}
\end{align}
The ultra-relativistic spin-3 algebra \eqref{adscar2N}-\eqref{adscar3N}, which we denote as $\mathfrak{hs_{3}adscar}^{\left(N\right)}$, appears after considering the multiplication law of the semigroup \eqref{mlSEN} and the relativistic commutation relations \eqref{AdS3}. In the flat limit $\ell\rightarrow\infty$ the algebra reduces to a generalized spin-3 Carroll algebra which we denote as $\mathfrak{hs_{3}car}^{\left(N\right)}$. For $N=1$ and $N=2$, we recover the previously obtained spin-3 AdS Carrol $\mathfrak{hs_{3}adscar}$ and its extended version $\mathfrak{hs_{3}eadscar}$, respectively. For $N=3$, the $\mathfrak{hs_{3}adscar}^{\left(N\right)}$ can be seen as an enhanced spin-3 AdS Carroll algebra which in the flat limit $\ell\rightarrow\infty$ reduces to a Carrollian analogue of the Newtonian algebra obtained in \cite{Hansen:2018ofj}. Let us note that an infinite-dimensional extension of the spin-3 AdS Carroll algebra is obtained considering a infinite-dimensional semigroup $S_{E}^{\left(\infty\right)}$ without zero element. In this case, the vanishing cosmological constant limit $\ell\rightarrow\infty$ leads to a spin-3 extension of the infinite-dimensional Carroll algebra \cite{Gomis:2019nih}. It is worth it to mention that the $\mathfrak{hs_{3}car}^{\left(N\right)}$ algebra can also be obtained by applying the S-expansion to the relativistic $\mathfrak{hs_{3}poi}$ algebra considering $S_{E}^{\left(N\right)}$ as the relevant semigroup. The following table summarize the ultra-relativistic counterparts of the spin-3 AdS algebra along their flat limits.
\begin{equation}
    \begin{tabular}{|m{6em}||c|c|c|c|c|}
\hline
 Relativistic spin-3 algebra & $N=1$ & $N=2$ & $N=3$  & $\cdots$ & $N$ \\ \hline\hline
$\mathfrak{hs_{3}AdS}$ & \,$\mathfrak{hs_{3}adscar}$ \, & \, $\mathfrak{hs_{3}eadscar}$ \,  & \, $\mathfrak{hs_{3}enhadscar}$ \,  &  $\cdots$ & \, $\mathfrak{hs_{3}adscar}^{\left(N\right)}$\,  \\ \hline
$\mathfrak{hs_{3}poi}$ &  $\mathfrak{hs_{3}car}$ & \ $\mathfrak{hs_{3}ecar}$\ & $\mathfrak{hs_{3}enhcar}$ & $\cdots$ & \,$\mathfrak{hs_{3}car}^{\left(N\right)}$\, \\ \hline
\end{tabular}\notag
\end{equation}
Unlike the non-relativistic version, the ultra-relativistic spin-3 symmetries does not suffer from degeneracy and admit a non-degenerate billinear invariant tensor for arbitrary values of $N$. Interestingly, the spin-3 extension of the infinite-dimensional AdS Carroll is not unique but can be done in a different consistent way. Indeed, a diverse subspace decomposition of the relativistic spin-3 AdS algebra can be considered involving new spin-3 AdS Carroll symmetries. Further details about other spin-3 extensions and their differences can be found in Appendix \ref{App2}.

\section{Non-relativistic Spin-3 Gravity Theories}\label{sec4}
In this section we present the explicit construction of three-dimensional non-relativistic spin-3 CS gravity actions based on the non-relativistic spin-3 algebras presented previously. In particular, we focus our attention in the particular case of CS action being gauge invariant under non-relativistic spin-3 algebra admitting non-degenerate bilinear invariant trace. The non-degeneracy of the invariant tensor ensures that the CS action involves a kinematical term for each gauge field.
\subsection{Spin-3 Extended Newton-Hooke Gravity}
The spin-3 extension of the extended Newton-Hooke algebra \eqref{enh3} admits the non-vanishing components of the invariant tensor for the spin-2 extended Newton-Hooke algebra,
\begin{align}
    \langle \texttt{G}_a\texttt{G}_{b} \rangle&=\alpha_{0}\delta_{ab}\,, &\langle \texttt{J}\texttt{S}\rangle &=-\alpha_{0}\,, & \langle \texttt{H}\texttt{S}\rangle &=-\alpha_{1}\,,  \notag \\
   \langle \texttt{G}_{a}\texttt{P}_{b} \rangle &=\alpha_{1}\delta_{ab}\,,  &\langle \texttt{H}\texttt{M}\rangle &=-\frac{\alpha_{0}}{\ell^2}\,, &  \langle \texttt{M}\texttt{J}\rangle &=-\alpha_{1}\,, \notag \\
    \langle \texttt{P}_a\texttt{P}_{b} \rangle&=\frac{\alpha_{0}}{\ell^{2}}\delta_{ab}\,, \label{ITNH1}
\end{align}
along its spin-3 extension,
\begin{align}
     \langle \texttt{G}_{ab}\texttt{G}_{cd} \rangle&=\alpha_{0}\left(\delta_{a\left(c\right.}\delta_{\left.d\right)b}-\frac{2}{3}\delta_{ab}\delta_{cd}\right)\,, &\langle \texttt{J}_{a}\texttt{S}_{b}\rangle &=-\alpha_{0}\delta_{ab}\,, & \langle  \texttt{H}_{a}\texttt{S}_{b}\rangle &=-\alpha_{1}\delta_{ab}\,,  \notag \\
     \langle \texttt{G}_{ab}\texttt{P}_{cd} \rangle&=\alpha_{1}\left(\delta_{a\left(c\right.}\delta_{\left.d\right)b}-\frac{2}{3}\delta_{ab}\delta_{cd}\right)\,, &\langle \texttt{H}_{a}\texttt{M}_{b}\rangle &=-\frac{\alpha_{0}}{\ell^2}\delta_{ab}\,, & \langle  \texttt{J}_{a}\texttt{M}_{b}\rangle &=-\alpha_{1}\delta_{ab}\,,  \notag \\
     \langle \texttt{P}_{ab}\texttt{P}_{cd} \rangle&=\frac{\alpha_{0}}{\ell^2}\left(\delta_{a\left(c\right.}\delta_{\left.d\right)b}-\frac{2}{3}\delta_{ab}\delta_{cd}\right)\,. \label{ITNH2}
\end{align}
Here, $\alpha_0$ and $\alpha_1$ are arbitrary constants and are related to the relativistic ones as
\begin{align}
    \alpha_0&=\lambda_2\hat{\alpha}_{0}\,, &\alpha_1&=\lambda_2\hat{\alpha}_{1}\,. \notag
\end{align}
One can notice that the non-degeneracy requires that $\alpha_0\neq\alpha_1/\ell^2$. In the vanishing cosmological constant limit $\ell\rightarrow\infty$, we obtain the invariant tensor for the $\mathfrak{hs_{3}ebarg}$ algebra presented in \cite{Bergshoeff:2016soe}.  Let us consider now the gauge connection one-form $A=A^{A}T_{A}$ taking values in the $\mathfrak{hs_{3}enh}$ algebra:
\begin{equation}
    A=\omega J+B^{a}G_{a}+sS+\tau H+e^{a}P_{a}+mM+\omega^{a} J_{a}+B^{ab}G_{ab}+s^{a}S_{a}+\tau^{a} H_{a}+e^{ab}P_{ab}+m^{a}M_{a}\,. \label{1FNH}
\end{equation}
Here, $\{\omega,B^{a},\tau,e^{a}\}$ are the usual spin-2 extended Newton-Hooke gauge field where $\tau$ is the time-like vielbein, $B^{a}$ is the spatial spin-connection, $e^{a}$ is the spatial vielbein and $\omega$ represents the spin-connection for boosts. On the other hand, $\{\omega^{a},B^{ab},\tau^{a},e^{ab}\}$ are the corresponding spin-3 versions of the spin-connections and vielbeins. The corresponding curvature two-form $F$ is given by
\begin{align}
    F=&R\left(\omega\right)J+R^{a}\left(B^{b}\right)G_{a}+R\left(s\right)S+R\left(\tau\right)H+R^{a}\left(e^{b}\right)P_{a}+R\left(m\right)M+R^{a}\left(\omega^{b}\right)J_{a}\notag \\
    &+R^{ab}\left(B^{cd}\right)G_{ab}+R^{a}\left(s^{b}\right)S_{a}+R^{a}\left(\tau^{b}\right)H_{a}+R^{ab}\left(e^{cd}\right)P_{ab}+R^{a}\left(m^{b}\right)M_{a}\,,\label{2FNHa}
\end{align}
where
\begin{eqnarray}
R\left(\omega\right)&=&d\omega-\frac{1}{2}\epsilon^{ac}\omega_a\omega_c-\frac{1}{2\ell^2}\epsilon^{ac}\tau_{a}\tau_{c}\,, \notag \\
R\left(\tau\right)&=&d\tau - \epsilon^{ac}\omega_{a}\tau_{c}\,, \notag \\
R^{a}\left(B^{b}\right)&=&dB^{a}+\epsilon^{ac}\omega B_c+\frac{1}{\ell^2}\epsilon^{ac}\tau e_c+\epsilon^{a\left(c\right.}\delta^{\left.b\right)d}\omega_{d}B_{bc}+\frac{1}{\ell^2}\epsilon^{a\left(c\right.}\delta^{\left.b\right)d}\tau_{d}e_{bc}\,, \notag\\
R^{a}\left(e^{b}\right)&=&de^{a}+\epsilon^{ac}\omega e_c+\epsilon^{ac}\tau B_c+\epsilon^{a\left(c\right.}\delta^{\left.b\right)d}\omega_{d}e_{bc}+\epsilon^{a\left(c\right.}\delta^{\left.b\right)d}\tau_{d}B_{bc}\,, \notag \\
R\left(s\right)&=&ds + \frac{1}{2} \epsilon^{ac} B_a B_c + \frac{1}{2\ell^2} \epsilon^{ac} e_a e_c - \epsilon^{ac}\omega_a s_c - \frac{1}{\ell^2} \epsilon^{ac} \tau_a m_c \notag\\
&&+ \frac{1}{2}\epsilon^{\left(a\right.\left(c\right.}\delta^{\left.d\right)\left.b\right)}B_{ad}B_{cb}+\frac{1}{2\ell^2}\epsilon^{\left(a\right.\left(c\right.}\delta^{\left.d\right)\left.b\right)}e_{ad}e_{cb}\,, \notag \\
R\left(m\right)&=&dm + \epsilon^{ac} B_a e_c - \epsilon^{ac} \omega_a m_c -\epsilon^{ac} \tau_a s_c + \epsilon^{\left(a\right.\left(c\right.}\delta^{\left.d\right)\left.b\right)}e_{ad}B_{cb}  \,, \notag \\
R^{a}\left(\omega^{b}\right) &=& d\omega^{a} + \epsilon^{ac}\omega\omega_{c}+\frac{1}{\ell^2}\epsilon^{ac}\tau\tau_{c} \,, \notag \\
R^{ab}\left(B^{cd}\right) &=& dB^{ac} + \epsilon^{\left(a\right|c}\omega B_{c}^{\ \left|b\right)} +\frac{1}{\ell^2}\epsilon^{\left(a\right|c}\tau e_{c}^{\ \left|b\right)}  + \epsilon^{ac}\omega^{b}B_{c}  + \frac{1}{\ell^2}\epsilon^{ac}\tau^{b}e_{c}\notag\\
&&+\delta^{ab}\epsilon^{cd}\omega_{c}B_{d}+\frac{1}{\ell^2}\delta^{ab}\epsilon^{cd}\tau_{c}e_{d}\,, \notag \\
R^{a}\left(s^{b}\right) &=& ds^{a} +\epsilon^{ac}\omega s_{c} + \frac{1}{\ell^2}\epsilon^{ac}\tau m_{c}+\epsilon^{ac} s \omega_{c}+ \frac{1}{\ell^2}\epsilon^{ac} m \tau_{c}+\epsilon^{c\left(d\right.}B_{c}B_{d}^{\ \left.a\right)}+\frac{1}{\ell^2}\epsilon^{c\left(d\right.}e_{c}e_{d}^{\ \left.a\right)}\,, \notag \\
R^{a}\left(\tau^{b}\right) &=&d\tau^{a} + \epsilon^{ac}\omega \tau_{c} +\epsilon^{ac}\tau\omega_{c} \,, \notag \\
R^{ab}\left(e^{cd}\right) &=& de^{ab}+\epsilon^{\left(a\right|c}\omega e_{c}^{\ \left|b\right)} + \epsilon^{\left(a\right|c}\tau B_{c}^{\ \left|b\right)} +\epsilon^{ac}\tau^{b}B_{c}+\epsilon^{ac}\omega^{b}e_{c}\notag \\
&&+\delta^{ab}\epsilon^{cd}\tau_{c}B_{d}+\delta^{ab}\epsilon^{cd}\omega_{c}e_{d}\,, \notag \\
R^{a}\left(m^{b}\right) &=& dm^{a}+\epsilon^{ac}\omega m_{c}+\epsilon^{ac}\tau s_{c}+\epsilon^{ac}m\omega_{c}+\epsilon^{ac} s\tau_{c}+\epsilon^{c\left(d\right.}e_{c}B_{d}^{\ \left.a\right)}+\epsilon^{c\left(d\right.}B_{c}e_{d}^{\ \left.a\right)}\,. \label{curvhs3enh}
\end{eqnarray}
The CS action based on the spin-3 extended Newton-Hooke algebra can be obtained considering the gauge connection one-form \eqref{1FNH} and the non-vanishing components of the invariant tensor \eqref{ITNH1} and \eqref{ITNH2} into the general expression of a three-dimensional CS action \eqref{CS}. In particular, we get
\begin{align}
    I_{\mathfrak{hs_{3}enh}}=\frac{k}{4\pi}\int & e_{a}\left[F^{a}\left(B^{b}\right)+2\epsilon^{a\left(c\right.}\delta^{\left.b\right)d}\omega_{d}B_{bc}\right]+B_{a}F^{a}\left(e^{b}\right)-\omega dm-sd\tau-m\left(d\omega-\epsilon^{ac}\omega_{a}\omega_{c}\right) \notag\\
    &\left. -\tau \left[ds+\epsilon^{ac}B_{a}B_{c}-2\epsilon^{ac}\omega_{a}s_{c}+\epsilon^{\left(a\right.\left(c\right.}\delta^{\left.d\right)\left.b\right)}B_{ad}B_{cb}\right]+B_{ab}F^{ab}\left(e^{cd}\right)\right. \notag\\
    &\left. +e_{ab}\left[F^{ab}\left(B^{cd}\right)+\epsilon^{ac}\omega^{b}B_{c}+\delta^{ab}\epsilon^{cd}\omega_{c}B_{d}\right]-s_{a}F^{a}\left(\tau^{b}\right)-\omega_{a}F^{a}\left(m^{b}\right)\right.\notag\\
    &\left. -\tau_{a}\left[F^{a}\left(s^{b}\right)+2\epsilon^{c\left(d\right.}B_{c}B_{d}^{\ \left.a\right)}\right]-m_{a}F^{a}\left(\omega^{b}\right)\right.\notag\\
    &+\frac{1}{\ell^2}\left(\epsilon^{ac}e_{a}\tau e_{c}+2\epsilon^{ac}\tau\tau_a m_{c}+\epsilon^{ac}m\tau_a\tau_c+2\epsilon^{a\left(c\right.}\delta^{\left.b\right)d}e_{a}\tau_{d}e_{bc}-\epsilon^{\left(a\right.\left(c\right.}\delta^{\left.d\right)\left.b\right)}\tau e_{ad}e_{cb}\right)\,,\label{CShs3enh}
\end{align}
where we have defined
\begin{eqnarray}
F^{a}\left(B^{b}\right)&=&dB^{a}+\epsilon^{ac}\omega B_{c}\,, \notag\\
F^{a}\left(e^{b}\right)&=&de^{a}+\epsilon^{ac}\omega e_{c}\,, \notag\\
F^{ab}\left(B^{cd}\right)&=&dB^{ab}+\epsilon^{\left(a\right|c}\omega B_{c}^{\ \left|b\right)}\,, \notag\\
F^{ab}\left(e^{cd}\right)&=&de^{ab}+\epsilon^{\left(a\right|c}\omega e_{c}^{\ \left|b\right)}\,, \notag \\
F^{a}\left(\omega^{b}\right)&=&d\omega^{a}+\epsilon^{ac}\omega\omega_{c}\,, \notag \\
F^{a}\left(s^{b}\right)&=&ds^{a}+\epsilon^{ac}\omega s_{c}+\epsilon^{ac}s\omega_{c}\,, \notag\\
F^{a}\left(\tau^{b}\right)&=&d\tau^{a}+\epsilon^{ac}\omega\tau_{c}\,,\notag\\
F^{a}\left(m^{b}\right)&=&dm^{a}+\epsilon^{ac}\omega m_{c}+\epsilon^{ac}s\tau_{c}\,. \label{defcurv}
\end{eqnarray}
Here, for the sake of simplicity, we have omitted the exotic term proportional to $\alpha_0$ and we have fixed $\alpha_{1}=1$. The spin-3 non-relativistic CS gravity action \eqref{CShs3enh} is gauge invariant under the $\mathfrak{hs_{3}enh}$ algebra \eqref{enh3} and reproduces the spin-3 extension of the extended Bargmann gravity in the vanishing cosmological constant limit $\ell\rightarrow\infty$. Let us note that the non-degeneracy of the invariant bilinear trace \eqref{ITNH1}-\eqref{ITNH2} ensures that the CS action \eqref{CShs3enh} involves a kinematical
term for each gauge field and the equations of motion imply that all curvatures two-forms \eqref{curvhs3enh} vanish. Naturally, in the vanishing cosmological constant limit $\ell\rightarrow\infty$, the equations of motion reproduces the vanishing of the curvatures of the $\mathfrak{hs_{3}ebarg}$ algebra. In the absence of spin-3 gauge field, the theory reduces to the usual extended Newton-Hooke gravity \cite{Papageorgiou:2010ud,Duval:2011mi,Hartong:2016yrf} and to the extended Bargmann one \cite{Bergshoeff:2016lwr} in the flat limit. 

Let us remark that the spin-3 non-relativistic CS gravity action \eqref{CShs3enh} for the spin-3 version of the extended Newton-Hooke symmetry can alternatively be obtained from the relativistic CS action for the $\mathfrak{hs_{3}AdS}$ algebra \eqref{CSADS3}. Indeed, the spin-3 non-relativistic CS action appears by expressing the non-relativistic gauge fields in terms of the relativistic ones through the semigroup elements:
\begin{align}
\omega&=\lambda_0 W^{0}\,, &B^{a}&=\lambda_1 W^{a}\,, &s&=\lambda_2 W^{0}\,, \notag \\
\tau&=\lambda_0 E^{0}\,, &e^{a}&=\lambda_1 E^{a}\,, &m&=\lambda_2 E^{0}\,, \notag \\
\omega^{a}&=\lambda_0 W^{0a}\,, &B^{ab}&=\lambda_1 W^{ab}\,, &s^{a}&=\lambda_2 W^{0a}\,, \notag \\
\tau^{a}&=\lambda_0 E^{0a}\,, &e^{ab}&=\lambda_1 E^{ab}\,, &m^{a}&=\lambda_2 E^{0a}\,. \label{gfexp}
\end{align}
\subsection{Infinite-dimensional Spin-3 Newton-Hooke Gravity}
The spin-3 extension of the infinite-dimensional Newton-Hooke algebra \eqref{idNH3} admits the non-vanishing components of the invariant tensor for the infinite-dimensional Newton-Hooke algebra:
\begin{align}
    \langle \texttt{G}_{a}^{(m)}\texttt{G}_{b}^{(n)} \rangle&=\sigma_{m+n+1}\delta_{ab}\,, &\langle \texttt{J}^{(m)}\texttt{J}^{(n)}\rangle&=-\sigma_{m+n}\,, \notag \\
    \langle \texttt{G}_{a}^{(m)}\texttt{P}_{b}^{(n)} \rangle&=\gamma_{m+n+1}\delta_{ab}\,, &\langle \texttt{J}^{(m)}\texttt{H}^{(n)}\rangle&=-\gamma_{m+n}\,, \notag \\
    \langle \texttt{P}_{a}^{(m)}\texttt{P}_{b}^{(n)} \rangle&=\frac{1}{\ell^2}\sigma_{m+n+1}\delta_{ab}\,, &\langle \texttt{H}^{(m)}\texttt{H}^{(n)}\rangle&=-\frac{1}{\ell^2}\sigma_{m+n}\,, \label{ITinfNH1}
\end{align}
along its spin-3 extension
\begin{align}
    \langle \texttt{G}_{ab}^{(m)}\texttt{G}_{cd}^{(n)}\rangle&=\sigma_{m+n+1}\left(\delta_{a\left(c\right.}\delta_{\left.d\right)b}-\frac{2}{3}\delta_{ab}\delta_{cd}\right)\,, &\langle\texttt{J}_{a}^{(m)}\texttt{J}_{b}^{(n)}\rangle&=-\sigma_{m+n}\delta_{ab}\,,\notag \\
    \langle \texttt{G}_{ab}^{(m)}\texttt{P}_{cd}^{(n)}\rangle&=\gamma_{m+n+1}\left(\delta_{a\left(c\right.}\delta_{\left.d\right)b}-\frac{2}{3}\delta_{ab}\delta_{cd}\right)\,, &\langle\texttt{J}_{a}^{(m)}\texttt{H}_{b}^{(n)}\rangle&=-\gamma_{m+n}\delta_{ab}\,,\notag \\
    \langle \texttt{P}_{ab}^{(m)}\texttt{P}_{cd}^{(n)}\rangle&=\frac{1}{\ell^2}\sigma_{m+n+1}\left(\delta_{a\left(c\right.}\delta_{\left.d\right)b}-\frac{2}{3}\delta_{ab}\delta_{cd}\right)\,, &\langle\texttt{H}_{a}^{(m)}\texttt{H}_{b}^{(n)}\rangle&=-\frac{1}{\ell^2}\sigma_{m+n}\delta_{ab}\,,\label{ITinfNH2}
\end{align}
where the $\sigma$'s
 and $\gamma$'s are defined in terms of the relativistic constants $\hat{\alpha}_{0}$ and $\hat{\alpha}_{1}$ through the elements of the semigroup $S_{E}^{(N)}$ as
 \begin{align}
     \sigma_{m+n}&=\lambda_{2(m+n)}\hat{\alpha}_{0}\,, \notag \\
     \gamma_{m+n}&=\lambda_{2(m+n)}\hat{\alpha}_{1}\,.
 \end{align}
Here $\sigma_{m+n}$ is related to an exotic non-relativistic term coming from the relativistic exotic coupling constant $\hat{\alpha}_{0}$. One can notice that $\sigma_1$ and $\gamma_1$ corresponds to the coupling constants of the $\mathfrak{hs_{3}enh}$ algebra $\alpha_0$ and $\alpha_{1}$, respectively. It is important to emphasize that the $\mathfrak{hs_{3}nh}^{(N)}$ algebra admits a non-degenerate invariant trace only for even values of $N$. Then, the first non-degenerate case for $N>2$ appears for the spin-3 extension of the Post-Newtonian extension of the spin-3 extended Newton-Hooke, which we have denoted as $\mathfrak{hs_{3}pne}$ and corresponds to $\mathfrak{hs_{3}nh}^{(4)}$. Let us construct the explicit general expression for the CS action based on the $\mathfrak{hs_{3}nh}^{(N)}$ algebra. To this end, let us first consider the gauge connection one-form taking values in the $\mathfrak{hs_{3}nh}^{(N)}$ algebra:
\begin{align}
    A&=\sum_{m=0}^{\left[\frac{N}{2}\right]}\left(\omega^{(m)}\texttt{J}^{(m)}+\tau^{(m)}\texttt{H}^{(m)}+\omega^{a(m)}\texttt{J}_{a}^{(m)}+\tau^{a(m)}\texttt{H}_{a}^{(m)}\right)\notag \\
    &+\sum_{m=0}^{\left[\frac{N+1}{2}\right]}\left(B^{a(m)}\texttt{G}_{a}^{(m)}+e^{a(m)}\texttt{P}_{a}^{(m)}+B^{ab(m)}\texttt{G}_{ab}^{(m)}+e^{ab(m)}\texttt{P}_{ab}^{(m)}\right)\,, \label{infA}
\end{align}
where $\{\omega^{(m)},B^{a(m)},\tau^{(m)},e^{a(m)}\}$ are spin-2 gauge fields being expansions of the spin-connections and vielbein. On the other hand, $\{\omega^{a(m)},B^{ab(m)},\tau^{a(m)},e^{ab(m)}\}$ represents spin-3 gauge fields corresponding to expansions of the spin-3 versions of the spin-connections and vielbeins. The curvature two-form $F$ reads
\begin{align}
    F=&R\left(\omega^{(m)}\right)\texttt{J}^{(m)}+R^{a}\left(B^{b(m)}\right)\texttt{G}_{a(m)}+R\left(\tau^{(m)}\right)\texttt{H}^{(m)}+R^{a}\left(e^{b(m)}\right)\texttt{P}_{a}^{(m)}\notag \\
    &+R^{a}\left(\omega^{b(m)}\right)\texttt{J}_{a}^{(m)}+R^{ab}\left(B^{cd(m)}\right)\texttt{G}_{ab}^{(m)}+R^{a}\left(\tau^{b(m)}\right)\texttt{H}_{a}^{(m)}+R^{ab}\left(e^{cd(m)}\right)\texttt{P}_{ab}^{(m)}\,,\label{2FinfNHa}
\end{align}
where the explicit expressions for each curvature can be found in appendix \eqref{App3}. Then, a $\mathfrak{hs_{3}nh}^{\left(N\right)}$ CS gravity action can be constructed considering the gauge connection one-form \eqref{infA} and the non-vanishing components of the invariant tensor for a finite value of N in the general expression of the CS form \eqref{CS}. For our purpose, we will focus only on the non-degenerate cases, mainly to the $\mathfrak{hs_{3}nh}^{(N)}$ algebras for even value of $N$. Thus, each gauge field appearing in the CS theories has a kinetic term and is then dynamical. Moreover, we will omit the exotic terms proportional to $\sigma$'s since they do not bring new information to the theory. Taking into account all our considerations, we get
\begin{align}
    I_{\mathfrak{hs_{3}nh}^{\left(N\right)}}=\frac{k}{4\pi}\int &\sum_{i=1}^{\left[N/2\right]}\gamma_{i} \, \, e_{a}^{(m)}\left[F^{a}\left(B^{b(n)}\right)\delta_{m+n+1}^{i}+2\epsilon^{a\left(c\right.}\delta^{\left.b\right)d}\omega_{d}^{(n)}B_{bc}^{(l)}\delta_{m+n+l+1}^{i}\right]-\omega^{(m)} d\tau^{(n)}\delta_{m+n}^{i}\notag\\
    &+\left[B_{a}^{(m)}F^{a}\left(e^{b(n)}\right)+B_{ab}^{(m)}F^{ab}\left(e^{cd(n)}\right)\right]\delta_{m+n+1}^{i} -\tau^{(m)} \left(d\omega^{(n)}\delta_{m+n}^{i}\right.\notag\\
    &\left.+\epsilon^{ac}B_{a}^{(n)}B_{c}^{(l)}\delta_{m+n+l+1}^{i}-\epsilon^{ac}\omega_{a}^{(n)}\omega_{c}^{(l)}\delta_{m+n+l}^{i}+\epsilon^{\left(a\right.\left(c\right.}\delta^{\left.d\right)\left.b\right)}B_{ad}^{(n)}B_{cb}^{(l)}\delta_{m+n+l+1}^{i}\right) \notag\\ &+e_{ab}^{(m)}\left[F^{ab}\left(B^{cd(n)}\right)\delta_{m+n+1}^{i}+\left(\epsilon^{ac}\omega^{b(n)}B_{c}^{(l)}+\delta^{ab}\epsilon^{cd}\omega_{c}^{(n)}B_{d}^{(l)}\right)\delta_{m+n+l+1}^{i}\right]\notag\\
    &-\tau_{a}^{(m)}\left[F^{a}\left(\omega^{b(n)}\right)\delta_{m+n}^{i}+2\epsilon^{c\left(d\right.}B_{c}^{(n)}B_{d}^{\ \left.a\right)\, (l)}\delta_{m+n+l+1}^{i}\right]-\omega_{a}^{(m)}F^{a}\left(\tau^{b(n)}\right)\delta_{m+n}^{i}\notag\\
    & +\frac{1}{\ell^2}\left[\left(\epsilon^{ac}e_{a}^{(m)}\tau^{(n)} e_{c}^{(l)}+2\epsilon^{a\left(c\right.}\delta^{\left.b\right)d}e_{a}^{(m)}\tau_{d}^{(n)}e_{bc}^{(l)}-\epsilon^{\left(a\right.\left(c\right.}\delta^{\left.d\right)\left.b\right)}\tau^{(m)} e_{ad}^{(n)}e_{cb}^{(l)}\right)\delta_{m+n+l+1}^{i}\right. \notag\\
    &\left.+\epsilon^{ac}\tau^{(m)}\tau_a^{(n)}\tau_{c}^{(l)}\delta_{m+n+l}^{i}\right]\,, \label{infCSNH}
\end{align}
where we have defined
\begin{eqnarray}
F^{a}\left(B^{b(m)}\right)&=&dB^{a(m)}+\sum_{n,l=0}^{\left[N/2\right]}\epsilon^{ac}\omega^{(n)} B_{c}^{(l)}\delta_{n+l}^{m}\,, \notag\\
F^{a}\left(e^{b(m)}\right)&=&de^{a(m)}+\sum_{n,l=0}^{\left[N/2\right]}\epsilon^{ac}\omega^{(n)} e_{c}^{(l)}\delta_{n+l}^{m}\,, \notag\\
F^{ab}\left(B^{cd(m)}\right)&=&dB^{ab(m)}+\sum_{n,l=0}^{\left[N/2\right]}\epsilon^{\left(a\right|c}\omega^{(n)} B_{c}^{\ \left|b\right)\,(l)}\delta_{n+l}^{m}\,, \notag\\
F^{ab}\left(e^{cd(m)}\right)&=&de^{ab(m)}+\sum_{n,l=0}^{\left[N/2\right]}\epsilon^{\left(a\right|c}\omega^{(n)} e_{c}^{\ \left|b\right)\,(l)}\delta_{n+l}^{m}\,, \notag \\
F^{a}\left(\omega^{b(m)}\right)&=&d\omega^{a(m)}+\sum_{n,l=0}^{\left[N/2\right]}\epsilon^{ac}\omega^{(n)}\omega_{c}^{(l)}\delta_{n+l}^{m}\,, \notag \\
F^{a}\left(\tau^{b(m)}\right)&=&d\tau^{a(m)}+\sum_{n,l=0}^{\left[N/2\right]}\epsilon^{ac}\omega^{(n)}\tau_{c}^{(l)}\delta_{n+l}^{m}\,. \label{defcurv2}
\end{eqnarray}
The CS action \eqref{infCSNH} is gauge invariant under the $\mathfrak{hs_{3}nh}^{\left(N\right)}$ algebra and can be seen as the cosmological extension of the generalized spin-3 Galilean algebra denoted as $\mathfrak{hs_{3}gal}^{\left(N\right)}$. Since the spin-3 extension of the infinite-dimensional Newton-Hooke algebra admits a non-degenerate bilinear invariant trace for even value of $N$, we have that the equations of motion of the theory are given by the vanishing of the curvatures \eqref{infhs3nh}. Such equations can be used to express not only the time-like and spatial spin-connections $\{\omega^{(0)},\omega^{(a)}\}$ but also their expansions $\{\omega^{(m)},\omega^{a(m)}\}$ in terms of the other gauge fields. One can see that the spin-3 extended Newton-Hooke CS gravity action \eqref{CShs3enh} is obtained for $N=2$. Let us remark that the CS action \eqref{infCSNH} can alternatively be obtained from the relativistic $\mathfrak{hs_{3}AdS}$ gravity action by expressing the non-relativistic gauge fields in term of the relativistic ones through the elements of $S_{E}^{\left(N\right)}$,
\begin{align}
    \omega^{(m)}&=\lambda_{2m}W^{0}\,, &B^{a(m)}&=\lambda_{2m+1}W^{a} \,, &\tau^{(m)}&=\lambda_{2m}E^{0}\,, &e^{a(m)}&=\lambda_{2m+1}E^{a}\,, \notag\\
    \omega^{a(m)}&=\lambda_{2m}W^{0a}\,, &B^{ab(m)}&=\lambda_{2m+1}W^{ab}\,, &\tau^{a(m)}&=\lambda_{2m}E^{0a}\,, &e^{ab(m)}&=\lambda_{2m+1}E^{ab}\,.
\end{align}
Interestingly, the CS action \eqref{infCSNH} can be rewritten as the sum of diverse spin-3 non-relativistic CS gravity actions as follows:
\begin{equation}
    I_{\mathfrak{hs_{3}nh}^{\left(N\right)}}=I_{\mathfrak{hs_{3}enh}}+I_{\mathfrak{hs_{3}pne}}+\sum_{i=3}^{N/2}I_{\mathfrak{hs_{3}nh}^{\left(2i\right)}}\,,
\end{equation}
where $I_{\mathfrak{hs_{3}enh}}$ is the CS action obtained in \eqref{CShs3enh}, while $I_{\mathfrak{hs_{3}pne}}$ is the CS action for the spin-3 extension of the Newton-Hooke version of the extended post-Newtonian algebra \eqref{pn3} whose explicit expression can be obtained from \eqref{infCSNH} for $N=4$. It is interesting to note that, in the vanishing cosmological constant limit $\ell\rightarrow\infty$, the $\mathfrak{hs_{3}gal}^{\left(N\right)}$ CS gravity action is given by
\begin{equation}
    I_{\mathfrak{hs_{3}gal}^{\left(N\right)}}=I_{\mathfrak{hs_{3}ebarg}}+I_{\mathfrak{hs_{3}eNewt}}+\sum_{i=3}^{N/2}I_{\mathfrak{hs_{3}gal}^{\left(2i\right)}}\,,
\end{equation}
with $I_{\mathfrak{hs_{3}eNewt}}$ being the CS action for the spin-3 extension of the so-called extended Newtonian symmetry \cite{Ozdemir:2019orp}. For completeness, we provide with the explicit CS expression for the $\mathfrak{hs_{3}pne}$:
\begin{align}
    I_{\mathfrak{hs_{3}pne}}=\frac{k}{4\pi}\int &\gamma_{2} \, \, e_{a}\left[F^{a}\left(C^{b}\right)+2\epsilon^{a\left(c\right.}\delta^{\left.b\right)d}\omega_{d}C_{bc}+2\epsilon^{a\left(c\right.}\delta^{\left.b\right)d}s_{d}B_{bc}\right]-\omega dy- s dm -zd\tau \notag\\
    &+t_{a}\left[F^{a}\left(B^{b}\right)+2\epsilon^{a\left(c\right.}\delta^{\left.b\right)d}\omega_{d}B_{bc}\right]+\left[C_{a}F^{a}\left(e^{b}\right)+B_{a}F^{a}\left(t^{b}\right)+C_{ab}F^{ab}\left(e^{cd}\right)\right.\notag\\
    &\left.+B_{ab}F^{ab}\left(t^{cd}\right)\right]-\tau \left(dz+2\epsilon^{ac}B_{a}C_{c}-2\epsilon^{ac}\omega_{a}z_{c}-\epsilon^{ac}s_{a}s_{c}+2\epsilon^{\left(a\right.\left(c\right.}\delta^{\left.d\right)\left.b\right)}B_{ad}C_{cb}\right) \notag\\
    &-m \left(ds+\epsilon^{ac}B_{a}B_{c}-2\epsilon^{ac}\omega_{a}s_{c}+\epsilon^{\left(a\right.\left(c\right.}\delta^{\left.d\right)\left.b\right)}B_{ad}B_{cb}\right)-y \left(d\omega-\epsilon^{ac}\omega_{a}\omega_{c}\right)\notag\\
    &+e_{ab}\left[F^{ab}\left(C^{cd}\right)+\epsilon^{ac}\omega^{b}C_{c}+\epsilon^{ac}s^{b}B_{c}+\delta^{ab}\epsilon^{cd}\omega_{c}C_{d}+\delta^{ab}\epsilon^{cd}s_{c}B_{d}\right]\notag\\
    &+t_{ab}\left[F^{ab}\left(B^{cd}\right)+\epsilon^{ac}\omega^{b}B_{c}+\delta^{ab}\epsilon^{cd}\omega_{c}B_{d}\right]\notag\\
    &-\tau_{a}\left[F^{a}\left(z^{b}\right)+2\epsilon^{c\left(d\right.}B_{c}C_{d}^{\ \left.a\right)}+2\epsilon^{c\left(d\right.}C_{c}B_{d}^{\ \left.a\right)}\right]-m_{a}\left[F^{a}\left(s^{b}\right)+2\epsilon^{c\left(d\right.}B_{c}B_{d}^{\ \left.a\right)}\right]\notag\\
    &-y_{a}F^{a}\left(\omega^{b}\right)-\omega_{a}F^{a}\left(y^{b}\right)-s_{a}F^{a}\left(m^{b}\right)-z_{a}F^{a}\left(\tau^{b}\right)\notag\\
    & +\frac{1}{\ell^2}\left[2\epsilon^{ac}e_{a}\tau t_{c}+\epsilon^{ac}e_{a}m e_{c}+2\epsilon^{a\left(c\right.}\delta^{\left.b\right)d}e_{a}\tau_{d}t_{bc}+2\epsilon^{a\left(c\right.}\delta^{\left.b\right)d}e_{a}m_{d}e_{bc}\right. \notag\\
    &\left.+2\epsilon^{a\left(c\right.}\delta^{\left.b\right)d}t_{a}\tau_{d}e_{bc}-2\epsilon^{\left(a\right.\left(c\right.}\delta^{\left.d\right)\left.b\right)}\tau e_{ad}t_{cb}-\epsilon^{\left(a\right.\left(c\right.}\delta^{\left.d\right)\left.b\right)}m e_{ad}e_{cb}+\epsilon^{ac}y\tau_a\tau_{c}\right.\notag\\
    &\left.+2\epsilon^{ac}\tau\tau_{a}y_{c}+2\epsilon^{ac}m\tau_am_{c}+\epsilon^{ac}\tau m_{a}m_{c}\right]\,, \label{CSpne}
\end{align}
where
\begin{align}
F^{a}\left(B^{b}\right)&=dC^{a}+\epsilon^{ac}\omega B_{c}\,, &F^{a}\left(C^{b}\right)&=dC^{a}+\epsilon^{ac}\omega C_{c}+\epsilon^{ac}s B_{c}\,, \notag\\ F^{a}\left(e^{b}\right)&=de^{a}+\epsilon^{ac}\omega e_{c}\,, &F^{a}\left(t^{b}\right)&=dt^{a}+\epsilon^{ac}\omega t_{c}+\epsilon^{ac}s e_{c}\,, \notag\\
F^{ab}\left(B^{cd}\right)&=dB^{ab}+\epsilon^{\left(a\right|c}\omega B_{c}^{\ \left|b\right)}\,, &F^{ab}\left(C^{cd}\right)&=dC^{ab}+\epsilon^{\left(a\right|c}\omega C_{c}^{\ \left|b\right)}++\epsilon^{\left(a\right|c}s B_{c}^{\ \left|b\right)}\,, \notag\\
F^{ab}\left(e^{cd}\right)&=de^{ab}+\epsilon^{\left(a\right|c}\omega e_{c}^{\ \left|b\right)}\,, &F^{ab}\left(t^{cd}\right)&=dt^{ab}+\epsilon^{\left(a\right|c}\omega t_{c}^{\ \left|b\right)}+\epsilon^{\left(a\right|c}s e_{c}^{\ \left|b\right)}\,, \notag \\
F^{a}\left(\omega^{b}\right)&=d\omega^{a}+\epsilon^{ac}\omega\omega_{c}\,, &F^{a}\left(s^{b}\right)&=ds^{a}+\epsilon^{ac}\omega s_{c}+\epsilon^{ac}s\omega_{c}\,, \notag\\
F^{a}\left(z^{b}\right)&=dz^{a}+\epsilon^{ac}\omega z_{c}+\epsilon^{ac}ss_{c}+\epsilon^{ac}z\omega_{c}\,, &F^{a}\left(\tau^{b}\right)&=d\tau^{a}+\epsilon^{ac}\omega\tau_{c}\,,\notag\\
F^{a}\left(m^{b}\right)&=dm^{a}+\epsilon^{ac}\omega m_{c}+\epsilon^{ac}s\tau_{c}\,, &F^{a}\left(y^{b}\right)&=dy^{a}+\epsilon^{ac}\omega y_{c}+\epsilon^{ac}sm_{c}+\epsilon^{ac}z\tau_{c}\,. \label{defcurv3}
\end{align}
Here, we have identified the gauge fields of the $\mathfrak{hs_{3}nh}^{\left(4\right)}$ algebra with the spin-3 post-Newtonian ones in the following way:
\begin{align}
\omega&=\omega^{(0)}\,, & B^{a}&=B^{a(0)} \,, &s&=\omega^{(1)}\,, &C^{a}&=B^{a(1)}\,, &z&=\omega^{(2)}\,, \notag\\
\tau&=\tau^{(0)}\,, &e^{a}&=e^{a(0)}\,, &m&=\tau^{(1)}\,, &t^{a}&=e^{a(1)}\,, &y&=\tau^{(2)}\,, \notag\\
\omega^{a}&=\omega^{a(0)}\,, &B^{ab}&=B^{ab(0)}\,, &s^{a}&=\omega^{a(1)}\,, &C^{ab}&=B^{ab(1)}\,, &z^{a}&=\omega^{a(2)}\,, \notag\\
\tau^{a}&=\tau^{a(0)}\,, &e^{ab}&=e^{ab(0)}\,, &m^{a}&=\tau^{a(1)}\,, &t^{ab}&=e^{ab(1)}\,, &y^{a}&=\tau^{a(2)}\,. 
\end{align}
Naturally in the vanishing cosmological constant limit $\ell\rightarrow\infty$, the theory reproduces the $\mathfrak{hs_{3}eNewt}$ CS gravity.
\section{Ultra-relativistic Spin-3 Gravity Theories}\label{sec5}
In this section, we present the construction of the ultra-relativistic spin-3 CS gravity actions in three-dimensions invariant under the ultra-relativistic spin-3 algebras introduced in Section \ref{sec3}. As we will see, unlike the non-relativistic case the infinite-dimensional spin-3 AdS Carroll symmetry admits non-degenerate invariant tensor for all values of $N$. Then, the ultra-relativistic CS actions will lead to the vanishing of all curvatures as equations of motion. Before approaching the infinite dimensional spin-3 AdS Carroll gravity, we will construct the simplest case, being the spin-3 extension of the AdS Carroll gravity.
\subsection{Spin-3 AdS Carroll Gravity}
The spin-3 AdS Carroll algebra admits the following non-vanishing components of an invariant tensor for the AdS Carroll algebra,
\begin{align}
    \langle \texttt{G}_a\texttt{P}_{b} \rangle&=\alpha_{1}\delta_{ab}\,, &\langle \texttt{J}\texttt{J}\rangle &=-\alpha_{0}\,, & \langle \texttt{J}\texttt{H}\rangle &=-\alpha_{1}\,, &
   \langle \texttt{P}_{a}\texttt{P}_{b} \rangle &=\frac{\alpha_{0}}{\ell^2}\delta_{ab}\,, \label{ITcar1}
\end{align}
along its spin-3 extension,
\begin{align}
     \langle \texttt{G}_{ab}\texttt{P}_{cd} \rangle&=\alpha_{1}\left(\delta_{a\left(c\right.}\delta_{\left.d\right)b}-\frac{2}{3}\delta_{ab}\delta_{cd}\right)\,, &\langle \texttt{J}_{a}\texttt{J}_{b}\rangle &=-\alpha_{0}\delta_{ab}\,, & \langle  \texttt{J}_{a}\texttt{H}_{b}\rangle &=-\alpha_{1}\delta_{ab}\,, \notag \\
     \langle \texttt{P}_{ab}\texttt{P}_{cd} \rangle&=\frac{\alpha_{0}}{\ell^2}\left(\delta_{a\left(c\right.}\delta_{\left.d\right)b}-\frac{2}{3}\delta_{ab}\delta_{cd}\right)\,,\label{ITcar2}
\end{align}
where $\alpha_0$ and $\alpha_1$ are arbitrary constants which are related to the relativistic ones as follows
\begin{align}
    \alpha_0&=\lambda_0\hat{\alpha}_{0}\,, &\alpha_1&=\lambda_1\hat{\alpha}_{1}\,. \notag
\end{align}
In order to build up the CS action, we first introduce the gauge field one-form $A$ 
\begin{equation}
    A=\omega J+B^{a}G_{a}+\tau H+e^{a}P_{a}+\omega^{a} J_{a}+B^{ab}G_{ab}+\tau^{a} H_{a}+e^{ab}P_{ab}\,.\label{Acar}
\end{equation}
Here, $\omega$ and $B^{a}$ can be interpreted as spin connections for spatial rotations and boosts, while  $\tau$ and $e^{a}$ corresponds to a time-like vielbein and a spatial vielbein, respectively. Analogously, $\omega^{a}$, $B^{ab}$, $\tau^{a}$, and $e^{ab}$ can be interpreted as the corresponding spin-3 versions of the aforementioned vielbeins and spin-connections. 
The corresponding curvature two-form is given by
\begin{align}
F&=R(\omega) J+R^{a}\left(B^{b}\right) G_{a}+R\left(\tau\right) H+R^{a}\left(e^{b}\right)P_{a}+R^{a}\left(\omega^{b}\right) J_{a}+R^{ab}\left(B^{cd}\right) G_{ab} \notag \\
&+R^{a}\left(\tau^{b}\right) H_{a}+R^{ab}\left(e^{cd}\right)P_{ab}\,,
\end{align}
where the components are
\begin{eqnarray}
R\left(\omega\right)&=&d\omega-\frac{1}{2}\epsilon^{ac}\omega_{a}\omega_{c}+\frac{1}{2\ell^{2}}\epsilon^{ac}e_{a}e_{c}+\frac{1}{2\ell^2}\epsilon^{\left(a\right.\left(c\right.}\delta^{\left.d\right)\left.b\right)}e_{ad}e_{cb}\,, \notag \\
R^{a}\left(B^{b}\right)&=& dB^{a}+\epsilon^{ac}\omega B_{c}+\frac{1}{\ell^{2}}\epsilon^{ac}\tau e_{c}+\epsilon^{a\left(c\right.}\delta^{\left.b\right)d}\omega_{d}B_{bc} + \frac{1}{\ell^2}\epsilon^{a\left(c\right.}\delta^{\left.b\right)d}\tau_{d}e_{bc}\,, \notag \\
R\left(\tau\right)&=&d\tau+\epsilon^{ac}B_{a}e_{c}-\epsilon^{ac}\omega_{a}\tau_{c}+\epsilon^{\left(a\right.\left(c\right.}\delta^{\left.d\right)\left.b\right)}e_{ad}B_{cb}\,, \notag \\
R^{a}\left(e^{b}\right)&=&de^{a}+\epsilon^{ac}\omega e_{c}+\epsilon^{a\left(c\right.}\delta^{\left.b\right)d}\omega_{d}e_{bc}\,, \notag \\
R^{a}\left(\omega^{b}\right)&=&d\omega^{a}+\epsilon^{ac}\omega \omega_{c}+\frac{1}{\ell^2}\epsilon^{c\left(d\right.}e_{c}e_{d}^{\ \left.a\right)}\,, \notag \\
R^{ab}\left(B^{cd}\right)&=& dB^{ab}+\epsilon^{\left(a\right|c}\omega B_{c}^{\ \left|b\right)}  +\frac{1}{\ell^2}\epsilon^{\left(a\right|c}\tau e_{c}^{\ \left|b\right)} +\epsilon^{ac}\omega^{b}B_{c}+\frac{1}{\ell^2}\epsilon^{ac}\tau^{b}e_{c}\notag\\
&&+\delta^{ab}\epsilon^{cd}\omega_{c}B_{d}+\frac{1}{\ell^2}\delta^{ab}\epsilon^{cd}\tau_{c}e_{d}\,, \notag \\
R^{a}\left(\tau^{b}\right)&=&d\tau^{a} + \epsilon^{ac}\omega \tau_{c} +\epsilon^{ac}\tau\omega_{c} + \epsilon^{c\left(d\right.}B_{c}e_{d}^{\ \left.a\right)} +\epsilon^{c\left(d\right.}e_{c}B_{d}^{\ \left.a\right)} \,, \notag\\
R^{ab}\left(e^{cd}\right)&=&de^{ab}+\epsilon^{\left(a\right|c}\omega e_{c}^{\ \left|b\right)}+\epsilon^{ac}\omega^{b}e_{c}+\delta^{ab}\epsilon^{cd}\omega_{c}e_{d}\,.\label{curadscar}
\end{eqnarray}
Then, the CS action invariant under the spin-3 AdS Carroll algebra is obtained considering the gauge connection one-form \eqref{Acar} and the non-vanishing components of the invariant tensor \eqref{ITcar1}-\eqref{ITcar2} in the three-dimensional CS action \eqref{CS}. Considering the terms along $\alpha_{1}$, we get in this case
\begin{align}
    I_{\mathfrak{hs_{3}adscar}}=\frac{k}{4\pi}\int & e_{a}\left[F^{a}\left(B^{b}\right)+2\epsilon^{a\left(c\right.}\delta^{\left.b\right)d}\omega_{d}B_{bc}\right]+B_{a}F^{a}\left(e^{b}\right)-\omega d\tau -\tau \left(d\omega-\epsilon^{ac}\omega_{a}\omega_{c}\right)\notag\\
    &\left.+B_{ab}F^{ab}\left(e^{cd}\right)+e_{ab}\left[F^{ab}\left(B^{cd}\right)+\epsilon^{ac}\omega^{b}B_{c}+\delta^{ab}\epsilon^{cd}\omega_{c}B_{d}\right]-\tau_{a}F^{a}\left(\omega^{b}\right)\right. \notag\\
    &\left. -\omega_{a}F^{a}\left(\tau^{b}\right)+\frac{1}{\ell^2}\left(\epsilon^{ac}e_{a}\tau e_{c}+2\epsilon^{a\left(c\right.}\delta^{\left.b\right)d}e_{a}\tau_{d}e_{bc}-\epsilon^{\left(a\right.\left(c\right.}\delta^{\left.d\right)\left.b\right)}\tau e_{ad}e_{cb}\right)\right.\,, \label{adscaract}
\end{align}
where 
we have used the definitions given in \eqref{defcurv}. 
The spin-3 ultra-relativistic CS gravity action \eqref{adscaract} is gauge invariant under the $\mathfrak{hs_{3}adscar}$ algebra \eqref{sp3adscar1} and can be seen as a spin-3 extension of the AdS Carroll CS gravity \cite{Matulich:2019cdo,Ali:2019jjp,Ravera:2019ize}. It is straightforward to see that the previous action leads to a spin-3 extension of the Carroll CS gravity \cite{Hartong:2015xda,Bergshoeff:2016soe} in the vanishing cosmological constant limit $\ell\rightarrow\infty$. Since the invariant tensor \eqref{ITcar1}-\eqref{ITcar2} is non-degenerate, the equations of motion reduce to the vanishing of all curvatures \eqref{curadscar}. Naturally, switching-off the spin-3 gauge fields, the previous theory reduces to the AdS Carroll gravity.

Let us note that the spin-3 ultra-relativistic CS gravity action \eqref{adscaract} based on the spin-3 extension of the AdS Carroll symmetry, can also be obtained directly from the relativistic CS action for the $\mathfrak{hs_{3}AdS}$ algebra \eqref{CSADS3}. This can be done by writing the ultra-relativistic gauge fields in terms of the relativistic ones appearing in \eqref{CSADS3}, through the semigroup elements as follows
\begin{align}
\omega&=\lambda_0 W^{0}\,, &B^{a}&=\lambda_1 W^{a}\,, \notag \\
e^{a}&=\lambda_0 E^{a}\,, &\tau&=\lambda_1 E^{0}\,, \notag \\
\omega^a&=\lambda_0 W^{0a}\,, &B^{ab}&=\lambda_1 W^{ab}\,, \notag \\
e^{ab}&=\lambda_0 E^{ab}\,, &\tau^{a}&=\lambda_1 E^{0a}\,. \label{scarexp}
\end{align}
\subsection{Infinite-dimensional Spin-3 AdS Carroll Gravity}
The spin-3 extension of the infinite-dimensional AdS Carroll algebra \eqref{adscar3N} admits the following non-vanishing components of the invariant tensor for the infinite-dimensional AdS Carroll algebra:
\begin{align}
    \langle \texttt{G}_{a}^{(m)}\texttt{G}_{b}^{(n)} \rangle&=\sigma_{m+n+1}\delta_{ab}\,, &\langle \texttt{J}^{(m)}\texttt{J}^{(n)}\rangle&=-\sigma_{m+n}\,, \notag \\
    \langle \texttt{G}_{a}^{(m)}\texttt{P}_{b}^{(n)} \rangle&=\gamma_{m+n}\delta_{ab}\,, &\langle \texttt{J}^{(m)}\texttt{H}^{(n)}\rangle&=-\gamma_{m+n}\,, \notag \\
    \langle \texttt{P}_{a}^{(m)}\texttt{P}_{b}^{(n)} \rangle&=\frac{1}{\ell^2}\sigma_{m+n}\delta_{ab}\,, &\langle \texttt{H}^{(m)}\texttt{H}^{(n)}\rangle&=-\frac{1}{\ell^2}\sigma_{m+n+1}\,, \label{ITinfcarads1}
\end{align}
along its spin-3 extension
\begin{align}
    \langle \texttt{G}_{ab}^{(m)}\texttt{G}_{cd}^{(n)}\rangle&=\sigma_{m+n+1}\left(\delta_{a\left(c\right.}\delta_{\left.d\right)b}-\frac{2}{3}\delta_{ab}\delta_{cd}\right)\,, &\langle\texttt{J}_{a}^{(m)}\texttt{J}_{b}^{(n)}\rangle&=-\sigma_{m+n}\delta_{ab}\,,\notag \\
    \langle \texttt{G}_{ab}^{(m)}\texttt{P}_{cd}^{(n)}\rangle&=\gamma_{m+n}\left(\delta_{a\left(c\right.}\delta_{\left.d\right)b}-\frac{2}{3}\delta_{ab}\delta_{cd}\right)\,, &\langle\texttt{J}_{a}^{(m)}\texttt{H}_{b}^{(n)}\rangle&=-\gamma_{m+n}\delta_{ab}\,,\notag \\
    \langle \texttt{P}_{ab}^{(m)}\texttt{P}_{cd}^{(n)}\rangle&=\frac{1}{\ell^2}\sigma_{m+n}\left(\delta_{a\left(c\right.}\delta_{\left.d\right)b}-\frac{2}{3}\delta_{ab}\delta_{cd}\right)\,, &\langle\texttt{H}_{a}^{(m)}\texttt{H}_{b}^{(n)}\rangle&=-\frac{1}{\ell^2}\sigma_{m+n+1}\delta_{ab}\,,\label{ITinfcarads2}
\end{align}
where the $\sigma$'s
 and $\gamma$'s are defined in terms of the relativistic constants $\hat{\alpha}_{0}$ and $\hat{\alpha}_{1}$ through the elements of the semigroup $S_{E}^{(N)}$ as
 \begin{align}
     \sigma_{m+n}&=\lambda_{2(m+n)}\hat{\alpha}_{0}\,, \notag \\
     \gamma_{m+n}&=\lambda_{2(m+n)+1}\hat{\alpha}_{1}\,.
 \end{align}
 In particular, $\sigma_{0}$ and $\gamma_{0}$ correspond to the arbitrary constants $\alpha_{0}$ and $\alpha_{1}$ appearing in the invariant tensor for the spin-3 extension of the AdS Carroll algebra. Note also that, unlike the $\mathfrak{hs_{3}nh}^{(N)}$, the $\mathfrak{hs_{3}adscar}^{(N)}$ admits non-degenerate invariant tensor for all values of $N$. Furthermore, in the flat limit $\ell\rightarrow\infty$, \eqref{ITinfcarads1}-\eqref{ITinfcarads2} reduce to the components of the invariant tensor for the spin-3 extension of the infinite-dimensional Carroll algebra.
 
 Let us move on now to the construction of the explicit CS gravity action based on the $\mathfrak{hs_{3}adscar}^{(N)}$ algebra. For this purpose, we define the gauge connection one-form valued on the aforesaid algebra as follows: 
 \begin{align}
    A&=\sum_{m=0}^{\left[\frac{N}{2}\right]}\left(\omega^{(m)}\texttt{J}^{(m)}+e^{a(m)}\texttt{P}_{a}^{(m)}+\omega^{a(m)}\texttt{J}_{a}^{(m)}+e^{ab(m)}\texttt{P}_{ab}^{(m)}\right)\notag \\
    &+\sum_{m=0}^{\left[\frac{N+1}{2}\right]}\left(B^{a(m)}\texttt{G}_{a}^{(m)}+\tau^{(m)}\texttt{H}^{(m)}+B^{ab(m)}\texttt{G}_{ab}^{(m)}+\tau^{a(m)}\texttt{H}_{a}^{(m)}\right)\,, \label{infAadscar}
\end{align}
where $\{\omega^{(m)},B^{a(m)},\tau^{(m)},e^{a(m)}\}$ are spin-2 gauge fields being expansions of the spin-connections and vielbeins. On the other hand, $\{\omega^{a(m)},B^{ab(m)},\tau^{a(m)},e^{ab(m)}\}$ represents spin-3 gauge fields corresponding to expansions of the spin-3 versions of the spin-connections and vielbeins. The curvature two-form $F$ reads
\begin{align}
    F=&R\left(\omega^{(m)}\right)\texttt{J}^{(m)}+R^{a}\left(B^{b(m)}\right)\texttt{G}_{a(m)}+R\left(\tau^{(m)}\right)\texttt{H}^{(m)}+R^{a}\left(e^{b(m)}\right)\texttt{P}_{a}^{(m)}\notag \\
    &+R^{a}\left(\omega^{b(m)}\right)\texttt{J}_{a}^{(m)}+R^{ab}\left(B^{cd(m)}\right)\texttt{G}_{ab}^{(m)}+R^{a}\left(\tau^{b(m)}\right)\texttt{H}_{a}^{(m)}+R^{ab}\left(e^{cd(m)}\right)\texttt{P}_{ab}^{(m)}\,,\label{2Finfadscara}
\end{align}
where the components are defined in Appendix \eqref{App3}. A CS action gauge invariant under the $\mathfrak{hs_{3}adscar}^{\left(N\right)}$ algebra can be constructed considering the gauge connection one-form \eqref{infAadscar} and the non-vanishing components of the invariant tensor \eqref{ITinfcarads1}-\eqref{ITinfcarads2} for a finite value of N in the general expression of the CS form \eqref{CS}. For the sake of simplicity, we will consider only the term along $\gamma$'s omitting the exotic terms. Then, the $\mathfrak{hs_{3}adscar}^{\left(N\right)}$ CS gravity action reads
\begin{align}
    I_{\mathfrak{hs_{3}adscar}^{\left(N\right)}}=\frac{k}{4\pi}\int &\sum_{i=0}^{\left[N/2\right]}\gamma_{i} \, \, e_{a}^{(m)}\left[F^{a}\left(B^{b(n)}\right)\delta_{m+n}^{i}+2\epsilon^{a\left(c\right.}\delta^{\left.b\right)d}\omega_{d}^{(n)}B_{bc}^{(l)}\delta_{m+n+l}^{i}\right]-\omega^{(m)} d\tau^{(n)}\delta_{m+n}^{i}\notag\\
    &+\left[B_{a}^{(m)}F^{a}\left(e^{b(n)}\right)+B_{ab}^{(m)}F^{ab}\left(e^{cd(n)}\right)\right]\delta_{m+n}^{i} -\tau^{(m)} \left(d\omega^{(n)}\delta_{m+n}^{i}\right.\notag\\
    &\left.+\epsilon^{ac}B_{a}^{(n)}B_{c}^{(l)}\delta_{m+n+l+1}^{i}-\epsilon^{ac}\omega_{a}^{(n)}\omega_{c}^{(l)}\delta_{m+n+l}^{i}+\epsilon^{\left(a\right.\left(c\right.}\delta^{\left.d\right)\left.b\right)}B_{ad}^{(n)}B_{cb}^{(l)}\delta_{m+n+l+1}^{i}\right) \notag\\ &+e_{ab}^{(m)}\left[F^{ab}\left(B^{cd(n)}\right)\delta_{m+n}^{i}+\left(\epsilon^{ac}\omega^{b(n)}B_{c}^{(l)}+\delta^{ab}\epsilon^{cd}\omega_{c}^{(n)}B_{d}^{(l)}\right)\delta_{m+n+l}^{i}\right]\notag\\
    &-\tau_{a}^{(m)}\left[F^{a}\left(\omega^{b(n)}\right)\delta_{m+n}^{i}+2\epsilon^{c\left(d\right.}B_{c}^{(n)}B_{d}^{\ \left.a\right)\, (l)}\delta_{m+n+l+2}^{i}\right]-\omega_{a}^{(m)}F^{a}\left(\tau^{b(n)}\right)\delta_{m+n}^{i}\notag\\
    & +\frac{1}{\ell^2}\left[\left(\epsilon^{ac}e_{a}^{(m)}\tau^{(n)} e_{c}^{(l)}+2\epsilon^{a\left(c\right.}\delta^{\left.b\right)d}e_{a}^{(m)}\tau_{d}^{(n)}e_{bc}^{(l)}-\epsilon^{\left(a\right.\left(c\right.}\delta^{\left.d\right)\left.b\right)}\tau^{(m)} e_{ad}^{(n)}e_{cb}^{(l)}\right)\delta_{m+n+l}^{i}\right. \notag\\
    &\left.+\epsilon^{ac}\tau^{(m)}\tau_a^{(n)}\tau_{c}^{(l)}\delta_{m+n+l+2}^{i}\right]\,, \label{infCSadscar}
\end{align}
where the curvatures $F^{a}\left(B^{b(m)}\right)$, $F^{a}\left(e^{b(m)}\right)$, $F^{ab}\left(B^{cd(m)}\right)$, $F^{ab}\left(e^{cd(m)}\right)$, $F^{a}\left(\omega^{b(m)}\right)$ and $F^{a}\left(\tau^{b(m)}\right)$ are given in \eqref{defcurv2}. For $N=\infty$, the CS action \eqref{infCSadscar} corresponds to the infinite-dimensional extension of the $\mathfrak{hs_{3}adscar}$ CS gravity which in the non-vanishing cosmological constant limit $\ell\rightarrow\infty$ reproduces the infinite-dimensional spin-3 Carrollian gravity. Unlike the non-relativistic spin-3 theory, the $\mathfrak{hs_{3}adscar}^{\left(N\right)}$ algebras admit non-degenerate invariant traces for any value of $N$. Then, the equations of motion are given by the vanishing of the $\mathfrak{hs_{3}adscar}^{\left(N\right)}$ curvature two-forms \eqref{infhs3adscar} allowing us to express the spin-connections in terms of the other gauge fields. Let us note that the CS action \eqref{infCSadscar} can alternatively been recovered from the relativistic $\mathfrak{hs_{3}AdS}$ CS action by expressing the ultra-relativistic gauge fields in terms of the relativistic ones and the elements of the $S_{E}^{\left(N\right)}$ semigroup,
\begin{align}
    \omega^{(m)}&=\lambda_{2m}W^{0}\,, &B^{a(m)}&=\lambda_{2m+1}W^{a} \,, &\tau^{(m)}&=\lambda_{2m+1}E^{0}\,, &e^{a(m)}&=\lambda_{2m}E^{a}\,, \notag\\
    \omega^{a(m)}&=\lambda_{2m}W^{0a}\,, &B^{ab(m)}&=\lambda_{2m+1}W^{ab}\,, &\tau^{a(m)}&=\lambda_{2m+1}E^{0a}\,, &e^{ab(m)}&=\lambda_{2m}E^{ab}\,.
\end{align}
For $N=1$, we recover the spin-3 extension of the AdS Carroll CS gravity action \eqref{adscaract} which in the flat limit $\ell\rightarrow\infty$ leads us to the spin-3 extension of the three-dimensional Carroll CS gravity. For $N=2$, we obtain the CS action for the $\mathfrak{hs_{3}eadscar}$ algebra. However, the new contributions from the spin-3 extended AdS Carroll algebra appears exclusively along the exotic term proportional to $\sigma_1$. It is necessary to consider $N=3$ in order to have a new CS term along $\gamma$'s. In particular, for odd values of $N$, the ultra-relativistic CS action \eqref{infCSadscar} can be written as
\begin{equation}
    I_{\mathfrak{hs_{3}adscar}^{\left(N\right)}}=I_{\mathfrak{hs_{3}adscar}}+I_{\mathfrak{hs_{3}enhadscar}}+\sum_{i=2}^{\left[N/2\right]}I_{\mathfrak{hs_{3}adscar}^{\left(2i+1\right)}}\,,
\end{equation}
where $I_{\mathfrak{hs_{3}adscar}}$ is the CS action given by \eqref{adscaract}, while $I_{\mathfrak{hs_{3}enhadscar}}$ can be obtained from the general expression \eqref{infCSadscar} setting $N=3$. Naturally, in the vanishing cosmological constant limit $\ell\rightarrow\infty$, the generalized spin-3 ultra-relativistic CS action is now given by
\begin{equation}
    I_{\mathfrak{hs_{3}car}^{\left(N\right)}}=I_{\mathfrak{hs_{3}car}}+I_{\mathfrak{hs_{3}enhcar}}+\sum_{i=2}^{\left[N/2\right]}I_{\mathfrak{hs_{3}car}^{\left(2i+1\right)}}\,,
\end{equation}
with $I_{\mathfrak{hs_{3}enhcar}}$ being the CS action gauge invariant under the spin-3 enhanced Carroll algebra.  For completeness, we present the explicit expression for the $\mathfrak{hs_{3}enhadscar}$ CS action:
\begin{align}
    I_{\mathfrak{hs_{3}enhadscar}}=\frac{k}{4\pi}\int &\gamma_{1} \, \, e_{a}\left[F^{a}\left(C^{b}\right)+2\epsilon^{a\left(c\right.}\delta^{\left.b\right)d}\omega_{d}C_{bc}+2\epsilon^{a\left(c\right.}\delta^{\left.b\right)d}s_{d}B_{bc}\right]-\omega dm-s d\tau\notag\\
    &+t_{a}\left[F^{a}\left(B^{b}\right)+2\epsilon^{a\left(c\right.}\delta^{\left.b\right)d}\omega_{d}B_{bc}\right]+B_{a}F^{a}\left(t^{b}\right)+C_{a}F^{a}\left(e^{b}\right)+B_{ab}F^{ab}\left(t^{cd}\right) \notag\\
    & +C_{ab}F^{ab}\left(e^{cd}\right)-\tau \left(ds+\epsilon^{ac}B_{a}B_{c}-2\epsilon^{ac}\omega_{a}s_{c}+\epsilon^{\left(a\right.\left(c\right.}\delta^{\left.d\right)\left.b\right)}B_{ad}B_{cb}\right)\notag\\
    &+e_{ab}\left[F^{ab}\left(C^{cd}\right)+\epsilon^{ac}\omega^{b}C_{c}+\epsilon^{ac}s^{b}B_{c}+\delta^{ab}\epsilon^{cd}\omega_{c}C_{d}+\delta^{ab}\epsilon^{cd}s_{c}B_{d}\right]\notag\\
    &+t_{ab}\left[F^{ab}\left(B^{cd}\right)+\epsilon^{ac}\omega^{b}B_{c}+\delta^{ab}\epsilon^{cd}\omega_{c}B_{d}\right]+m\left(d\omega-\epsilon^{ac}\omega_{a}\omega_{c}\right)\notag\\
    &-\tau_{a}\left[F^{a}\left(s^{b}\right)+2\epsilon^{c\left(d\right.}B_{c}B_{d}^{\ \left.a\right)\, }\right]-m_{a}F^{a}\left(\omega^{b}\right)-\omega_{a}F^{a}\left(m^{b}\right)\notag\\
    & +\frac{1}{\ell^2}\left[\epsilon^{ac}e_{a}m e_{c}+2\epsilon^{ac}e_{a}\tau t_{c}+2\epsilon^{a\left(c\right.}\delta^{\left.b\right)d}e_{a}m_{d}e_{bc}+2\epsilon^{a\left(c\right.}\delta^{\left.b\right)d}e_{a}\tau_{d}t_{bc}\right. \notag\\
    &\left.+2\epsilon^{a\left(c\right.}\delta^{\left.b\right)d}t_{a}\tau_{d}e_{bc}-\epsilon^{\left(a\right.\left(c\right.}\delta^{\left.d\right)\left.b\right)}m e_{ad}e_{cb}-2\epsilon^{\left(a\right.\left(c\right.}\delta^{\left.d\right)\left.b\right)}\tau e_{ad}t_{cb}+\epsilon^{ac}\tau\tau_a\tau_{c}\right]\,, \label{CSenhadscar}
\end{align}
where the curvatures $F^{a}$ and $F^{ab}$ are given by \eqref{defcurv3}. Moreover, we have identified the gauge fields of the $\mathfrak{hs_{3}adscar}^{\left(3\right)}$ algebra with the spin-3 enhanced AdS Carroll fields as:
\begin{align}
\omega&=\omega^{(0)}\,, & B^{a}&=B^{a(0)} \,, &s&=\omega^{(1)}\,, &C^{a}&=B^{a(1)}\,, \notag\\
\tau&=\tau^{(0)}\,, &e^{a}&=e^{a(0)}\,, &m&=\tau^{(1)}\,, &t^{a}&=e^{a(1)}\,, \notag\\
\omega^{a}&=\omega^{a(0)}\,, &B^{ab}&=B^{ab(0)}\,, &s^{a}&=\omega^{a(1)}\,, &C^{ab}&=B^{ab(1)}\,,  \notag\\
\tau^{a}&=\tau^{a(0)}\,, &e^{ab}&=e^{ab(0)}\,, &m^{a}&=\tau^{a(1)}\,, &t^{ab}&=e^{ab(1)}\,. 
\end{align}
In the vanishing cosmological constant limit $\ell\rightarrow\infty$ the CS action \eqref{CSenhadscar} reproduces the $\mathfrak{hs_{3}enhcar}$ CS gravity action which can be seen as the spin-3 Carrollian analogue of the Newtonian gravity \cite{Hansen:2018ofj}.

\section{Discussions}\label{concl}

In this work we present novel spin-3 extensions of known non- and ultra-relativistic gravity theories. We first focus our study on the derivation of spin-3 non- and ultra-relativistic algebras by considering the semigroup expansion method to the spin-3 AdS algebra. We show that, analogously to \cite{Gomis:2019nih}, the expansion procedure based on the $S_E$ semigroup \cite{Izaurieta:2006zz} can be seen as a non- or ultra-relativistic expansion of a given relativistic algebra which turns out to be useful to derive the non-vanishing components of a invariant tensor of the expanded algebra. 

At the non-relativistic level, we first obtain the spin-3 extension of the extended Newton-Hooke gravity theory which in the vanishing cosmological constant limit reproduces a CS gravity theory invariant under the spin-3 extended Bargmann algebra introduced in \cite{Bergshoeff:2016soe}. By considering a bigger semigroup, we obtain a post-Newtonian extension of the spin-3 extended Newton-Hooke which can be viewed as the spin-3 extension of the enhanced Bargmann-Newton-Hooke gravity \cite{Gomis:2019nih,Concha:2019dqs,Bergshoeff:2020fiz}. We then extend our analysis to an infinite-dimensional semigroup $S_E^{\left(N\right)}$ allowing us to reproduce a spin-3 infinite-dimensional Newton-Hooke gravity which in the flat limit leads us to a spin-3 extension of the infinite-dimensional Galilean gravity \cite{Gomis:2019nih,Concha:2022you}. We focus our construction only on even value of $N$ which allows us to derive spin-3 non-relativistic symmetries admitting a non-degenerate bilinear invariant trace and ensuring the proper construction of a well-defined CS action. Subsequently, we generalize our procedure at the ultra-relativistic realm which does not suffer from degeneracy. We first present a spin-3 AdS Carroll CS gravity theory which in the flat limit reproduces a CS theory based on the spin-3 Carroll symmetry \cite{Bergshoeff:2016soe}. We end our study with a spin-3 extension of the infinite-dimensional AdS Carroll gravity.

The results obtained along the procedure considered here could be used to approach several open questions. It would be interesting to explore first if the expansion method based on the $S_E$ semigroup can be used to elucidate the corresponding non- and ultra-relativistic versions of the asymptotic symmetries of the three-dimensional AdS CS gravity and its Poincaré limit. One could argue that the $S_{E}^{\left(2\right)}$-expansion of the conformal and $\mathfrak{bms}_{3}$ algebra \cite{Brown:1986nw,Barnich:2006av} correspond to infinite-dimensional lifts of the extended Newton-Hooke and extended Bargmann algebra, respectively. To verify our conjecture, it would be necessary to consider a direct asymptotic symmetry analysis by imposing suitable boundary conditions [work in progress]. If our conjecture is correct, it would be useful to obtain non- and ultra-relativistic versions of the $\mathcal{W}_{3}$ algebra which should be an infinite-dimensional lift of the spin-3 Nappi-Witten structure presented here.

Another aspect that it would be worth it to study is the extension of our results to spin higher than $3/2$ by analyzing the non- and ultra-relativistic limit of the hypersymmetric extension of gravity \cite{Aragone:1979hx,Henneaux:2015ywa}. To our knowledge a non- or ultra-relativistic version of the so-called hyper-Poincaré symmetry \cite{Fuentealba:2015jma,Fuentealba:2015wza} is unknown and, as in the relativistic theory, we expect the presence of spin-4 generators [work in progress]. The method used here could be useful to avoid the difficulty appearing in presence of spin-$3/2$. As it was shown in \cite{Concha:2020tqx,Concha:2021jos}, the expansion procedure based on semigroups allows us to derive in a straightforward way the corresponding non-relativistic superalgebra from a relativistic one. 

On the other hand, one could explore if the spin-3 extended Bargmann gravity theory discussed here could allows us to approach a spin-3 generalization of the Ho\v{r}ava-Lifshitz gravity \cite{Hartong:2015zia,Griffin:2011xs,Griffin:2012qx}. Such guessing is motivated by the fact that the extended Bargmann gravity can be seen as
a particular kinetic term of the Ho\v{r}ava-Lifshitz gravity \cite{Bergshoeff:2016lwr,Hartong:2016yrf}. Moreover, one could explore the effects arising from the presence of the additional gauge field appearing in the post-Newtonian extensions.

Finally, it would be interesting to extend our results to other symmetries being of interest in gravity. For instance, one could explore the non- and ultra-relativistic regime of the spin-3 Maxwell CS gravity theory discussed in \cite{Caroca:2017izc} along its cosmological extension\footnote{A spin-3 non-relativistic Maxwell theory can be found in \cite{Caroca:2022byi} which has appeared in the literature simultaneously to our results.}. The Maxwell algebra has been introduced to describe a constant Minskowski spacetime in the presence of an electromagnetic background \cite{Schrader:1972zd,Bacry:1970ye,Gomis:2017cmt}and have found several application in the gravity context \cite{Duval:2008tr,Durka:2011nf,Salgado:2014jka,Hoseinzadeh:2014bla,Concha:2014vka,Concha:2018zeb,Concha:2018jxx,Chernyavsky:2020fqs,Adami:2020xkm,Caroca:2021bjo,Cebecioglu:2021dqb}. One could expect to get a spin-3 extension of the Maxwellian extended Bargmann gravity \cite{Aviles:2018jzw} and the enlarged extended Bargmann gravity \cite{Concha:2019lhn}.


\section*{Acknowledgment}

This work was funded by the National Agency for Research and Development ANID - PAI grant No. 77190078, ANID - SIA grant No. SA77210097 and FONDECYT grants No. 1211077, 11220328 and 11220486.  P.C. and E.R. would like to thank to the Dirección de Investigación and Vice-rectoría de Investigación of the Universidad Católica de la Santísima Concepción, Chile, for their constant support. 


\appendix

\section{Alternative Spin-3 extensions of the Newton-Hooke family}\label{App1}

An alternative spin-3 extension of the infinite-dimensional Newton-Hooke algebra can be obtained considering a different subspace decomposition of the $\mathfrak{hs_{3}AdS}$. Let $V_0$ and $V_1$ be two subspaces of the relativistic spin-3 AdS algebra defined as
\begin{align}
   V_0&=\{\hat{\texttt{J}},\hat{\texttt{H}},\hat{\texttt{G}}_{ab},\hat{\texttt{P}}_{ab}\} \,,\notag \\
   V_1&=\{\hat{\texttt{G}}_{a},\hat{\texttt{P}}_{a},\hat{\texttt{J}}_{a},\hat{\texttt{H}}_{a}\}\,.\label{sda}
\end{align}
Such subspace decomposition, although it is different to \eqref{sd}, also satisfies a $\mathbb{Z}_2$ graded-Lie algebra. On the other hand, one can consider the same semigroup $S_{E}^{\left(N\right)}$ with the resonant subset decomposition \eqref{sdN} satisfying the same structure than the subspace \eqref{sda}. Then, after applying a resonant $S_{E}^{\left(N\right)}$-expansion of the $\mathfrak{hs_{3}AdS}$ algebra and extracting a $0_s$-reduction we get an alternative generalized spin-3 Newton-Hooke algebra whose commutators are given by \eqref{idNH} and
\begin{align}
\left[\texttt{J}^{(m)},\texttt{J}_a^{(n)}\right]&=\epsilon_{ab}\texttt{J}_{b}^{(m+n)}\,,   &\left[\texttt{H}^{(m)},\texttt{J}_a^{(n)}\right]&=\epsilon_{ab}\texttt{H}_{b}^{(m+n)}\,, \notag\\
\left[\texttt{J}^{(m)},\texttt{H}_a^{(n)}\right]&=\epsilon_{ab}\texttt{H}_{b}^{(m+n)}\,,   &\left[\texttt{H}^{(m)},\texttt{H}_a^{(n)}\right]&=\frac{1}{\ell^2}\epsilon_{ab}\texttt{J}_{b}^{(m+n)}\,, \notag\\
\left[\texttt{J}^{(m)},\texttt{G}_{ab}^{(n)}\right]&=-\epsilon_{c\left(a\right.}\texttt{G}_{\left.b\right)c}^{(m+n)}\,,  &\left[\texttt{H}^{(m)},\texttt{G}_{ab}^{(n)}\right]&=-\epsilon_{c\left(a\right.}\texttt{P}_{\left.b\right)c}^{(m+n)}\,, \notag \\
\left[\texttt{J}^{(m)},\texttt{P}_{ab}^{(n)}\right]&=-\epsilon_{c\left(a\right.}\texttt{P}_{\left.b\right)c}^{(m+n)}\,,   &\left[\texttt{H}^{(m)},\texttt{P}_{ab}^{(n)}\right]&=-\frac{1}{\ell^2}\epsilon_{c\left(a\right.}\texttt{G}_{\left.b\right)c}^{(m+n)}\,, \notag \\
\left[\texttt{G}_a^{(m)},\texttt{G}_{bc}^{(n)}\right]&=-\epsilon_{a\left(b\right.}\texttt{J}_{\left.c\right)}^{(m+n)}\,,  &\left[\texttt{P}_a^{(m)},\texttt{G}_{bc}^{(n)}\right]&=-\epsilon_{a\left(b\right.}\texttt{H}_{\left.c\right)}^{(m+n)}\,, \notag \\
\left[\texttt{G}_{a}^{(m)},\texttt{J}_{b}^{(n)}\right]&=-\left(\epsilon_{ac}\texttt{G}_{cb}^{(m+n+1)}+\epsilon_{ab}\texttt{G}_{cc}^{(m+n+1)}\right)\,, &\left[\texttt{G}_a^{(m)},\texttt{P}_{bc}^{(n)}\right]&=-\epsilon_{a\left(b\right.}\texttt{H}_{\left.c\right)}^{(m+n)}\,, \notag \\
\left[\texttt{P}_{a}^{(m)},\texttt{J}_{b}^{(n)}\right]&=-\left(\epsilon_{ac}\texttt{P}_{cb}^{(m+n+1)}+\epsilon_{ab}\texttt{P}_{cc}^{(m+n+1)}\right)\,, &\left[\texttt{P}_a^{(m)},\texttt{P}_{bc}^{(n)}\right]&=-\frac{1}{\ell^2}\epsilon_{a\left(b\right.}\texttt{J}_{\left.c\right)}^{(m+n)}\,, \notag\\
\left[\texttt{G}_{a}^{(m)},\texttt{H}_{b}^{(n)}\right]&=-\left(\epsilon_{ac}\texttt{P}_{cb}^{(m+n+1)}+\epsilon_{ab}\texttt{P}_{cc}^{(m+n+1)}\right)\,, &\left[\texttt{J}_a^{(m)},\texttt{G}_{bc}^{(n)}\right]&=\delta_{a\left(b\right.}\epsilon_{\left.c\right)d}\texttt{G}_{d}^{(m+n)}\,,  \notag \\
\left[\texttt{P}_{a}^{(m)},\texttt{H}_{b}^{(n)}\right]&=-\frac{1}{\ell^2}\left(\epsilon_{ac}\texttt{G}_{cb}^{(m+n+1)}+\epsilon_{ab}\texttt{G}_{cc}^{(m+n+1)}\right)\,,  &\left[\texttt{H}_a^{(m)},\texttt{G}_{bc}^{(n)}\right]&=\delta_{a\left(b\right.}\epsilon_{\left.c\right)d}\texttt{P}_{d}^{(m+n)}\,, \notag \\
\left[\texttt{J}_a^{(m)},\texttt{P}_{bc}^{(n)}\right]&=\delta_{a\left(b\right.}\epsilon_{\left.c\right)d}\texttt{P}_{d}^{(m+n)}\,,   &\left[\texttt{H}_a^{(m)},\texttt{P}_{bc}^{(n)}\right]&=\frac{1}{\ell^2}\delta_{a\left(b\right.}\epsilon_{\left.c\right)d}\texttt{G}_{d}^{(m+n)}\,, \notag \\
\left[\texttt{G}_{ab}^{(m)},\texttt{G}_{bc}^{(n)}\right]&=\delta_{\left(a\right.\left(c\right.}\epsilon_{\left.d\right)\left.b\right)}\texttt{J}^{(m+n)}\,,  &\left[\texttt{G}_{ab}^{(m)},\texttt{P}_{bc}^{(n)}\right]&=\delta_{\left(a\right.\left(c\right.}\epsilon_{\left.d\right)\left.b\right)}\texttt{H}^{(m+n)}\,, \notag\\ \left[\texttt{P}_{ab}^{(m)},\texttt{P}_{bc}^{(n)}\right]&=\frac{1}{\ell^2}\delta_{\left(a\right.\left(c\right.}\epsilon_{\left.d\right)\left.b\right)}\texttt{J}^{(m+n)}\,, 
&\left[\texttt{J}_{a}^{(m)},\texttt{J}_{b}^{(n)}\right]&=\epsilon_{ab}\texttt{J}^{(m+n+1)}\,,\notag\\ \left[\texttt{J}_{a}^{(m)},\texttt{H}_{b}^{(n)}\right]&=\epsilon_{ab}\texttt{H}^{(m+n+1)}\,, 
&\left[\texttt{H}_{a}^{(m)},\texttt{H}_{b}^{(n)}\right]&=\frac{1}{\ell^2}\epsilon_{ab}\texttt{J}^{(m+n+1)}\,. \label{idNH3b}
\end{align}
Here the expanded generators are related to the spin-3 AdS ones through the semigroup elements as
\begin{align}
\texttt{J}^{(m)} &=\lambda _{2m}\hat{\texttt{J}}\,, \quad 
&\texttt{G}_{a}^{(m)} &=\lambda _{2m+1}\hat{\texttt{G}}_{a}\,,  \quad &\texttt{G}_{ab}^{(m)} &=\lambda _{2m}\hat{\texttt{G}}_{ab}\,, \quad &\texttt{J}_{a}^{(m)} &=\lambda _{2m+1}\hat{\texttt{J}}_{a}\,,\notag \\
\texttt{H}^{(m)} &=\lambda _{2m}\hat{\texttt{H}}\,,  \quad
&\texttt{P}_{a}^{(m)} &=\lambda_{2m+1}\hat{\texttt{P}}_{a}\,, \quad &\texttt{P}_{ab}^{(m)} &=\lambda _{2m}\hat{\texttt{P}}_{ab}\,, \quad &\texttt{H}_{a}^{(m)} &=\lambda _{2m+1}\hat{\texttt{H}}_{a}\,. \label{expgen2}
\end{align}
This alternative infinite-dimensional non-relativistic algebras family, which we denote as $\mathfrak{hs_{3}nh2}^{\left(N\right)}$, naturally contains the same spin-2 subalgebra given by \eqref{idNH}. However, it presents various differences with the $\mathfrak{hs_{3}nh}^{\left(N\right)}$ one \eqref{idNH3} mainly due to the fact that the relativistic spin-3 generators have been interchanged of subspaces. The differences appears more clearly at finite level. For instance, for $N=1$, the $\mathfrak{hs_{3}nh2}^{\left(N\right)}$ algebra reproduces the alternative spin-3 Newton-Hooke $\mathfrak{hs_{3}nh2}$ introduced in \cite{Bergshoeff:2016soe}. For $N=2$, an alternative spin-3 extension of the extended Newton-Hooke algebra appears. The expanded algebra corresponds to an extension of the $\mathfrak{hs_{3}nh2}$ algebra presented in \cite{Bergshoeff:2016soe} and is then denoted as $\mathfrak{hs_{3}enh2}$.
\begin{align}
\left[\texttt{J},\texttt{G}_a\right]&=\epsilon_{ab}\texttt{G}_{b}\,,\quad \quad   &\left[\texttt{G}_{a},\texttt{G}_{b}\right]&=-\epsilon_{ab}\texttt{S}\,,\quad  &\left[\texttt{H},\texttt{G}_a\right]&=\epsilon_{ab}\texttt{P}_{b}\,, \notag\\
\left[\texttt{J},\texttt{P}_a\right]&=\epsilon_{ab}\texttt{P}_{b}\,,\quad \quad   &\left[\texttt{G}_{a},\texttt{P}_{b}\right]&=-\epsilon_{ab}\texttt{M}\,,\quad  &\left[\texttt{H},\texttt{P}_a\right]&=\frac{1}{\ell^2}\epsilon_{ab}\texttt{G}_{b}\,, \notag\\
\left[\texttt{J},\texttt{J}_a\right]&=\epsilon_{ab}\texttt{J}_{b}\,,\quad \quad   &\left[\texttt{P}_{a},\texttt{P}_{b}\right]&=-\frac{1}{\ell^2}\epsilon_{ab}\texttt{S}\,,\quad   &\left[\texttt{H},\texttt{J}_a\right]&=\epsilon_{ab}\texttt{H}_{b}\,, \notag\\
\left[\texttt{J},\texttt{H}_a\right]&=\epsilon_{ab}\texttt{H}_{b}\,,\quad \quad   &\left[\texttt{G}_{a},\texttt{J}_{b}\right]&=-\left(\epsilon_{am}\texttt{S}_{mb}+\epsilon_{ab}\texttt{S}_{mm}\right)\,,\quad  &\left[\texttt{H},\texttt{H}_a\right]&=\frac{1}{\ell^2}\epsilon_{ab}\texttt{J}_{b}\,, \notag\\
\left[\texttt{J},\texttt{G}_{ab}\right]&=-\epsilon_{m\left(a\right.}\texttt{G}_{\left.b\right)m}\,, \quad \quad &\left[\texttt{G}_{a},\texttt{H}_{b}\right]&=-\left(\epsilon_{am}\texttt{M}_{mb}+\epsilon_{ab}\texttt{M}_{mm}\right)\,,\quad  &\left[\texttt{H},\texttt{G}_{ab}\right]&=-\epsilon_{m\left(a\right.}\texttt{P}_{\left.b\right)m}\,, \notag \\
\left[\texttt{J},\texttt{P}_{ab}\right]&=-\epsilon_{m\left(a\right.}\texttt{P}_{\left.b\right)m}\,, \quad \quad &\left[\texttt{P}_{a},\texttt{J}_{b}\right]&=-\left(\epsilon_{am}\texttt{M}_{mb}+\epsilon_{ab}\texttt{M}_{mm}\right)\,,\quad  &\left[\texttt{H},\texttt{P}_{ab}\right]&=-\frac{1}{\ell^2}\epsilon_{m\left(a\right.}\texttt{G}_{\left.b\right)m}\,, \notag \\
\left[\texttt{J},\texttt{S}_{ab}\right]&=-\epsilon_{m\left(a\right.}\texttt{S}_{\left.b\right)m}\,, \quad \quad &\left[\texttt{P}_{a},\texttt{H}_{b}\right]&=-\frac{1}{\ell^2}\left(\epsilon_{am}\texttt{S}_{mb}+\epsilon_{ab}\texttt{S}_{mm}\right)\,,\quad  &\left[\texttt{H},\texttt{S}_{ab}\right]&=-\frac{1}{\ell^2}\epsilon_{m\left(a\right.}\texttt{M}_{\left.b\right)m}\,, \notag \\
\left[\texttt{J},\texttt{M}_{ab}\right]&=-\epsilon_{m\left(a\right.}\texttt{M}_{\left.b\right)m}\,, \quad \quad &\left[\texttt{S},\texttt{G}_{ab}\right]&=-\epsilon_{m\left(a\right.}\texttt{S}_{\left.b\right)m}\,,\quad  &\left[\texttt{H},\texttt{M}_{ab}\right]&=-\frac{1}{\ell^2}\epsilon_{m\left(a\right.}\texttt{S}_{\left.b\right)m}\,, \notag \\
\left[\texttt{S},\texttt{P}_{ab}\right]&=-\epsilon_{m\left(a\right.}\texttt{M}_{\left.b\right)m}\,, \quad \quad &\left[\texttt{M},\texttt{G}_{ab}\right]&=-\epsilon_{m\left(a\right.}\texttt{M}_{\left.b\right)m}\,,\quad  &\left[\texttt{M},\texttt{P}_{ab}\right]&=-\frac{1}{\ell^2}\epsilon_{m\left(a\right.}\texttt{S}_{\left.b\right)m}\,, \notag \\
\left[\texttt{G}_a,\texttt{G}_{bc}\right]&=-\epsilon_{a\left(b\right.}\texttt{J}_{\left.c\right)}\,, \quad \quad &\left[\texttt{J}_{a},\texttt{J}_{b}\right]&=\epsilon_{ab}\texttt{S}\,,\quad  &\left[\texttt{P}_a,\texttt{G}_{bc}\right]&=-\epsilon_{a\left(b\right.}\texttt{H}_{\left.c\right)}\,, \notag \\
\left[\texttt{G}_a,\texttt{P}_{bc}\right]&=-\epsilon_{a\left(b\right.}\texttt{H}_{\left.c\right)}\,, \quad \quad &\left[\texttt{J}_{a},\texttt{H}_{b}\right]&=\epsilon_{ab}\texttt{M}\,,\quad  &\left[\texttt{P}_a,\texttt{P}_{bc}\right]&=-\frac{1}{\ell^2}\epsilon_{a\left(b\right.}\texttt{J}_{\left.c\right)}\,, \notag \\
\left[\texttt{J}_a,\texttt{G}_{bc}\right]&=\delta_{a\left(b\right.}\epsilon_{\left.c\right)m}\texttt{G}_{m}\,, \quad \quad &\left[\texttt{H}_{a},\texttt{H}_{b}\right]&=\frac {1}{\ell^2}\epsilon_{ab}\texttt{S}\,,\quad  &\left[\texttt{H}_a,\texttt{G}_{bc}\right]&=\delta_{a\left(b\right.}\epsilon_{\left.c\right)m}\texttt{P}_{m}\,, \notag \\
\left[\texttt{J}_a,\texttt{P}_{bc}\right]&=\delta_{a\left(b\right.}\epsilon_{\left.c\right)m}\texttt{P}_{m}\,, \quad \quad &\left[\texttt{S}_{ab},\texttt{G}_{bc}\right]&=\delta_{\left(a\right.\left(c\right.}\epsilon_{\left.d\right)\left.b\right)}\texttt{S}\,, \quad &\left[\texttt{H}_a,\texttt{P}_{bc}\right]&=\frac{1}{\ell^2}\delta_{a\left(b\right.}\epsilon_{\left.c\right)m}\texttt{G}_{m}\,, \notag \\
\left[\texttt{G}_{ab},\texttt{G}_{cd}\right]&=\delta_{\left(a\right.\left(c\right.}\epsilon_{\left.d\right)\left.b\right)}\texttt{J}\,, \quad \quad &\left[\texttt{S}_{ab},\texttt{P}_{cd}\right]&=\delta_{\left(a\right.\left(c\right.}\epsilon_{\left.d\right)\left.b\right)}\texttt{M}\,,\quad  &\left[\texttt{P}_{ab},\texttt{P}_{cd}\right]&=\frac{1}{\ell^2}\delta_{\left(a\right.\left(c\right.}\epsilon_{\left.d\right)\left.b\right)}\texttt{J}\,, \notag \\
\left[\texttt{G}_{ab},\texttt{P}_{cd}\right]&=\delta_{\left(a\right.\left(c\right.}\epsilon_{\left.d\right)\left.b\right)}\texttt{H}\,, \quad \quad &\left[\texttt{M}_{ab},\texttt{G}_{cd}\right]&=\delta_{\left(a\right.\left(c\right.}\epsilon_{\left.d\right)\left.b\right)}\texttt{M}\,,\quad  &\left[\texttt{M}_{ab},\texttt{P}_{cd}\right]&=\frac{1}{\ell^2}\delta_{\left(a\right.\left(c\right.}\epsilon_{\left.d\right)\left.b\right)}\texttt{S}\,.\label{enh32}
\end{align}
One can notice that there are several differences with the commutation relations of the $\mathfrak{hs_{3}enh}$ algebra \eqref{enh3}. Such differences are due to the fact that the spin-3 content of the $\mathfrak{hs_{3}enh2}$ is given by $\{\texttt{J}_{a},\texttt{H}_{a},\texttt{G}_{ab},\texttt{P}_{ab},\texttt{S}_{ab},\texttt{M}_{ab}\}$, while the $\mathfrak{hs_{3}enh}$ algebra contain $\{\texttt{J}_{a},\texttt{H}_{a},\texttt{G}_{ab},\texttt{P}_{ab},\texttt{S}_{a},\texttt{M}_{a}\}$ as spin-3 generators. Although they are quite different both spin-3 non-relativistic algebras admits a non-degenerate invariant tensor and both structures can be written as two copies of a spin-3 extension of the Nappi-Witten algebra \cite{Nappi:1993ie,Figueroa-OFarrill:1999cmq}.  In particular, the $\mathfrak{hs_{3}enh2}$ algebra can be written as:
\begin{align}
\left[\texttt{J}^{\pm},\texttt{G}^{\pm}_a\right]&=\epsilon_{ab}\texttt{G}^{\pm}_{b}\,,\quad \quad   &\left[\texttt{G}^{\pm}_a,\texttt{G}^{\pm}_{bc}\right]&=-\epsilon_{a\left(b\right.}\texttt{J}^{\pm}_{\left.c\right)}\,,\quad  &\left[\texttt{G}^{\pm}_{a},\texttt{G}^{\pm}_{b}\right]&=-\epsilon_{ab}\texttt{S}^{\pm}\,, \notag\\
\left[\texttt{J}^{\pm},\texttt{J}^{\pm}_a\right]&=\epsilon_{ab}\texttt{J}^{\pm}_{b}\,,\quad \quad   &\left[\texttt{J}^{\pm}_a,\texttt{G}^{\pm}_{bc}\right]&=\delta_{a\left(b\right.}\epsilon_{\left.c\right)m}\texttt{G}^{\pm}_{m}\,,\quad  &\left[\texttt{J}^{\pm}_{a},\texttt{J}^{\pm}_{b}\right]&=\epsilon_{ab}\texttt{S}^{\pm}\,, \notag\\
\left[\texttt{J}^{\pm},\texttt{G}^{\pm}_{ab}\right]&=-\epsilon_{a\left(b\right.}\texttt{G}^{\pm}_{\left.b\right)m} \,,\quad \quad   &\left[\texttt{G}^{\pm}_{ab},\texttt{G}^{\pm}_{cd}\right]&=\delta_{\left(a\right.\left(c\right.}\epsilon_{\left.d\right)\left.b\right)}\texttt{J}^{\pm}\,,\quad &\left[\texttt{G}^{\pm}_{a},\texttt{J}^{\pm}_{b}\right]&=-\left(\epsilon_{am}\texttt{S}^{\pm}_{mb}+\epsilon_{ab}\texttt{S}^{\pm}_{mm}\right)\,, \notag\\
\left[\texttt{J}^{\pm},\texttt{S}^{\pm}_{ab}\right]&=-\epsilon_{m\left(a\right.}\texttt{S}^{\pm}_{\left.b\right)m}\,, \quad \quad &\left[\texttt{S}^{\pm}_{ab},\texttt{G}^{\pm}_{cd}\right]&=\delta_{\left(a\right.\left(c\right.}\epsilon_{\left.d\right)\left.b\right)}\texttt{S}^{\pm}\,, \quad \quad &\left[\texttt{S}^{\pm},\texttt{G}^{\pm}_{ab}\right]&=-\epsilon_{m\left(a\right.}\texttt{S}^{\pm}_{\left.b\right)m}\,. \label{hs3nw2}
\end{align}
These two copies of the spin-3 Nappi-Witten algebra, which we have denoted as $\mathfrak{hs_{3}nw2}$ differ from the $\mathfrak{hs_{3}nw}$ obtained in section \ref{sec311} not only in various specific commutators but also in the spin-3 generators.

An alternative Post-Newtonian extension of the spin-3 extended Newton-Hooke appears considering a resonant $S_{E}^{\left(4\right)}$-expansion of the $\mathfrak{hs_{3}AdS}$ algebra, the latter being decomposed as in \eqref{sda}. The alternative non-relativistic algebra is denoted as $\mathfrak{hs_{3}pne2}$ and obeys \eqref{enh32} along the following commutation relations:
\begin{align}
\left[\texttt{J},\texttt{B}_a\right]&=\epsilon_{ab}\texttt{B}_{b}\,,\quad \quad   &\left[\texttt{G}_{a},\texttt{B}_{b}\right]&=-\epsilon_{ab}\texttt{Z}\,,\quad  &\left[\texttt{H},\texttt{B}_a\right]&=\epsilon_{ab}\texttt{T}_{b}\,, \notag\\
\left[\texttt{J},\texttt{T}_a\right]&=\epsilon_{ab}\texttt{T}_{b}\,,\quad \quad   &\left[\texttt{G}_{a},\texttt{T}_{b}\right]&=-\epsilon_{ab}\texttt{Y}\,,\quad  &\left[\texttt{H},\texttt{T}_a\right]&=\frac{1}{\ell^2}\epsilon_{ab}\texttt{B}_{b}\,, \notag\\
\left[\texttt{J},\texttt{S}_a\right]&=\epsilon_{ab}\texttt{S}_{b}\,,\quad \quad   &\left[\texttt{P}_{a},\texttt{B}_{b}\right]&=-\epsilon_{ab}\texttt{Y}\,,\quad   &\left[\texttt{H},\texttt{S}_a\right]&=\epsilon_{ab}\texttt{M}_{b}\,, \notag\\
\left[\texttt{J},\texttt{M}_a\right]&=\epsilon_{ab}\texttt{M}_{b}\,,\quad \quad   &\left[\texttt{P}_{a},\texttt{T}_{b}\right]&=-\frac{1}{\ell^2}\epsilon_{ab}\texttt{Z}\,,\quad   &\left[\texttt{H},\texttt{M}_a\right]&=\frac{1}{\ell^2}\epsilon_{ab}\texttt{S}_{b}\,, \notag\\
\left[\texttt{S},\texttt{G}_a\right]&=\epsilon_{ab}\texttt{B}_{b}\,,\quad \quad   &\left[\texttt{G}_{a},\texttt{S}_{b}\right]&=-\left(\epsilon_{am}\texttt{Z}_{mb}+\epsilon_{ab}\texttt{Z}_{mm}\right)\,,\quad  &\left[\texttt{M},\texttt{G}_a\right]&=\epsilon_{ab}\texttt{T}_{b}\,, \notag\\
\left[\texttt{S},\texttt{P}_a\right]&=\epsilon_{ab}\texttt{T}_{b}\,,\quad \quad   &\left[\texttt{G}_{a},\texttt{M}_{b}\right]&=-\left(\epsilon_{am}\texttt{Y}_{mb}+\epsilon_{ab}\texttt{Y}_{mm}\right)\,,\quad  &\left[\texttt{M},\texttt{P}_a\right]&=\frac{1}{\ell^2}\epsilon_{ab}\texttt{B}_{b}\,, \notag\\
\left[\texttt{S},\texttt{J}_a\right]&=\epsilon_{ab}\texttt{S}_{b}\,,\quad \quad   &\left[\texttt{P}_{a},\texttt{S}_{b}\right]&=-\left(\epsilon_{am}\texttt{Y}_{mb}+\epsilon_{ab}\texttt{Y}_{mm}\right)\,,\quad  &\left[\texttt{M},\texttt{J}_a\right]&=\epsilon_{ab}\texttt{M}_{b}\,, \notag\\
\left[\texttt{S},\texttt{H}_a\right]&=\epsilon_{ab}\texttt{M}_{b}\,,\quad \quad   &\left[\texttt{P}_{a},\texttt{M}_{b}\right]&=-\frac{1}{\ell^2}\left(\epsilon_{am}\texttt{Z}_{mb}+\epsilon_{ab}\texttt{Z}_{mm}\right)\,,\quad  &\left[\texttt{M},\texttt{H}_a\right]&=\frac{1}{\ell^2}\epsilon_{ab}\texttt{S}_{b}\,, \notag\\
\left[\texttt{J},\texttt{Z}_{ab}\right]&=-\epsilon_{m\left(a\right.}\texttt{Z}_{\left.b\right)m}\,, \quad \quad &\left[\texttt{B}_{a},\texttt{J}_{b}\right]&=-\left(\epsilon_{am}\texttt{Z}_{mb}+\epsilon_{ab}\texttt{Z}_{mm}\right)\,,\quad  &\left[\texttt{H},\texttt{Z}_{ab}\right]&=-\epsilon_{m\left(a\right.}\texttt{Y}_{\left.b\right)m}\,, \notag \\
\left[\texttt{J},\texttt{Y}_{ab}\right]&=-\epsilon_{m\left(a\right.}\texttt{Y}_{\left.b\right)m}\,, \quad \quad &\left[\texttt{B}_{a},\texttt{H}_{b}\right]&=-\left(\epsilon_{am}\texttt{Y}_{mb}+\epsilon_{ab}\texttt{Y}_{mm}\right)\,,\quad  &\left[\texttt{H},\texttt{Y}_{ab}\right]&=-\frac{1}{\ell^2}\epsilon_{m\left(a\right.}\texttt{Z}_{\left.b\right)m}\,, \notag \\
\left[\texttt{Z},\texttt{G}_{ab}\right]&=-\epsilon_{m\left(a\right.}\texttt{Z}_{\left.b\right)m}\,, \quad \quad &\left[\texttt{T}_{a},\texttt{J}_{b}\right]&=-\left(\epsilon_{am}\texttt{Y}_{mb}+\epsilon_{ab}\texttt{Y}_{mm}\right)\,,\quad  &\left[\texttt{Y},\texttt{G}_{ab}\right]&=-\epsilon_{m\left(a\right.}\texttt{Y}_{\left.b\right)m}\,, \notag \\
\left[\texttt{Z},\texttt{P}_{ab}\right]&=-\epsilon_{m\left(a\right.}\texttt{Y}_{\left.b\right)m}\,, \quad \quad &\left[\texttt{T}_{a},\texttt{H}_{b}\right]&=-\frac{1}{\ell^2}\left(\epsilon_{am}\texttt{Z}_{mb}+\epsilon_{ab}\texttt{Z}_{mm}\right)\,,\quad  &\left[\texttt{Y},\texttt{P}_{ab}\right]&=-\frac{1}{\ell^2}\epsilon_{m\left(a\right.}\texttt{Z}_{\left.b\right)m}\,, \notag \\
\left[\texttt{S},\texttt{S}_{ab}\right]&=-\epsilon_{m\left(a\right.}\texttt{Z}_{\left.b\right)m}\,, \quad \quad &\left[\texttt{J}_{a},\texttt{S}_{b}\right]&=\epsilon_{ab}\texttt{Z}\,,\quad  &\left[\texttt{M},\texttt{S}_{ab}\right]&=-\epsilon_{m\left(a\right.}\texttt{Y}_{\left.b\right)m}\,, \notag \\
\left[\texttt{S},\texttt{M}_{ab}\right]&=-\epsilon_{m\left(a\right.}\texttt{Y}_{\left.b\right)m}\,, \quad \quad &\left[\texttt{J}_{a},\texttt{M}_{b}\right]&=\epsilon_{ab}\texttt{Y}\,,\quad  &\left[\texttt{M},\texttt{M}_{ab}\right]&=-\frac{1}{\ell^2}\epsilon_{m\left(a\right.}\texttt{Z}_{\left.b\right)m}\,, \notag \\
\left[\texttt{G}_a,\texttt{S}_{bc}\right]&=-\epsilon_{a\left(b\right.}\texttt{S}_{\left.c\right)}\,, \quad \quad &\left[\texttt{H}_{a},\texttt{S}_{b}\right]&=\epsilon_{ab}\texttt{Y}\,,\quad  &\left[\texttt{P}_a,\texttt{S}_{bc}\right]&=-\epsilon_{a\left(b\right.}\texttt{M}_{\left.c\right)}\,, \notag \\
\left[\texttt{G}_a,\texttt{M}_{bc}\right]&=-\epsilon_{a\left(b\right.}\texttt{M}_{\left.c\right)}\,, \quad \quad &\left[\texttt{H}_{a},\texttt{M}_{b}\right]&=\frac{1}{\ell^2}\epsilon_{ab}\texttt{Z}\,,\quad  &\left[\texttt{P}_a,\texttt{M}_{bc}\right]&=-\frac{1}{\ell^2}\epsilon_{a\left(b\right.}\texttt{S}_{\left.c\right)}\,, \notag \\
\left[\texttt{B}_a,\texttt{G}_{bc}\right]&=-\epsilon_{a\left(b\right.}\texttt{S}_{\left.c\right)}\,, \quad \quad &\left[\texttt{J}_a,\texttt{S}_{bc}\right]&=\delta_{a\left(b\right.}\epsilon_{\left.c\right)m}\texttt{B}_{m}\,,\quad  &\left[\texttt{T}_a,\texttt{G}_{bc}\right]&=-\epsilon_{a\left(b\right.}\texttt{M}_{\left.c\right)}\,, \notag \\
\left[\texttt{B}_a,\texttt{P}_{bc}\right]&=-\epsilon_{a\left(b\right.}\texttt{M}_{\left.c\right)}\,, \quad \quad &\left[\texttt{J}_a,\texttt{M}_{bc}\right]&=\delta_{a\left(b\right.}\epsilon_{\left.c\right)m}\texttt{T}_{m}\,,\quad  &\left[\texttt{T}_a,\texttt{P}_{bc}\right]&=-\frac{1}{\ell^2}\epsilon_{a\left(b\right.}\texttt{S}_{\left.c\right)}\,, \notag \\
\left[\texttt{S}_a,\texttt{G}_{bc}\right]&=\delta_{a\left(b\right.}\epsilon_{\left.c\right)m}\texttt{B}_{m}\,, \quad \quad &\left[\texttt{H}_a,\texttt{S}_{bc}\right]&=\delta_{a\left(b\right.}\epsilon_{\left.c\right)m}\texttt{T}_{m}\,,\quad  &\left[\texttt{M}_a,\texttt{G}_{bc}\right]&=\delta_{a\left(b\right.}\epsilon_{\left.c\right)m}\texttt{T}_{m}\,, \notag \\
\left[\texttt{S}_a,\texttt{P}_{bc}\right]&=\delta_{a\left(b\right.}\epsilon_{\left.c\right)m}\texttt{T}_{m}\,, \quad \quad &\left[\texttt{H}_a,\texttt{M}_{bc}\right]&=\frac{1}{\ell^2}\delta_{a\left(b\right.}\epsilon_{\left.c\right)m}\texttt{B}_{m}\,,\quad  &\left[\texttt{M}_a,\texttt{P}_{bc}\right]&=\frac{1}{\ell^2}\delta_{a\left(b\right.}\epsilon_{\left.c\right)m}\texttt{B}_{m}\,, \notag \\
\left[\texttt{S}_{ab},\texttt{S}_{cd}\right]&=\delta_{\left(a\right.\left(c\right.}\epsilon_{\left.d\right)\left.b\right)}\texttt{Z}\,, \quad \quad &\left[\texttt{S}_{ab},\texttt{M}_{cd}\right]&=\delta_{\left(a\right.\left(c\right.}\epsilon_{\left.d\right)\left.b\right)}\texttt{Y}\,,\quad  &\left[\texttt{M}_{ab},\texttt{M}_{cd}\right]&=\frac{1}{\ell^2}\delta_{\left(a\right.\left(c\right.}\epsilon_{\left.d\right)\left.b\right)}\texttt{Z}\,, \notag \\
\left[\texttt{G}_{ab},\texttt{Z}_{cd}\right]&=\delta_{\left(a\right.\left(c\right.}\epsilon_{\left.d\right)\left.b\right)}\texttt{Z}\,, \quad \quad &\left[\texttt{G}_{ab},\texttt{Y}_{cd}\right]&=\delta_{\left(a\right.\left(c\right.}\epsilon_{\left.d\right)\left.b\right)}\texttt{Y}\,,\quad  &\left[\texttt{P}_{ab},\texttt{Z}_{cd}\right]&=\delta_{\left(a\right.\left(c\right.}\epsilon_{\left.d\right)\left.b\right)}\texttt{Y}\,, \notag \\
\left[\texttt{P}_{ab},\texttt{Y}_{cd}\right]&=\frac{1}{\ell^2}\delta_{\left(a\right.\left(c\right.}\epsilon_{\left.d\right)\left.b\right)}\texttt{Z}\,.\label{pn32}
\end{align}
The non-relativistic algebra \eqref{pn32} can be seen as a Post-Newtonian extension of the $\mathfrak{hs_{3}enh2}$ algebra \eqref{enh32}. Interestingly, for $\texttt{Z}=\texttt{Y}=\texttt{Z}_{ab}=\texttt{Y}_{ab}=0$ we obtain the $\mathfrak{hs{3}nhNewt2}$ algebra being an alternative spin-3 extension of the Newton-Hooke version of the Newtonian algebra introduced in \cite{Hansen:2018ofj}.
\section{Alternative Spin-3 extensions of the AdS Carroll family}\label{App2}
An additional spin-3 extensions of the infinite-dimensional AdS Carroll algebra can alternatively be recovered from the expansion of the $\mathfrak{hs_{3}AdS}$ algebra. Indeed, there is an alternative subspace decomposition of the relativistic spin-3 AdS algebra diverse to \eqref{uvsd} that allows us to obtain spin-3 versions of the AdS Carroll symmetries. The two possible subspace decompositions are listed in the following table:
\begin{equation}
    \begin{tabular}{|c|c|}
\hline
 Subspace decomposition 1 & Subspace decomposition 2  \\ \hline\hline
$V_0=\{\hat{\texttt{J}},\hat{\texttt{P}}_{a},\hat{\texttt{J}}_{a},\hat{\texttt{P}}_{ab}\}$ & 
   $V_0=\{\hat{\texttt{J}},\hat{\texttt{P}}_{a},\hat{\texttt{H}}_{a},\hat{\texttt{G}}_{ab}\}$  \\
   $V_1=\{\hat{\texttt{G}}_{a},\hat{\texttt{H}},\hat{\texttt{H}}_{a},\hat{\texttt{G}}_{ab}\}$ &
   $V_1=\{\hat{\texttt{G}}_{a},\hat{\texttt{H}},\hat{\texttt{J}}_{a},\hat{\texttt{P}}_{ab}\}$  \\ \hline
\end{tabular}\notag
\end{equation}
The alternative version of the infinite-dimensional spin-3 AdS Carroll algebra appears by considering the subspace decomposition 2 which, as the subspace decomposition 1 in \eqref{uvsd}, satisfy a $\mathbb{Z}_2$ graded Lie algebra. Then, after considering a resonant $S_{E}^{\left(N\right)}$-expansion with the subset decomposition \eqref{sdN} and applying a $0_S$-reduction we get an alternative spin-3 extension of the infinite-dimensional AdS Carroll algebra. In particular, the ultra-relativistic generators are related to the relativistic ones through the semigroup elements as
\begin{equation}
    \begin{tabular}{|m{10em}|m{10em}|}
\hline
 Expanded generators 1 & Expanded generators 2  \\ \hline\hline
$\texttt{J}^{(m)} =\lambda _{2m}\hat{\texttt{J}}$ & $\texttt{J}^{(m)} =\lambda _{2m}\hat{\texttt{J}}$ \\ $\texttt{H}^{(m)} =\lambda _{2m+1}\hat{\texttt{H}}$ &  $\texttt{H}^{(m)} =\lambda _{2m+1}\hat{\texttt{H}}$
   \\
  $\texttt{G}_{a}^{(m)} =\lambda _{2m+1}\hat{\texttt{G}}_{a}$ & $\texttt{G}_{a}^{(m)} =\lambda _{2m+1}\hat{\texttt{G}}_{a}$ \\ $\texttt{P}_{a}^{(m)} =\lambda_{2m}\hat{\texttt{P}}_{a}$ & $\texttt{P}_{a}^{(m)} =\lambda_{2m}\hat{\texttt{P}}_{a}$
   \\
  $\texttt{G}_{ab}^{(m)} =\lambda _{2m+1}\hat{\texttt{G}}_{ab}$ & $\texttt{G}_{ab}^{(m)} =\lambda _{2m}\hat{\texttt{G}}_{ab}$\\ $\texttt{P}_{ab}^{(m)} =\lambda _{2m}\hat{\texttt{P}}_{ab}$  & $\texttt{P}_{ab}^{(m)} =\lambda _{2m+1}\hat{\texttt{P}}_{ab}$
   \\
   $\texttt{J}_{a}^{(m)} =\lambda _{2m}\hat{\texttt{J}}_{a}$ & $\texttt{J}_{a}^{(m)} =\lambda _{2m+1}\hat{\texttt{J}}_{a}$ \\ $\texttt{H}_{a}^{(m)} =\lambda _{2m+1}\hat{\texttt{H}}_{a}$ & $\texttt{H}_{a}^{(m)} =\lambda _{2m}\hat{\texttt{H}}_{a}$
    \\ \hline
\end{tabular}\notag
\end{equation}
In both resonant expansions, the expanded spin-2 generators satisfy the infinite-dimensional extension of the AdS Carroll algebra \eqref{adscar2N}. However, the expanded spin-3 generators obtained from the subspace decomposition 2 satisfy commutation relations diverse to those obtained from the subspace decomposition 1 in \eqref{adscar3N}. Both spin-3 expanded generators satisfy the following commutators:
\begin{align}
      \left[\texttt{J}^{(m)},\texttt{J}_{a}^{(n)}\right]&=\epsilon_{ab}\texttt{J}_{b}^{(m+n)}\,,&
      \left[\texttt{J}^{(m)},\texttt{H}_{a}^{(n)}\right]&=\epsilon_{ab}\texttt{H}_{b}^{(m+n)}\,, \notag\\
      \left[\texttt{J}_{a}^{(m)},\texttt{G}_{bc}^{(n)}\right]&=\delta_{a\left(b\right.}\epsilon_{\left.c\right)d}\texttt{G}_{d}^{(m+n)}\,, & \left[\texttt{H}_{a}^{(m)},\texttt{P}_{bc}^{(n)}\right]&=\frac{1}{\ell^2}\delta_{a\left(b\right.}\epsilon_{\left.c\right)d}\texttt{G}_{d}^{(m+n)}\,, \notag\\  \left[\texttt{P}_{a}^{(m)},\texttt{J}_{b}^{(n)}\right]&=-\left(\epsilon_{ac}\texttt{P}_{cb}^{(m+n)}+\epsilon_{ab}\texttt{P}_{cc}^{(m+n)}\right)\,, &
       \left[\texttt{J}^{(m)},\texttt{G}_{ab}^{(n)}\right]&=-\epsilon_{c\left(a\right.}\texttt{G}_{\left.b\right)c}^{(m+n)}\,, \notag\\ \left[\texttt{P}_{a}^{(m)},\texttt{H}_{b}^{(n)}\right]&=-\frac{1}{\ell^2}\left(\epsilon_{ac}\texttt{G}_{cb}^{(m+n)}+\epsilon_{ab}\texttt{G}_{cc}^{(m+n)}\right)\,, & \left[\texttt{J}^{(m)},\texttt{P}_{ab}^{(n)}\right]&=-\epsilon_{c\left(a\right.}\texttt{P}_{\left.b\right)c}^{(m+n)}\,,
        \notag\\   \left[\texttt{P}_{a}^{(m)},\texttt{G}_{bc}^{(n)}\right]&=-\epsilon_{a\left(b\right.}\texttt{H}_{\left.c\right)}^{(m+n)}\,, &
       \left[\texttt{J}_{a}^{(m)},\texttt{H}_{b}^{(n)}\right]&=\epsilon_{ab}\texttt{H}^{(m+n)} \,, \notag\\ \left[\texttt{P}_{a}^{(m)},\texttt{P}_{bc}^{(n)}\right]&=-\frac{1}{\ell^2}\epsilon_{a\left(b\right.}\texttt{J}_{\left.c\right)}^{(m+n)}\,, &
       \left[\texttt{G}_{ab}^{(m)},\texttt{P}_{cd}^{(n)}\right]&=\delta_{\left(a\right.\left(c\right.}\epsilon_{\left.d\right)\left.b\right)}\texttt{H}^{(m+n)}\,, \label{adscar3N2}
\end{align}
but differ in the following commutation relations:
\begin{equation}
    \begin{tabular}{|m{19,5em}|m{20em}|}
\hline
 $\mathfrak{hs_{3}adscar1}^{\left(N\right)}$ & $\mathfrak{hs_{3}adscar2}^{\left(N\right)}$  \\ \hline\hline
$[\texttt{H}^{(m)},\texttt{J}_{a}^{(n)}]=\epsilon_{ab}\texttt{H}_{b}^{(m+n)}$ & $[\texttt{H}^{(m)},\texttt{J}_{a}^{(n)}]=\epsilon_{ab}\texttt{H}_{b}^{(m+n+1)}$ \\
  $[\texttt{H}^{(m)},\texttt{H}_{a}^{(n)}]=\frac{1}{\ell^2}\epsilon_{ab}\texttt{J}_{b}^{(m+n+1)}$ & $[\texttt{H}^{(m)},\texttt{H}_{a}^{(n)}]=\frac{1}{\ell^2}\epsilon_{ab}\texttt{J}_{b}^{(m+n)}$ \\ $[\texttt{G}_{a}^{(m)},\texttt{J}_{b}^{(n)}]=-\left(\epsilon_{ac}\texttt{G}_{cb}^{(m+n)}+\epsilon_{ab}\texttt{G}_{cc}^{(m+n)}\right)$ & $\left[\texttt{G}_{a}^{(m)},\texttt{J}_{b}^{(n)}\right]=-\left(\epsilon_{ac}\texttt{G}_{cb}^{(m+n+1)}+\epsilon_{ab}\texttt{G}_{cc}^{(m+n+1)}\right)$
   \\
  $[\texttt{G}_{a}^{(m)},\texttt{H}_{b}^{(n)}]=-\left(\epsilon_{ac}\texttt{P}_{cb}^{(m+n+1)}+\epsilon_{ab}\texttt{P}_{cc}^{(m+n+1)}\right)$ & $[\texttt{G}_{a}^{(m)},\texttt{H}_{b}^{(n)}]=-\left(\epsilon_{ac}\texttt{P}_{cb}^{(m+n)}+\epsilon_{ab}\texttt{P}_{cc}^{(m+n)}\right) $ \\ $[\texttt{J}_{a}^{(m)},\texttt{P}_{bc}^{(n)}]=\delta_{a\left(b\right.}\epsilon_{\left.c\right)d}\texttt{P}_{d}^{(m+n)}$  & $[\texttt{J}_{a}^{(m)},\texttt{P}_{bc}^{(n)}]=\delta_{a\left(b\right.}\epsilon_{\left.c\right)d}\texttt{P}_{d}^{(m+n+1)}$
   \\
   $[\texttt{H}_{a}^{(m)},\texttt{G}_{bc}^{(n)}]=\delta_{a\left(b\right.}\epsilon_{\left.c\right)d}\texttt{P}_{d}^{(m+n+1)}$ & $[\texttt{H}_{a}^{(m)},\texttt{G}_{bc}^{(n)}]=\delta_{a\left(b\right.}\epsilon_{\left.c\right)d}\texttt{P}_{d}^{(m+n)}$ \\ $[\texttt{G}_{a}^{(m)},\texttt{G}_{bc}^{(n)}]=-\epsilon_{a\left(b\right.}\texttt{J}_{\left.c\right)}^{(m+n+1)}$ & $[\texttt{G}_{a}^{(m)},\texttt{G}_{bc}^{(n)}]=-\epsilon_{a\left(b\right.}\texttt{J}_{\left.c\right)}^{(m+n)}$
    \\ 
    $[\texttt{G}_{a}^{(m)},\texttt{P}_{bc}^{(n)}]=-\epsilon_{a\left(b\right.}\texttt{H}_{\left.c\right)}^{(m+n)}$ & $[\texttt{G}_{a}^{(m)},\texttt{P}_{bc}^{(n)}]=-\epsilon_{a\left(b\right.}\texttt{H}_{\left.c\right)}^{(m+n+1)}$ \\
    $[\texttt{H}^{(m)},\texttt{G}_{ab}^{(n)}]=-\epsilon_{c\left(a\right.}\texttt{P}_{\left.b\right)c}^{(m+n+1)}$ & $[\texttt{H}^{(m)},\texttt{G}_{ab}^{(n)}]=-\epsilon_{c\left(a\right.}\texttt{P}_{\left.b\right)c}^{(m+n)}$ \\
    $[\texttt{H}^{(m)},\texttt{P}_{ab}^{(n)}]=-\frac{1}{\ell^2}\epsilon_{c\left(a\right.}\texttt{G}_{\left.b\right)c}^{(m+n)}$ & $[\texttt{H}^{(m)},\texttt{P}_{ab}^{(n)}]=-\frac{1}{\ell^2}\epsilon_{c\left(a\right.}\texttt{G}_{\left.b\right)c}^{(m+n+1)}$ \\
    $[\texttt{J}_{a}^{(m)},\texttt{J}_{b}^{(n)}]=\epsilon_{ab}\texttt{J}^{(m+n)}$ & $[\texttt{J}_{a}^{(m)},\texttt{J}_{b}^{(n)}]=\epsilon_{ab}\texttt{J}^{(m+n+1)}$ \\
    $[\texttt{G}_{ab}^{(m)},\texttt{G}_{cd}^{(n)}]=\delta_{\left(a\right.\left(c\right.}\epsilon_{\left.d\right)\left.b\right)}\texttt{J}^{(m+n+1)}$ & $[\texttt{G}_{ab}^{(m)},\texttt{G}_{cd}^{(n)}]=\delta_{\left(a\right.\left(c\right.}\epsilon_{\left.d\right)\left.b\right)}\texttt{J}^{(m+n)}$ \\
    $[\texttt{H}_{a}^{(m)},\texttt{H}_{b}^{(n)}]=\frac{1}{\ell^2}\epsilon_{ab}\texttt{J}^{(m+n+1)}$ & $[\texttt{H}_{a}^{(m)},\texttt{H}_{b}^{(n)}]=\frac{1}{\ell^2}\epsilon_{ab}\texttt{J}^{(m+n)}$ \\
    $[\texttt{P}_{ab}^{(m)},\texttt{P}_{cd}^{(n)}]=\frac{1}{\ell^2}\delta_{\left(a\right.\left(c\right.}\epsilon_{\left.d\right)\left.b\right)}\texttt{J}^{(m+n)}$ & $[\texttt{P}_{ab}^{(m)},\texttt{P}_{cd}^{(n)}]=\frac{1}{\ell^2}\delta_{\left(a\right.\left(c\right.}\epsilon_{\left.d\right)\left.b\right)}\texttt{J}^{(m+n+1)}$ \\
    \hline
\end{tabular}\notag
\end{equation}
For finite values of $N$ the difference between both spin-3 extensions of the generalized AdS Carroll implies that distinct sectors of the algebra are abelian. In particular, for $N=1$, the resonant $S_{E}^{\left(1\right)}$-expansion of the spin-3 AdS algebra with the subspace decomposition 2 produces an alternative spin-3 AdS Carroll which coincides with the $\mathfrak{hs_{3}ppoi2}$ introduced in \cite{Bergshoeff:2016soe}. For $N\geq 2$, the spin-3 algebras are new and are extension of the $\mathfrak{hs_{3}adscar2}$ algebra.  For instance, for $N=2$, the $\mathfrak{hs_{3}adscar2}^{\left(2\right)}$ is spanned by the set $\{\texttt{J},\texttt{P}_a,\texttt{H}_a,\texttt{G}_{ab},\texttt{G}_{a},\texttt{H},\texttt{J}_a,\texttt{P}_{ab},\texttt{S},\texttt{T}_a,\texttt{Y}_a,\texttt{S}_{ab}\}$ which satisfy
\begin{align}
    \left[\texttt{J},\texttt{G}_{a}\right]&=\epsilon_{ab}\texttt{G}_{b}\,, & \left[\texttt{G}_{a},\texttt{P}_{b}\right]&=-\epsilon_{ab}\texttt{H}\,, & 
    \left[\texttt{H},\texttt{P}_{a}\right]&=\frac{1}{\ell^{2}}\epsilon_{ab}\texttt{G}_{b}\,,\notag \\
     \left[\texttt{J},\texttt{P}_{a}\right]&=\epsilon_{ab}\texttt{P}_{b}\,, & 
     \left[\texttt{P}_{a},\texttt{P}_{b}\right]&=-\frac{1}{\ell^{2}}\epsilon_{ab}\texttt{J}\,, & 
     \left[\texttt{H},\texttt{J}_{a}\right]&=\epsilon_{ab}\texttt{Y}_{b}\,, \notag\\
     \left[\texttt{J},\texttt{T}_{a}\right]&=\epsilon_{ab}\texttt{T}_{b}\,, & \left[\texttt{G}_{a},\texttt{G}_{b}\right]&=-\epsilon_{ab}\texttt{S}\,, & 
    \left[\texttt{H},\texttt{H}_{a}\right]&=\frac{1}{\ell^{2}}\epsilon_{ab}\texttt{J}_{b}\,,\notag \\
     \left[\texttt{J},\texttt{Y}_{a}\right]&=\epsilon_{ab}\texttt{Y}_{b}\,, & 
     \left[\texttt{P}_{a},\texttt{T}_{b}\right]&=-\frac{1}{\ell^{2}}\epsilon_{ab}\texttt{S}\,, & 
     \left[\texttt{H},\texttt{G}_{a}\right]&=\epsilon_{ab}\texttt{T}_{b}\,, \notag\\
      \left[\texttt{J},\texttt{H}_{a}\right]&=\epsilon_{ab}\texttt{H}_{b}\,, & 
      \left[\texttt{G}_{a},\texttt{J}_{b}\right]&=-\left(\epsilon_{am}\texttt{S}_{mb}+\epsilon_{ab}\texttt{S}_{mm}\right)\,, & 
      \left[\texttt{H},\texttt{P}_{ab}\right]&=-\frac{1}{\ell^2}\epsilon_{m\left(a\right.}\texttt{S}_{\left.b\right)m}\,,\notag \\
      \left[\texttt{J},\texttt{J}_{a}\right]&=\epsilon_{ab}\texttt{J}_{b}\,, & 
      \left[\texttt{P}_{a},\texttt{H}_{b}\right]&=-\frac{1}{\ell^2}\left(\epsilon_{am}\texttt{G}_{mb}+\epsilon_{ab}\texttt{G}_{mm}\right)\,, & \left[\texttt{P}_{a},\texttt{P}_{bc}\right]&=-\frac{1}{\ell^2}\epsilon_{a\left(b\right.}\texttt{J}_{\left.c\right)}\,,\notag \\
      \left[\texttt{J},\texttt{P}_{ab}\right]&=-\epsilon_{m\left(a\right.}\texttt{P}_{\left.b\right)m}\,, & \left[\texttt{P}_{a},\texttt{J}_{b}\right]&=-\left(\epsilon_{am}\texttt{P}_{mb}+\epsilon_{ab}\texttt{P}_{mm}\right)\,, & \left[\texttt{P}_{a},\texttt{G}_{bc}\right]&=-\epsilon_{a\left(b\right.}\texttt{H}_{\left.c\right)}\,, \notag \\
      \left[\texttt{S},\texttt{H}_{a}\right]&=\epsilon_{ab}\texttt{Y}_{b}\,, & 
      \left[\texttt{G}_{a},\texttt{H}_{b}\right]&=-\left(\epsilon_{am}\texttt{P}_{mb}+\epsilon_{ab}\texttt{P}_{mm}\right)\,, & 
      \left[\texttt{H},\texttt{G}_{ab}\right]&=-\epsilon_{m\left(a\right.}\texttt{P}_{\left.b\right)m}\,,\notag \\
      \left[\texttt{S},\texttt{P}_{a}\right]&=\epsilon_{ab}\texttt{T}_{b}\,, & 
      \left[\texttt{P}_{a},\texttt{Y}_{b}\right]&=-\frac{1}{\ell^2}\left(\epsilon_{am}\texttt{S}_{mb}+\epsilon_{ab}\texttt{S}_{mm}\right)\,, & \left[\texttt{G}_{a},\texttt{G}_{bc}\right]&=-\epsilon_{a\left(b\right.}\texttt{J}_{\left.c\right)}\,,\notag \\
      \left[\texttt{J},\texttt{S}_{ab}\right]&=-\epsilon_{m\left(a\right.}\texttt{S}_{\left.b\right)m}\,, & \left[\texttt{T}_{a},\texttt{H}_{b}\right]&=-\frac{1}{\ell^2}\left(\epsilon_{am}\texttt{S}_{mb}+\epsilon_{ab}\texttt{S}_{mm}\right)\,, & \left[\texttt{T}_{a},\texttt{G}_{bc}\right]&=-\frac{1}{\ell^2}\epsilon_{a\left(b\right.}\texttt{Y}_{\left.c\right)}\,, \notag \\
      \left[\texttt{J},\texttt{G}_{ab}\right]&=-\epsilon_{m\left(a\right.}\texttt{G}_{\left.b\right)m}\,, &  \left[\texttt{J}_{a},\texttt{H}_{b}\right]&=\epsilon_{ab}\texttt{H}\,, & \left[\texttt{J}_{a},\texttt{P}_{bc}\right]&=\delta_{a\left(b\right.}\epsilon_{\left.c\right)m}\texttt{T}_{m}\,, \notag \\
       \left[\texttt{G}_{a},\texttt{P}_{bc}\right]&=-\epsilon_{a\left(b\right.}\texttt{Y}_{\left.c\right)}\,, &  \left[\texttt{J}_{a},\texttt{J}_{b}\right]&=\epsilon_{ab}\texttt{S}\,, &
       \left[\texttt{J}_{a},\texttt{G}_{bc}\right]&=\delta_{a\left(b\right.}\epsilon_{\left.c\right)m}\texttt{G}_{m}\,, \notag \\
      \left[\texttt{S},\texttt{G}_{ab}\right]&=-\epsilon_{m\left(a\right.}\texttt{S}_{\left.b\right)m}\,, &  \left[\texttt{H}_{a},\texttt{Y}_{b}\right]&=\frac{1}{\ell^2}\epsilon_{ab}\texttt{S}\,, & \left[\texttt{H}_{a},\texttt{S}_{bc}\right]&=\delta_{a\left(b\right.}\epsilon_{\left.c\right)m}\texttt{T}_{m}\,, \notag \\
       \left[\texttt{P}_{a},\texttt{S}_{bc}\right]&=-\frac{1}{\ell^2}\epsilon_{a\left(b\right.}\texttt{Y}_{\left.c\right)}\,, &  \left[\texttt{H}_{a},\texttt{H}_{b}\right]&=\frac{1}{\ell^2}\epsilon_{ab}\texttt{J}\,, &
       \left[\texttt{H}_{a},\texttt{G}_{bc}\right]&=\delta_{a\left(b\right.}\epsilon_{\left.c\right)m}\texttt{P}_{m}\,, \notag \\
       \left[\texttt{H}_{a},\texttt{P}_{bc}\right]&=\frac{1}{\ell^2}\delta_{a\left(b\right.}\epsilon_{\left.c\right)m}\texttt{G}_{m}\,, &  \left[\texttt{G}_{ab},\texttt{P}_{cd}\right]&=\delta_{\left(a\right.\left(c\right.}\epsilon_{\left.d\right)\left.b\right)}\texttt{H}\,, & \left[\texttt{P}_{ab},\texttt{P}_{cd}\right]&=\frac{1}{\ell^2}\delta_{\left(a\right.\left(c\right.}\epsilon_{\left.d\right)\left.b\right)}\texttt{S}\,, \notag\\
       \left[\texttt{Y}_{a},\texttt{G}_{bc}\right]&=\delta_{a\left(b\right.}\epsilon_{\left.c\right)m}\texttt{T}_{m}\,, &  \left[\texttt{G}_{ab},\texttt{G}_{cd}\right]&=\delta_{\left(a\right.\left(c\right.}\epsilon_{\left.d\right)\left.b\right)}\texttt{J}\,, & \left[\texttt{G}_{ab},\texttt{S}_{cd}\right]&=\delta_{\left(a\right.\left(c\right.}\epsilon_{\left.d\right)\left.b\right)}\texttt{S}\,.\label{sp3adscar2}
\end{align}
The present algebra, which we denote as $\mathfrak{hs_{3}eadscar2}$, contains several different commutators regarding the $\mathfrak{hs_{3}eadscar}$ obtained in \eqref{sp3adscar1} and \eqref{sp3adscar}. In the vanishing cosmological constant limit $\ell\rightarrow\infty$, the algebra reduces to an alternative spin-3 extended Carroll algebra which can be seen as an extension of the $\mathfrak{hs_{3}car2}$ introduced in \cite{Bergshoeff:2016soe}.
\section{Explicit curvature two-forms for infinite-dimensional extensions of spin-3 Newton-Hooke and spin-3 AdS Carroll}\label{App3}
This appendix contains the explicit expressions of the curvature two-forms for the $\mathfrak{hs_{3}nh}^{(N)}$ and the $\mathfrak{hs_{3}adscar}^{(N)}$ algebras. For the spin-3 extension of the infinite-dimensional Newton-Hooke algebra, the curvature two-forms read
\begin{eqnarray}
R\left(\omega^{(m)}\right)&=& d\omega^{(m)} + \frac{1}{2}\sum_{n,l=0}^{\left[\frac{N+1}{2}\right]}\left(\epsilon^{ac} B_a^{(n)} B_c^{(l)} + \frac{1}{\ell^2}\epsilon^{ac}e_{a}^{(n)}e_{c}^{(l)}+\epsilon^{\left(a\right.\left(c\right.}\delta^{\left.d\right)\left.b\right)}B_{ad}^{(n)}B_{cb}^{(l)}\right.\notag \\
&&\left.+\frac{1}{\ell^2}\epsilon^{\left(a\right.\left(c\right.}\delta^{\left.d\right)\left.b\right)}e_{ad}^{(n)}e_{cb}^{(l)}\right)\delta_{n+l+1}^{m}-\frac{1}{2}\sum_{n,l=0}^{\left[\frac{N}{2}\right]}\left(\epsilon^{ac}\omega_{a}^{(n)}\omega_{c}^{(l)}-\frac{1}{\ell^2}\epsilon^{ac}\tau_{a}^{(n)}\tau_{c}^{(l)}\right)\delta_{n+l}^{m}\notag\\
R^{a}\left(B^{b(m)}\right)&=&dB^{a(m)}+\sum_{n,l=0}^{\left[\frac{N}{2}\right]}\left(\epsilon^{ac}\omega^{(n)} B_{c}^{(l)}+\frac{1}{\ell^2}\epsilon^{ac}\tau^{(n)} e_{c}^{(l)}+\epsilon^{a\left(c\right.}\delta^{\left.b\right)d}\omega_{d}^{(n)}B_{bc}^{(l)}\right.\notag \\
&&\left.+ \frac{1}{\ell^2}\epsilon^{a\left(c\right.}\delta^{\left.b\right)d}\tau_{d}^{(n)}e_{bc}^{(l)}\right)\delta_{n+l}^{m} \,, \notag \\
R\left(\tau^{(m)}\right)&=&d\tau^{(m)}+\sum_{n,l=0}^{\left[\frac{N+1}{2}\right]}\left(\epsilon^{ac}B_{a}^{(n)}e_{c}^{(l)}+\epsilon^{\left(a\right.\left(c\right.}\delta^{\left.d\right)\left.b\right)}e_{ad}^{(n)}B_{cb}^{(l)}\right)\delta_{n+l+1}^{m}\notag\\&&-\sum_{n,l=0}^{\left[\frac{N}{2}\right]}\left(\epsilon^{ac}\omega_{a}^{(n)}\tau_{c}^{(l)}\right)\delta_{n+l}^{m} \,, \notag \\
R^{a}\left(e^{b(m)}\right)&=&de^{a(m)}+\sum_{n,l=0}^{\left[\frac{N}{2}\right]}\left(\epsilon^{ac}\omega^{(n)} e_{c}^{(l)}+\epsilon^{ac}\tau^{(n)} B_{c}^{(l)} + \epsilon^{a\left(c\right.}\delta^{\left.b\right)d}\tau_{d}^{(n)}B_{bc}^{(l)}\right.\notag\\
&&\left.+\epsilon^{a\left(c\right.}\delta^{\left.b\right)d}\omega_{d}^{(n)}e_{bc}^{(l)}\right)\delta_{n+l}^{m}\,, \notag\\
R^{a}\left(\omega^{b(m)}\right)&=&d\omega^{a(m)}+\sum_{n,l=0}^{\left[\frac{N+1}{2}\right]}\left(\epsilon^{c\left(d\right.}B_{c}^{(n)}B_{d}^{\ \left.a\right)\, (l)}+\frac{1}{\ell^2}\epsilon^{c\left(d\right.}e_{c}^{(n)}e_{d}^{\ \left.a\right)\,(l)}\right)\delta_{n+l+1}^{m}\notag\\
&&+\sum_{n,l=0}^{\left[\frac{N}{2}\right]}\left(\epsilon^{ac}\omega^{(n)}\omega_{c}^{(l)}+\frac{1}{\ell^2}\epsilon^{ac}\tau^{(n)}\tau_{c}^{(l)} \right)\delta_{n+l}^{m} \,, \notag \\
R^{ab}\left(B^{cd(m)}\right)&=& dB^{ab(m)}+\sum_{n,l=0}^{\left[\frac{N}{2}\right]}\left(\epsilon^{\left(a\right|c}\omega^{(n)} B_{c}^{\ \left|b\right)\,(l)}  +\frac{1}{\ell^2}\epsilon^{\left(a\right|c}\tau^{(n)} e_{c}^{\ \left|b\right)\,(l)} +\epsilon^{ac}\omega^{b(n)}B_{c}^{(l)}\right.\notag\\
&&\left.+\frac{1}{\ell^2}\epsilon^{ac}\tau^{b(n)}e_{c}^{(l)}+\delta^{ab}\epsilon^{cd}\omega_{c}^{(n)}B_{d}^{(l)}+\frac{1}{\ell^2}\delta^{ab}\epsilon^{cd}\tau_{c}^{(n)}e_{d}^{(l)}\right)\delta_{n+l}^{m}\,, \notag \\
R^{a}\left(\tau^{b(m)}\right)&=&d\tau^{a(m)} +\sum_{n,l=0}^{\left[\frac{N+1}{2}\right]}\left(\epsilon^{c\left(d\right.}B_{c}^{(n)}e_{d}^{\ \left.a\right)\,(l)} +\epsilon^{c\left(d\right.}e_{c}^{(n)}B_{d}^{\ \left.a\right)\,(l)}\right)\delta_{n+l+1}^{m}\notag\\
&&+\sum_{n,l=0}^{\left[\frac{N}{2}\right]}\left( \epsilon^{ac}\omega^{(n)} \tau_{c}^{(l)} +\epsilon^{ac}\tau^{(n)}\omega_{c}^{(l)}\right)\delta_{n+l}^{m} \,, \notag\\
R^{ab}\left(e^{cd(m)}\right)&=& de^{ab(m)}+\sum_{n,l=0}^{\left[\frac{N}{2}\right]}\left(\epsilon^{\left(a\right|c}\omega^{(n)} e_{c}^{\ \left|b\right)\,(l)} + \epsilon^{\left(a\right|c}\tau^{(n)} B_{c}^{\ \left|b\right)\,(l)} +\epsilon^{ac}\tau^{b(n)}B_{c}^{(l)}+\epsilon^{ac}\omega^{b(n)}e_{c}^{(l)}\right. \notag\\
&&\left.+\delta^{ab}\epsilon^{cd}\tau_{c}^{(n)}B_{d}^{(l)}+\delta^{ab}\epsilon^{cd}\omega_{c}^{(n)}e_{d}^{(l)}\right)\delta_{n+l}^{m}\,.  \label{infhs3nh}
\end{eqnarray}
One can note that the vanishing cosmological constant limit $\ell\rightarrow\infty$ leads us to the curvature two-forms for the $\mathfrak{hs_{3}gal}^{(N)}$ algebra. Naturally, for $N=\infty$ we obtain the curvature for the infinite-dimensional spin-3 Newton-Hooke. The curvatures for the spin-3 extended Newton-Hooke along its flat limit are recovered for $N=2$. For $N=4$, we obtain the curvature two-forms for the post-Newtonian extension of the spin-3 extended Newton-Hooke and its flat limit: the spin-3 extension of the extended Newtonian symmetry which we have denoted as $\mathfrak{hs_{3}eNewt}$.

In the ultra-relativistic regime, the curvature two-forms for the spin-3 extension of the infinite-dimensional AdS Carroll algebra are given by
\begin{eqnarray}
R\left(\omega^{(m)}\right)&=& d\omega^{(m)} + \frac{1}{2}\sum_{n,l=0}^{\left[\frac{N+1}{2}\right]}\left(\epsilon^{ac} B_a^{(n)} B_c^{(l)}-\frac{1}{\ell^2}\epsilon^{ac}\tau_{a}^{(n)}\tau_{c}^{(l)}+\epsilon^{\left(a\right.\left(c\right.}\delta^{\left.d\right)\left.b\right)}B_{ad}^{(n)}B_{cb}^{(l)}\right)\delta_{n+l+1}^{m}\notag \\
&&-\frac{1}{2}\sum_{n,l=0}^{\left[\frac{N}{2}\right]}\left(\epsilon^{ac}\omega_{a}^{(n)}\omega_{c}^{(l)}+\frac{1}{\ell^2}\epsilon^{ac}e_{a}^{(n)}e_{c}^{(l)}+\frac{1}{\ell^2}\epsilon^{\left(a\right.\left(c\right.}\delta^{\left.d\right)\left.b\right)}e_{ad}^{(n)}e_{cb}^{(l)}\right)\delta_{n+l}^{m}\notag\\
R^{a}\left(B^{b(m)}\right)&=&dB^{a(m)}+\sum_{n,l=0}^{\left[\frac{N}{2}\right]}\left(\epsilon^{ac}\omega^{(n)} B_{c}^{(l)}+\frac{1}{\ell^2}\epsilon^{ac}\tau^{(n)} e_{c}^{(l)}+\epsilon^{a\left(c\right.}\delta^{\left.b\right)d}\omega_{d}^{(n)}B_{bc}^{(l)}\right.\notag \\
&&\left.+ \frac{1}{\ell^2}\epsilon^{a\left(c\right.}\delta^{\left.b\right)d}\tau_{d}^{(n)}e_{bc}^{(l)}\right)\delta_{n+l}^{m} \,, \notag \\
R\left(\tau^{(m)}\right)&=&d\tau^{(m)}+\sum_{n,l=0}^{\left[\frac{N}{2}\right]}\left(\epsilon^{ac}B_{a}^{(n)}e_{c}^{(l)}+\epsilon^{\left(a\right.\left(c\right.}\delta^{\left.d\right)\left.b\right)}e_{ad}^{(n)}B_{cb}^{(l)}-\epsilon^{ac}\omega_{a}^{(n)}\tau_{c}^{(l)}\right)\delta_{n+l}^{m}\notag\\
R^{a}\left(e^{b(m)}\right)&=&de^{a(m)}+\sum_{n,l=0}^{\left[\frac{N+1}{2}\right]}\left(\epsilon^{ac}\tau^{(n)} B_{c}^{(l)} + \epsilon^{a\left(c\right.}\delta^{\left.b\right)d}\tau_{d}^{(n)}B_{bc}^{(l)}\right)\delta_{n+l+1}^{m}\notag\\
&&+\sum_{n,l=0}^{\left[\frac{N}{2}\right]}\left(\epsilon^{ac}\omega^{(n)} e_{c}^{(l)}+\epsilon^{a\left(c\right.}\delta^{\left.b\right)d}\omega_{d}^{(n)}e_{bc}^{(l)}\right)\delta_{n+l}^{m}\,, \notag\\
R^{a}\left(\omega^{b(m)}\right)&=&d\omega^{a(m)}+\sum_{n,l=0}^{\left[\frac{N+1}{2}\right]}\left(\epsilon^{c\left(d\right.}B_{c}^{(n)}B_{d}^{\ \left.a\right)\, (l)}+\frac{1}{\ell^2}\epsilon^{ac}\tau^{(n)}\tau_{c}^{(l)}\right)\delta_{n+l+1}^{m}\notag\\
&&+\sum_{n,l=0}^{\left[\frac{N}{2}\right]}\left(\epsilon^{ac}\omega^{(n)}\omega_{c}^{(l)}+\frac{1}{\ell^2}\epsilon^{c\left(d\right.}e_{c}^{(n)}e_{d}^{\ \left.a\right)\,(l)} \right)\delta_{n+l}^{m} \,, \notag \\
R^{ab}\left(B^{cd(m)}\right)&=& dB^{ab(m)}+\sum_{n,l=0}^{\left[\frac{N}{2}\right]}\left(\epsilon^{\left(a\right|c}\omega^{(n)} B_{c}^{\ \left|b\right)\,(l)}  +\frac{1}{\ell^2}\epsilon^{\left(a\right|c}\tau^{(n)} e_{c}^{\ \left|b\right)\,(l)} +\epsilon^{ac}\omega^{b(n)}B_{c}^{(l)}\right.\notag\\
&&\left.+\frac{1}{\ell^2}\epsilon^{ac}\tau^{b(n)}e_{c}^{(l)}+\delta^{ab}\epsilon^{cd}\omega_{c}^{(n)}B_{d}^{(l)}+\frac{1}{\ell^2}\delta^{ab}\epsilon^{cd}\tau_{c}^{(n)}e_{d}^{(l)}\right)\delta_{n+l}^{m}\,, \notag \\
R^{a}\left(\tau^{b(m)}\right)&=&d\tau^{a(m)} +\sum_{n,l=0}^{\left[\frac{N}{2}\right]}\left(\epsilon^{c\left(d\right.}B_{c}^{(n)}e_{d}^{\ \left.a\right)\,(l)} +\epsilon^{c\left(d\right.}e_{c}^{(n)}B_{d}^{\ \left.a\right)\,(l)}+\epsilon^{ac}\omega^{(n)} \tau_{c}^{(l)}+\epsilon^{ac}\tau^{(n)}\omega_{c}^{(l)}\right)\delta_{n+l}^{m}\notag\\
R^{ab}\left(e^{cd(m)}\right)&=& de^{ab(m)}+\sum_{n,l=0}^{\left[\frac{N+1}{2}\right]}\left(\epsilon^{\left(a\right|c}\tau^{(n)} B_{c}^{\ \left|b\right)\,(l)} +\epsilon^{ac}\tau^{b(n)}B_{c}^{(l)}+\delta^{ab}\epsilon^{cd}\tau_{c}^{(n)}B_{d}^{(l)}\right)\delta_{n+l+1}^{m}\notag\\
&&+\sum_{n,l=0}^{\left[\frac{N}{2}\right]}\left(\epsilon^{\left(a\right|c}\omega^{(n)} e_{c}^{\ \left|b\right)\,(l)}+\epsilon^{ac}\omega^{b(n)}e_{c}^{(l)}+\delta^{ab}\epsilon^{cd}\omega_{c}^{(n)}e_{d}^{(l)}\right)\delta_{n+l}^{m}\,.  \label{infhs3adscar}
\end{eqnarray}
 Here, in the vanishing cosmological constant limit $\ell\rightarrow\infty$, we get the curvatures for the spin-3 extension of the infinite-dimensional Carroll algebra. For $N=1$, the curvature two-forms correspond to the spin-3 AdS Carroll ones introduced in \eqref{curadscar} along its flat limit. 

Although the differences between the curvatures for the $\mathfrak{hs_{3}adscar}^{(N)}$ and $\mathfrak{hs_{3}nh}^{(N)}$ algebras seems subtle, they are quite different leading to diverse field equations. For $N=2$, in which the $\mathfrak{hs_{3}nh}^{(N)}$ algebra admits a non-degenerate invariant trace, the equations of motion of the CS actions for both non-relativistic and ultra-relativistic spin-3 symmetries are given by the vanishing of the curvatures. However, as we can notice, there are several differences with various physical implications at the dynamical level. For instance, we have that the cosmological term $\frac{1}{\ell^2}\epsilon^{ac}e_{a}e_{c}$ contributes to the curvature $R\left(\omega\right)$ of the spin-3 extended AdS Carroll symmetry which in the spin-3 extended Newton-Hooke case does not occur (see \eqref{curvhs3enh}).

\bibliographystyle{fullsort.bst}
\bibliography{NR_and_UR_spin3_rev}

\end{document}